\documentclass[amsmath,amssymb,reprint,aps,prd,eqsecnum,balancelastpage]{revtex4-2}
 
\usepackage{graphicx}
\usepackage{dcolumn}
\usepackage{bm}
\usepackage{mathrsfs}
\usepackage{color}
\usepackage[normalem]{ulem}
\usepackage{amsmath}
\usepackage{hyperref}
\usepackage{comment}

\allowdisplaybreaks

\DeclareMathOperator{\cosech}{cosech}

\begin{document}

\title{Quantum scalar field theory on equal-angular-momenta Myers-Perry-AdS black holes}

\author{Alessandro Monteverdi}

\email{AMonteverdi1@sheffield.ac.uk}

\author{Elizabeth Winstanley}

\email{E.Winstanley@sheffield.ac.uk}

\affiliation{
School of Mathematical and Physical Sciences,
The University of Sheffield,
Hicks Building,
Hounsfield Road,
Sheffield. S3 7RH United Kingdom}

\date{\today}

\begin{abstract}
We study the canonical quantization of a massive scalar field on a five dimensional, rotating black hole space-time. 
We focus on the case where the space-time is asymptotically anti-de Sitter and the black hole's two angular momentum parameters are equal.
In this situation the geometry possesses additional symmetries which simplify both the mode solutions of the scalar field equation and the stress-energy tensor.
When the angular momentum of the black hole is sufficiently small that there is no speed-of-light surface, there exists a Killing vector which is time-like in the region exterior to the event horizon.
In this case classical superradiance is absent and we construct analogues of the usual Boulware and Hartle-Hawking quantum states for the quantum scalar field.
We compute the differences in expectation values of the square of the quantum scalar field operator and the stress-energy tensor operator between these two quantum states. 
\end{abstract}

\maketitle

\section{Introduction}
\label{sec:intro}

Before compelling observational evidence demonstrated the existence of rotating black holes in our Universe, theorists had obtained not only a classical metric describing a rotating black hole \cite{Kerr:1963ud} but also investigated quantum processes on this space-time background \cite{Starobinsky:1973aij,Unruh:1974bw}. 
Notably, even before the remarkable discovery that {\em {all}} nonextremal black holes emit thermal quantum radiation \cite{Hawking:1975vcx}, it was known that quantum particles spontaneously emanate from rotating black holes \cite{Starobinsky:1973aij,Unruh:1974bw}. 

These profound discoveries continue, more than fifty years later, to stimulate research into the behaviour of quantum fields on rotating black hole space-times. 
However, any such investigation is technically extremely challenging,
even though the Kerr black hole \cite{Kerr:1963ud} is axisymmetric and the Teukolsky equation governing linear, classical perturbations of this space-time is separable \cite{Teukolsky:1973ha}.

Of central importance in studying quantum processes on black hole space-times is the expectation value of the stress-energy tensor (SET) operator, which 
encodes detailed information about the Hawking emission and
governs the back-reaction of the quantum field on the space-time geometry. 
Computations of the SET expectation value on static, spherically symmetric black holes have been undertaken since the 1980s \cite{Howard:1984qp, Howard:1984ttx, Anderson:1993if, Anderson:1994hg} and have received fresh impetus in the past decade with the development of new methodologies \cite{Levi:2016quh, Levi:2016paz, Taylor:2022sly, Arrechea:2024cnv} which both improve the efficiency of the numerical calculations and permit the examination of the SET on a  wider range of black hole space-times.  

It is striking that, despite the long history of the subject and the deep motivation for studies of the SET expectation value, its full computation on a Kerr black hole has been performed only comparatively recently \cite{Levi:2016exv}. 
This is because the original techniques \cite{Howard:1984qp, Howard:1984ttx, Anderson:1993if, Anderson:1994hg} were adapted to static and spherically symmetric space-times and were not easily extended to the reduced symmetry of a rotating black hole geometry.
The sophisticated methodology employed in \cite{Levi:2016exv} still faces formidable practical challenges, for example some four million scalar field modes are required to produce those results. 

These difficulties provide compelling motivation for the study of quantum fields on alternative rotating black hole geometries which reduce the technical complexities. 
One line of attack is to lower the number of space-time dimensions to three. 
The rotating BTZ black hole \cite{Banados:1992wn,Banados:1992gq} is constructed by identifying points in three-dimensional anti-de Sitter (AdS) space-time. 
As a result, the SET expectation value of a quantum scalar field on this background can be found using the method of images \cite{Steif:1993zv}. 
The consequent SET expectation value is sufficiently simple in form to enable the back-reaction of the quantum field on the space-time to be studied in depth \cite{Casals:2016odj,Casals:2019jfo}. 

Naively, one would anticipate that increasing the number of space-time dimensions only aggravates the problem of the complexity of rotating black hole geometries. 
However, when the number of space-time dimensions is odd, there exist rotating black hole metrics with {\em {enhanced}} symmetries \cite{Kunduri:2006qa}. 
Black holes in $2N+3$ space-time dimensions have $N+1$ independent angular momentum parameters \cite{Myers:1986un,Myers:2011yc} and the augmented symmetry arises on setting these parameters to be equal.

Our principal concept is therefore to explore quantum scalar field theory on these equal-angular-momenta higher dimensional black holes, anticipating that the additional symmetry will ameliorate some of the technical difficulties of working on a four-dimensional Kerr black hole. 
However, as well as computational challenges, there remain additional complexities in studying quantum field theory on rotating space-times.

Central to these are the subtleties which arise in the definition of suitable quantum states on rotating black holes.  
The Unruh state \cite{Unruh:1976db} is the most relevant for the study of Hawking radiation; it is the analogue on an eternal black hole geometry of the quantum state pertinent to a black hole formed by gravitational collapse.
The physical definition of the Unruh state on a rotating black hole space-time is uncontroversial \cite{Ottewill:2000qh}.  Although a rigorous construction of this state for a neutral scalar field on Kerr is absent from the literature, such a construction has been performed rigorously on Kerr-de Sitter space-time \cite{Klein:2022jtb} and for a quantum fermion field on Kerr space-time \cite{Gerard:2020tdo}.
In addition, the SET expectation value has been computed for a neutral scalar field in the Unruh state on Kerr \cite{Levi:2016exv}.

The Unruh state does not preserve all the symmetries of the underlying space-time (in particular, it is not symmetric under the simultaneous inversion of the time and azimuthal angular coordinate).
For this reason, both the original methods \cite{Howard:1984qp, Howard:1984ttx, Anderson:1993if, Anderson:1994hg}  for computing the SET expectation value and a more recent approach \cite{Taylor:2022sly} have used the Hartle-Hawking state \cite{Hartle:1976tp}. 
This latter state is particularly well-adapted to SET computations, being a thermal equilibrium state, regular across both the future and past event horizons, and respecting the symmetries of the underlying black hole geometry. 
Furthermore, since differences in expectation values between two Hadamard quantum states do not require renormalization, they are comparatively straightforward to compute.

Unfortunately, the Hartle-Hawking state does not exist on a Kerr black hole background for a quantum scalar field \cite{Kay:1988mu}; there is no state representing a Kerr black hole in equilibrium with a thermal heat bath at the Hawking temperature.
This can be understood heuristically by considering the toy model of a rotating thermal state in Minkowski space-time \cite{Duffy:2002ss}. 
In unbounded Minkowski space-time, a rigidly-rotating thermal state does not exist for a quantum scalar field \cite{Duffy:2002ss}, but such a state can be constructed if Minkowski space-time is bounded by an infinite cylinder, symmetrical about the axis of rotation, of sufficiently small radius that there is no speed-of-light surface \cite{Duffy:2002ss}. 
The speed-of-light surface is the surface on which rigidly-rotating observers must travel at the speed of light.
Correspondingly, if a Kerr black hole is surrounded by a perfectly reflecting mirror inside the speed-of-light surface, then a Hartle-Hawking state can be defined for a quantum scalar field \cite{Duffy:2005mz}.
One disadvantage of introducing a reflecting boundary in order to construct rotating thermal states is that Casimir divergences result on the boundary \cite{Duffy:2002ss,Deutsch:1978sc}.

The issue of Casimir divergences can be circumvented by considering black holes which are asymptotically AdS rather than asymptotically flat. 
For example, the BTZ black hole \cite{Banados:1992wn,Banados:1992gq} does not have a speed-of-light surface, enabling the construction of a well-defined Hartle-Hawking state.
Indeed, this is the state for which the SET expectation values are computed in \cite{Steif:1993zv}.
In four space-time dimensions, Kerr-AdS black holes \cite{Carter:1968ks} do not have a speed-of-light surface if their angular momentum is sufficiently small.  
In this case 
there are no superradiant instabilities \cite{Cardoso:2004nk,Cardoso:2004hs} and
the black hole can be in thermal equilibrium with a heat bath at the Hawking temperature \cite{Hawking:1999dp}. 
In the absence of a speed-of-light surface, the space-time possesses a Killing vector which is timelike outside the event horizon.
This Killing vector plays a central role in the rigorous construction of the quantum scalar field Hartle-Hawking state on a four-dimensional stationary black hole without a speed-of-light surface \cite{Gerard:2021yoi}.

We are therefore motivated to study quantum fields on asymptotically AdS equal-angular-momenta black holes.
While such black holes have enhanced symmetry in any number of odd space-time dimensions, for simplicity we consider only the five-dimensional case.
This has the advantage that there are no ``ultra-spinning'' instabilities \cite{Dias:2009iu}.
We assume that the angular momentum of the black hole is sufficiently small that there is no speed-of-light surface. 
In this case there is an elegant argument \cite{Hawking:1999dp} that the black holes are classically stable (see also \cite{Kunduri:2006qa,Murata:2008yx,Murata:2008xr,Dias:2010eu,Murata:2011zz} for studies of the gravitational perturbations of these black holes, confirming the absence of instabilities).
The arguments in \cite{Hawking:1999dp}, valid for a general five-dimensional asymptotically AdS rotating black hole without speed-of-light surface, reveal that the black holes we study can be in thermal equilibrium with a heat bath at the Hawking temperature.
Accordingly, we conjecture that it will be possible to construct a Hartle-Hawking state in this scenario.

While the Hartle-Hawking state is our primary interest, we also seek to define a second quantum state,  in order to study differences in expectation values between two Hadamard quantum states without recourse to renormalization. 
The existence of an Unruh-like state on asymptotically-AdS black holes is complex.
For an asymptotically-AdS black hole formed by gravitational collapse, the nature of the quantum state at late times depends on the details of the collapse process and the boundary conditions applied to the quantum field far from the black hole \cite{Hemming:2000as,Giddings:2001ii,Abel:2015aqe}.
On an eternal, asymptotically-AdS black hole, if reflecting boundary conditions are applied to the quantum field (as will be the case in our study), then it is not possible to define an Unruh-like state. 

We therefore seek an analogue of the Boulware state \cite{Boulware:1974dm}, in other words a zero temperature state, which is expected to be a Hadamard state away from the event horizon. 
For a nonrotating black hole, this state is as empty as possible as seen by a static observer far from the black hole, although it is divergent at the event horizon \cite{Candelas:1980zt}.
On Kerr space-time, for a quantum scalar field there is no corresponding state which is empty at both future and past null infinity \cite{Ottewill:2000qh}.
This is because a rotating black hole spontaneously emits particles even if its temperature is zero \cite{Starobinsky:1973aij,Unruh:1974bw}.
For a bosonic field (such as quantum scalar field on which we focus in this paper), this spontaneous nonthermal  quantum emission occurs in precisely those field modes which are subject to classical superradiance \cite{Brito:2015oca}.
For a scalar field, such superradiant modes are absent on Kerr-AdS black holes when reflective boundary conditions are applied \cite{Winstanley:2001nx}. 
Similar conclusions can be drawn for a scalar field on the asymptotically-AdS equal-angular-momenta black holes \cite{Jorge:2014kra}.
We therefore expect that a Boulware-like state can be constructed in our set-up.

Our paper is structured as follows. 
In Sec.~\ref{sec:geometry} we review the geometry of five dimensional, equal-angular-momenta black holes in asymptotically AdS space-time, paying particular attention to the symmetries of the geometry.
The properties of a classical massive scalar field on this background are examined in Sec.~\ref{sec:scalar}, where we derive separable mode solutions of the scalar field equation.
In Sec.~\ref{sec:quantum} we turn to the canonical quantization of the scalar field, constructing analogues of the standard Boulware and Hartle-Hawking states. 
Differences in expectation values of the square of the quantum scalar field and the SET between these two states are computed in Sec.~\ref{sec:observables}.
Finally, Sec.~\ref{sec:conc} contains further discussion and our conclusions.

\section{Equal-angular-momenta black holes in five dimensions}
\label{sec:geometry}

The higher-dimensional, asymptotically flat, rotating Myers-Perry black hole solutions  \cite{Myers:1986un,Myers:2011yc} of the vacuum Einstein equations are generalizations of the four-dimensional Kerr metric \cite{Kerr:1963ud}.
Including a cosmological constant further complicates the metric in higher dimensions \cite{Hawking:1998kw,Gibbons:2004js,Gibbons:2004uw,Gibbons:2004ai}. 
Working in five space-time dimensions, the general rotating black hole geometry \cite{Hawking:1998kw} has, in addition to the cosmological constant, a mass parameter and two independent angular momentum parameters. 
Setting these two angular momentum parameters equal results in a space-time geometry which has enhanced symmetry \cite{Kunduri:2006qa} and these equal-angular-momenta black holes are the focus of our study.

\begin{figure}
    \centering
    \includegraphics[scale=1.2]{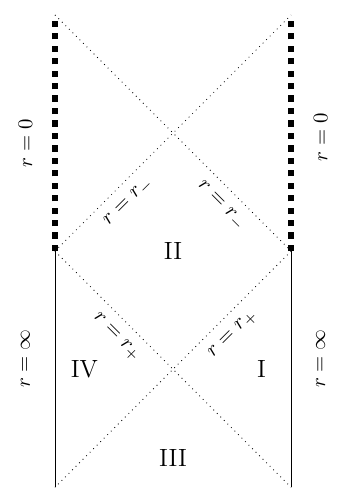}
    \caption{Penrose diagram of an equal-angular-momenta, asymptotically-anti-de Sitter, Myers-Perry black hole \cite{AlBalushi:2020rqe}. Dotted lines denote horizons. The event horizon is located at $r=r_{+}$, and the inner horizon at $r=r_{-}$. Null infinity is the time-like surface at $r=\infty $, while there is a curvature singularity at $r=0$ (denoted by the dashed lines).}
    \label{fig:penrose}
\end{figure}

In terms of the coordinates $(t,r,\theta,\phi,\psi)$ which are adapted to the enhanced symmetry, the black hole metric takes the form \cite[Eq.~(1)]{Kunduri:2006qa}:
\begin{subequations}
\label{eq:metric}
\begin{align}
\label{eq:DRS-Metric}
		ds^2=&-f(r)^2dt^2+g(r)^2dr^2+\frac{r^2}{4}\left[ d\theta^2+\sin^2\theta \, d\varphi^2\right] 
 \nonumber \\ & +h(r)^2\left[ d\psi+\frac{1}{2}\cos\theta \, d\varphi-\Omega(r)\, dt\right] ^2,
	\end{align}
where $ \theta \in [0,\pi) $, $ \varphi \in [0,4\pi) $, $ \psi \in [0,2\pi) $, and
	\begin{align}
		\label{eq:metricfunctionsstart}
		g(r)^2&=\left(1+\frac{r^2}{L^2}-\frac{2M}{r^2}+\frac{2Ma^2}{r^2L^2}+\frac{2Ma^2}{r^4}\right)^{-1}, \\ h(r)^2&=r^2\left(1+\frac{2Ma^2}{r^4}\right),
  \\ \Omega(r)&=\frac{2Ma}{r^2h(r)^2},  \label{eq:Omega}
  \\ f(r)& =\frac{r}{g(r)h(r)} .
  \label{eq:metricfunctionsend}
	\end{align}
 \end{subequations}
Throughout this paper, the space-time signature is $(-,+,+,+,+)$  and we use units in which $8\pi G=c=\hbar = k_{B}=1$.
The constants in the metric (\ref{eq:metric}) have the following interpretation: $M$ is the mass parameter, $a$ is the angular momentum parameter,  and $L$ is the AdS length scale, related to the cosmological constant $\Lambda $ by $\Lambda = - 6L^{-2}$. 
Taking the limit $L\rightarrow \infty $ gives an asymptotically flat black hole, considered (using different coordinates) in \cite{Frolov:2002xf,Frolov:2003en}.
The mass $E$ and angular momentum $J$ of the black hole are given in terms of the parameters $M$ and $J$ by \cite[Eq.~(5)]{Kunduri:2006qa}:
\begin{equation}
    E = 
    2\pi ^{2}M\left(  3+ \frac{a^{2}}{L^{2}}\right) ,
    \qquad
    J =  8\pi ^{2} aM .
\end{equation}
The  square root of minus the determinant of the metric (\ref{eq:metric}) is
\begin{equation}
    {\mathsf {g}} = {\sqrt { - \det g_{\mu \nu } }} = \frac{1}{4}r^{3} \sin \theta .
    \label{eq:metricdet}
\end{equation}
The inverse metric $g^{\mu \nu }$ has the following nonzero components:
\begin{subequations}
\label{eq:inversemetric}
    \begin{align}
        g^{tt} & = -f(r)^{-2} , \\
        g^{t\psi } & = -\Omega (r) f(r)^{-2} , \\
        g^{rr} & = g(r)^{-2}, \\
        g^{\theta \theta } & = 4r^{-2}, \\
     g^{\varphi \varphi } & = 4r^{-2} \csc ^{2} \theta , \\
     g^{\varphi \psi } & = -2r^{-2}\cot \theta \csc \theta  , \\
     g^{\psi \psi } & = r^{-2}\cot ^{2} \theta + h(r)^{-2} - \Omega (r)^{2} f(r)^{-2}  .
    \end{align}
\end{subequations}
By evaluating the Kretschmann scalar $R_{\alpha \beta \gamma \delta }R^{\alpha \beta \gamma \delta }$, where $R_{\alpha \beta \gamma \delta }$ is the Riemann tensor, we find that the metric (\ref{eq:metric}) has a singularity at $r=0$. 
The Penrose diagram for this black hole geometry can be found in Fig.~\ref{fig:penrose} \cite{AlBalushi:2020rqe}.

It is striking that the black hole metric (\ref{eq:metric}) essentially depends only on the radial coordinate $r$. 
The angular coordinates $(\theta,  \varphi , \psi )$ are coordinates on the three-sphere ${\mathbb{S}}^{3}$, here written as an ${\mathbb{S}}^{1}$ fibre (parameterized by the angle $\psi $) over ${\mathbb{CP}}^{1}$ \cite{Kunduri:2006qa}. 
With this interpretation, the part of the metric
\begin{equation}
    \frac{r^2}{4}\left[ d\theta^2+\sin^2\theta \, d\varphi^2\right] 
\end{equation}
corresponds to the Fubini-Study metric on ${\mathbb{CP}}^{1}$, while 
\begin{equation}
   \frac{1}{2}\cos\theta \, d\varphi  
   \label{eq:CPgauge}
\end{equation}
is the K\"ahler form on ${\mathbb{CP}}^{1}$ \cite{Kunduri:2006qa}. 

The metric \eqref{eq:metric} possesses the following independent Killing vectors \cite[Sec.~IV and Eq.~(8)]{Delsate:2015ina} (see also \cite{Frolov:2002xf}):
\begin{subequations}
\label{eq:Killing}
    \begin{align}
{\mathcal {V}}_1&=\frac{\partial}{\partial t}, \qquad {\mathcal {V}}_2=\frac{\partial}{\partial \varphi}, \qquad {\mathcal {V}}_3=\frac{\partial}{\partial \psi},
\label{eq:xi123}
\\
{\mathcal {V}}_4&=2\cos\varphi\frac{\partial}{\partial \theta}-2\sin\varphi\cot\theta\frac{\partial}{\partial \varphi}+\sin\varphi\csc\theta\frac{\partial}{\partial \psi},
\label{eq:xi4}
\\
{\mathcal {V}}_5&=-2\sin\varphi\frac{\partial}{\partial \theta}-2\cos\varphi\cot\theta\frac{\partial}{\partial \varphi}+\cos\varphi\csc\theta\frac{\partial}{\partial \psi} .
\label{eq:xi5}
\end{align}
\end{subequations}
In addition, the metric (\ref{eq:metric}) has ``hidden symmetries'' (Killing-Yano tensors) \cite{Krtous:2006qy,Frolov:2017kze} which we do not need to consider in detail, but which become manifest in the separability of the Klein-Gordon equation, as studied in Sec.~\ref{sec:scalar}.

Horizons occur at the zeros of the polynomial $g(r)^{-2}$. 
This gives a cubic in $x=r^{2}$:
\begin{equation}
    x^{3} + L^{2}x^{2} + 2 M \left(  a^{2} - L^{2} \right) x + 2a^{2}L^{2}M = 0,
    \label{eq:cubic}
\end{equation}
the discriminant of which is always positive, so that there are three real roots, which we will denote by $r_{0}^{2}$, $r_{-}^{2}$ and $r_{+}^{2}$.
One of these roots (taken to be $r_{0}^{2}$ without loss of generality) is negative, and the other two are positive. 
The larger of the two positive roots, $r_{+}^{2}$, corresponds to the (outer) event horizon of the black hole, while $r_{-}^{2}$ corresponds to the inner horizon.
The event horizon radius $r_{+}$ is given by 
\begin{subequations}
\label{eq:horizon}
    \begin{equation}
    r_{+}^{2} = \frac{1}{3} \Bigg\{  2{\sqrt{ \mathfrak {A}}}\cos \left(\frac{1}{3}  \cos^{-1}\left[ -\frac{{\sqrt {\mathfrak {B}} }}{{\mathfrak {A}}^{\frac{3}{2}}} \right]   \right)  -L^{2} \Bigg\}  ,
\end{equation}
where
\begin{align}
    {\mathfrak {A}} & = L^{4} - 6M\left( a^{2}-L^{2} \right) ,
    \nonumber \\ 
    {\mathfrak {B}} & = L^{6} +9L^{2}M \left( 2a^{2} + L^{2} \right) .
\end{align}
\end{subequations}
Using (\ref{eq:cubic}), we can write the mass parameter $M$ in terms of $a$, $L$ and $r_{+}$:
\begin{equation}
    M = -\frac{r_{+}^{4}\left(r_{+}^{2}+L^{2}\right) }{2\left( r_{+}^{2} \left[ a^{2}-L^{2} \right]  +a^{2}L^{2}\right) }.
\end{equation}
Substituting for $M$ in the cubic (\ref{eq:cubic}) then gives a quadratic having roots $x=r_{-}^{2}$ and $x=r_{0}^{2}$:
\begin{equation}
    x^{2} + \left( r_{+}^{2}+L^{2} \right) x + \frac{a^{2}L^{2}r_{+}^{2} \left( r_{+}^{2}+L^{2} \right)}{r_{+}^{2} \left( a^{2}-L^{2} \right)  +a^{2}L^{2} }  = 0 .
\end{equation}
The roots of this quadratic are then:
\begin{subequations}
    \begin{align}
    r_{-}^{2} & = - \frac{1}{2} \left( r_{+}^{2} + L^{2} \right)
    \nonumber \\ & \quad \times \left[ 1 -{\sqrt { \frac{ a^{2}\left( L^{2}-r_{+}^{2}\right)^{2}-L^{2}r_{+}^{2}\left( r_{+}^{2}+L^{2} \right)}{\left( r_{+}^{2} \left[ a^{2}-L^{2} \right]  +a^{2}L^{2} \right) \left( r_{+}^{2}+L^{2} \right)} }}  \right]  ,
    \\
    r_{0}^{2} & = -\frac{1}{2} \left( r_{+}^{2} + L^{2} \right)
    \nonumber \\ & \quad \times \left[ 1 + {\sqrt { \frac{ a^{2}\left( L^{2}-r_{+}^{2}\right)^{2}-L^{2}r_{+}^{2}\left( r_{+}^{2}+L^{2} \right)}{\left( r_{+}^{2} \left[ a^{2}-L^{2} \right]  +a^{2}L^{2} \right) \left( r_{+}^{2}+L^{2} \right)} }}  \right]  .
\end{align}
\end{subequations}
The following relations involving the roots of the cubic (\ref{eq:cubic}) will be useful in Sec.~\ref{sec:Heun}:
\begin{subequations}
\begin{align}
    r_{+}^{2} + r_{-}^{2} + r_{0}^{2} &  =  ~ -L^{2},
    \\
    r_{+}^{2}r_{-}^{2}+ r_{-}^{2}r_{0}^{2}+r_{0}^{2}r_{+}^{2} & = ~ 2M \left( a^{2}- L^{2} \right) ,
    \\
    r_{+}^{2}r_{-}^{2}r_{0}^{2} & = ~ -2a^{2}ML^{2},
\end{align}
and from these we can write $M$ and $a^{2}$ as follows:
\begin{align}
    a^{2} & = ~ \frac{r_{+}^{2}r_{-}^{2}r_{0}^{2}\left( r_{+}^{2} + r_{-}^{2} + r_{0}^{2} \right)}{\left( r_{+}^{2}+r_{-}^{2}\right) \left( r_{-}^{2}+r_{0}^{2}\right) \left( r_{0}^{2}+r_{+}^{2} \right)} ,
    \\
    M & = ~ \frac{\left( r_{+}^{2}+r_{-}^{2}\right) \left( r_{-}^{2}+r_{0}^{2}\right) \left( r_{0}^{2}+r_{+}^{2} \right)}{2\left( r_{+}^{2}+r_{-}^{2}+r_{0}^{2}\right) ^{2}} .
\end{align}
\end{subequations}

The outer and inner horizons coincide (so that $r_{+}=r_{-}$ and the black hole becomes extremal) when the spin parameter $a=a_{\rm {ext}}$, which is given, in terms of the event horizon radius, by 
\begin{equation}
    a_{\rm {ext}}^{2}  = \frac{L^{2}r_{+}^{2}\left( 2r_{+}^{2} + L^{2} \right) }{2\left( r_{+}^{2}+L^{2} \right) ^{2}}.
    \label{eq:aext}
\end{equation}
For values of the spin parameter $a$ above $a_{\rm {ext}}$, there is a naked singularity.

Since the black hole is rotating, it possesses an ergosphere, inside which an observer cannot remain at rest relative to infinity. 
The boundary of this region is the stationary limit surface, given by 
\begin{equation}
    0 = -g_{tt} = f(r)^{2} - h(r)^{2} \Omega (r)^{2} = \frac{r^{2}}{L^{2}} - \frac{2M}{r^{2}}+1 .
\end{equation}
This has two roots for $x=r^{2}$, only one of which is positive. 
This gives the radius $r_{{\rm {S}}}$ of the stationary limit surface:
\begin{equation}
    r_{{\rm {S}}}^{2} = \frac{L^{2}}{2}\left[ {\sqrt {  1 + \frac{8M}{L^{2}} }} - 1  \right] . 
    \label{eq:rs}
\end{equation}
Thus, unlike the situation in four-dimensional Kerr space-time, the radius of the stationary limit surface is a constant and does not depend on any of the angular coordinates. 
This is a result of the enhanced symmetry of equal-angular-momenta black holes.
Surprisingly, the radius (\ref{eq:rs}) is independent of the spin parameter $a$ and depends only on the mass parameter $M$ and the AdS length scale $L$. 

The event horizon at $r=r_{+}$ is a Killing horizon for the Killing vector
\begin{equation}
{\mathcal {V}}_{\rm {+}}=\frac{\partial}{\partial t}+\Omega_{\rm {+}}\frac{\partial}{\partial \psi},
\label{eq:xiH}
\end{equation}
where $\Omega _{+}$ is the angular speed of the event horizon, given by \cite[Eq.~(4)]{Kunduri:2006qa}
\begin{equation}
    \Omega _{+} =\Omega (r_{+})= \frac{2Ma}{r_{+}^{4} + 2Ma^{2}}.
    \label{eq:OmegaH}
\end{equation}
The surface gravity of the event horizon $\kappa _{+}$  is given by $\kappa _{+}^{2} = \left( \nabla _{\alpha }|{\mathcal {V}}_{+}| \right) \left( \nabla ^{\alpha } | {\mathcal {V}}_{+}| \right) $ \cite[Eq.~(14)]{Gibbons:2004uw} where $|{\mathcal {V}}_{+}| ^{2}= -g_{\alpha \gamma }{\mathcal {V}}_{+}^{\alpha }{\mathcal {V}}_{+}^{\gamma }$, and takes the form 
\begin{equation}
    \kappa _{+} = \frac{2r_{+}^{6}+r_{+}^{4}L^{2}-2Ma^{2}L^{2}}{L^{2}r_{+}^{3}{\sqrt { r_{+}^{4}+2Ma^{2}}} }.
    \label{eq:kappa+}
\end{equation}
Kruskal coordinates regular across the event horizon can be constructed by first defining a corotating coordinate $\psi _{+}$:
\begin{equation}
    \psi _{+} = \psi - \Omega _{+} t,
    \label{eq:psi+}
\end{equation}
in terms of which the metric (\ref{eq:metric}) becomes
\begin{align}
		ds^2=&-f(r)^2dt^2+g(r)^2dr^2+\frac{r^2}{4}\left[ d\theta^2+\sin^2\theta \, d\varphi^2\right] 
 \nonumber \\ & +h(r)^2\left[ d\psi _{+}+\frac{1}{2}\cos\theta \, d\varphi-\left\{ \Omega(r) - \Omega _{+} \right\} dt\right] ^2 .
 \label{eq:rotatingmetric}
	\end{align}
 Next, the ``tortoise'' coordinate $r_{\star }$ is given by \cite[Eq.~(14)]{Kunduri:2006qa}
 \begin{equation}
     \frac{dr_{\star }}{dr} = \frac{g(r)}{f(r)} = \frac{1}{r}g(r)^{2}h(r),
     \label{eq:tortoise}
 \end{equation}
and the usual double-null coordinates are
\begin{equation}
    u = t- r_{\star }, \qquad  v= t+ r_{\star } .
    \label{eq:null}
\end{equation}
We then define the Kruskal coordinates $U$, $V$ as:
\begin{equation}
    U = - \frac{1}{\kappa _{+}} e^{-\kappa _{+}u} , \qquad 
    V = \frac{1}{\kappa _{+}} e^{\kappa _{+}v}.
    \label{eq:Kruskal}
\end{equation}
In terms of these Kruskal coordinates, the metric is
\begin{align}
		ds^2_{\rm {K}}=&-\frac{f(r)^2}{\kappa _{+}^{2}UV} \, dU \, dV
  +\frac{r^2}{4}\left[ d\theta^2+\sin^2\theta \, d\varphi^2\right] 
 \nonumber \\ & 
 +h(r)^2\left[ d\psi _{+}+\frac{1}{2}\cos\theta \, d\varphi
 \right. \nonumber \\ & \left. \qquad 
 -\frac{1}{2\kappa _{+}}\left\{ \Omega(r) - \Omega _{+} \right\} \left( \frac{dV}{V} + \frac{dU}{U} \right)  \right] ^2 . 
 \label{eq:metricKruskal}
	\end{align}

Close to the event horizon, the Killing vector ${\mathcal {V}} _{+}$ (\ref{eq:xiH}) is timelike. 
It becomes spacelike on the speed-of-light surface, where
\begin{equation}
    0 = g_{tt} + 2\Omega _{+}g_{t\psi } + \Omega _{+}^{2} g_{\psi \psi }.
    \label{eq:soleqnfirst}
\end{equation}
This gives a quartic equation for $r_{\rm {L}}^{2}$, the square of the radius of the speed-of-light surface (details of the derivation can be found in App.~\ref{sec:solderiv}):
\begin{multline}
\left( r_{\rm {L}}^{2}-r_{+}^{2}\right) \left( r_{\rm {L}}^{4} + 2Ma^{2}\right) \left[ \left( 1- \Omega _{+}^{2}L^{2} \right) r_{\rm {L}}^{2} + \frac{\Omega _{+}^{2}r_{+}^{6}L^{2}}{2Ma^{2}}\right]  \\ = 0,
\label{eq:soleqn}
\end{multline}
from which we deduce that
\begin{equation}
r_{\rm {L}}^{2} = \frac{\Omega _{+}^{2}r_{+}^{6}L^{2}}{2Ma^{2}\left( \Omega _{+}^{2}L^{2}-1\right)}.
\label{eq:rL}
\end{equation}
In a four-dimensional Kerr space-time, the speed-of-light surface has a complicated structure and its radius depends on the angular coordinate $\theta $ \cite[Appendix]{Duffy:2005mz}.
In our set-up, the radius $r_{\rm {L}}$ of the speed-of-light surface is independent of all the angular variables.
When $\Omega _{+}<L^{-1}$, the right-hand-side of (\ref{eq:rL}) is negative and the speed-of-light surface is absent; this is the situation in which we are interested.
We can find the maximum value of the spin parameter $a$ for which there is no speed-of-light surface as follows. 
We first use the cubic (\ref{eq:cubic}) to write the mass parameter $M$ in terms of the event horizon radius $r_{+}$ and $\Omega _{+}$ \cite[Eq.~(6)]{Kunduri:2006qa}:
\begin{equation}
    M = \frac{r_{+}^{2}\left( 1 + r_{+}^{2}/L^{2} \right)^{2}}{2\left(  1+r_{+}^{2}/L^{2} - r_{+}^{2}\Omega _{+}^{2} \right) } ,
\end{equation}
and then the angular speed of the event horizon $\Omega _{+}$ (\ref{eq:OmegaH}) takes the form
\begin{equation}
    \Omega _{+} = a \left( \frac{1}{L^{2}} + \frac{1}{r_{+}^{2}} \right) .
    \label{eq:OmegaHalt}
\end{equation}
Thus $\Omega _{+} = L^{-1}$ when $a=a_{\rm {max}}$, where
\begin{equation}
    a_{\rm {max}} = \frac{r_{+}^{2}L}{r_{+}^{2}+L^{2}}. 
    \label{eq:amax}
\end{equation}

\begin{figure}
    \centering
    \includegraphics[scale=0.5]{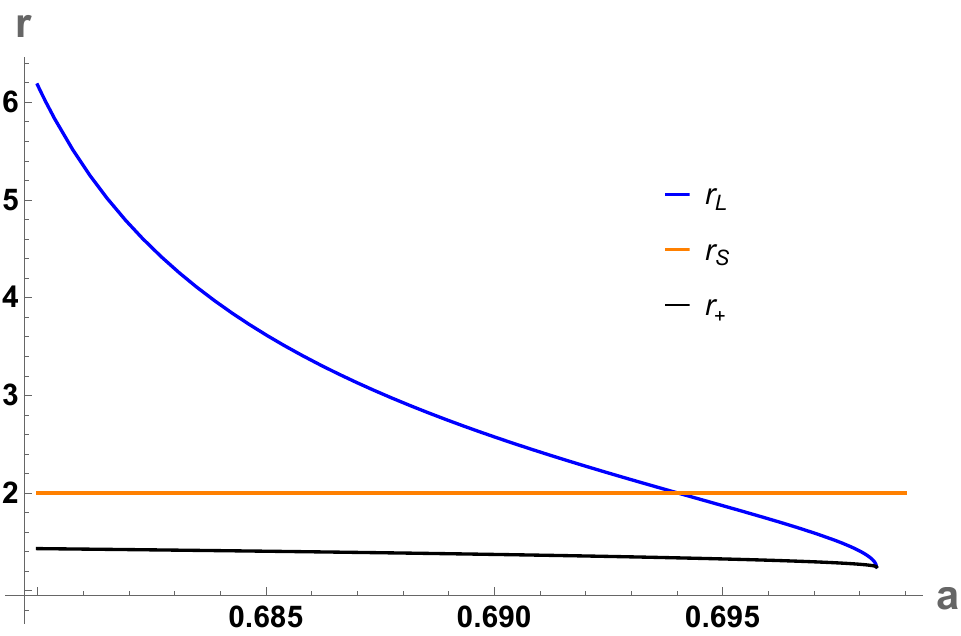}
    \caption{Locations of the event horizon $r_{+}$ (\ref{eq:horizon}), stationary limit surface $r_{{\rm {S}}}$ (\ref{eq:rs}) and speed-of-light surface $r_{\rm {L}}$ (\ref{eq:rL}) as functions of the spin parameter $a$ for fixed mass parameter $M=10L^{2}$. We use units in which $L=1$.}
    \label{fig:rvalues}
\end{figure}

\begin{figure}
    \centering
    \includegraphics[scale=0.5]{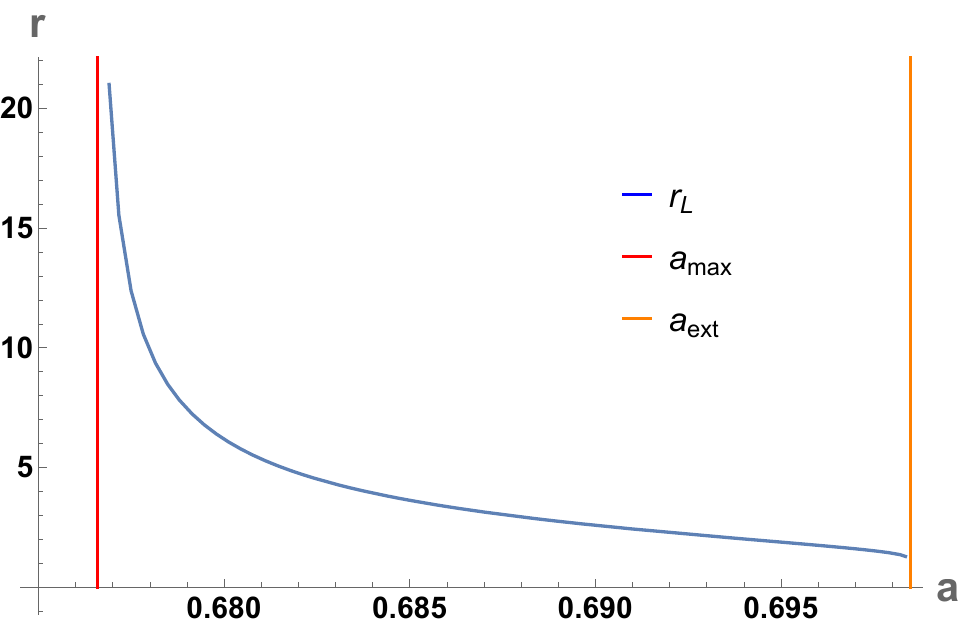}
    \caption{Location of the speed-of-light surface $r_{\rm {L}}$ (\ref{eq:rL}) as a function of the spin parameter $a$ for fixed mass parameter $M=10L^{2}$. The vertical lines give the values of $a_{\rm {max}}$ (\ref{eq:amax}) and $a_{\rm {ext}}$ (\ref{eq:aext}). We use units in which $L=1$.}
    \label{fig:avalues}
\end{figure}

Our equal-angular-momenta black hole space-time therefore has a ``shell-like'' structure, with three key surfaces at constant values of the radial coordinate $r$, namely: the event horizon $r=r_{+}$ (\ref{eq:horizon}), the stationary limit surface $r=r_{\rm {S}}$ (\ref{eq:rs}) and the speed-of-light surface $r=r_{\rm {L}}$ (\ref{eq:rL}).
For fixed values of the mass parameter $M$ and AdS length $L$, the location of the stationary limit surface (\ref{eq:rs}) is independent of the spin parameter $a$, but the event horizon radius and speed-of-light surface (if it exists) depend on $a$. 

In Fig.~\ref{fig:rvalues} we show these three radii as functions of $a$ for fixed $M=10L^{2}$, using units in which $L=1$ (we shall use these units for the rest of the paper). 
The behaviour is qualitatively similar for other values of $M$.
The orange line gives the value of $r_{\rm {S}}$ (\ref{eq:rs}), which is independent of the spin parameter $a$. 
The black line is the event horizon radius $r_{+}$ (\ref{eq:horizon}), which decreases slowly as $a$ increases for the range of values of $a$ shown in the plot. 
The blue curve is the location of the speed-of-light surface $r_{\rm {L}}$ (\ref{eq:rL}).
This changes rapidly with varying $a$ and decreases as $a$ increases. 
When $a=a_{\rm {ext}}$, we find that $r_{+}=r_{\rm {L}}$ and the speed-of-light surface coincides with the event horizon.
For sufficiently large values of $a$ (below $a_{\rm {ext}}$) the speed-of-light surface lies inside the stationary limit surface; similar behaviour is seen near the equatorial plane for near-extremal Kerr black holes \cite{Duffy:2005mz}.
The behaviour of $r_{\rm {L}}$ on decreasing $a$ further can be seen in Fig.~\ref{fig:avalues}.
In particular, $r_{\rm {L}}$ increases rapidly as $a$ decreases, and the speed-of-light surface moves away from the black hole event horizon.
There is an asymptote in $r_{\rm {L}}$ as $a\rightarrow a_{\rm {max}}$; below this value of $a$ there is no speed-of-light surface.

For the remainder of this paper we shall be concerned only with values of $0<a<a_{\rm {max}}$, so that the speed-of-light surface is absent. 
In this situation the Killing vector ${\mathcal {V}} _{+}$ (\ref{eq:xiH}) is timelike everywhere outside the event horizon.
This will be important for the quantization of the scalar field in Sec.~\ref{sec:quantum}.

\section{Classical scalar field}
\label{sec:scalar}

We consider a scalar field  $\Phi $ propagating on the background geometry discussed in Sec.~\ref{sec:geometry} and satisfying the Klein-Gordon equation
\begin{equation}
   \left[  \Box - \left(  \mu ^{2} + \xi R \right) \right] \Phi =0 ,
   \label{eq:KG}
\end{equation}
where $\Box =\nabla _{\alpha } \nabla ^{\alpha }$ with $\nabla _{\alpha }$  the usual covariant derivative, $\mu $ is the mass of the scalar field, and $\xi $ is a constant coupling the scalar field to the Ricci scalar $R$ of the geometry.
In five space-time dimensions, the scalar field is minimally coupled for $\xi = 0$ and conformally coupled when $\xi = 3/16$.
Since the Ricci scalar $R=-20/L^2$ is a constant for equal-angular-momenta black holes, we define a constant $\nu $ (the scalar field ``effective mass'') by 
\begin{equation}
    \nu ^{2} = \mu ^{2} + \xi R .
    \label{eq:nu}
\end{equation}
The classical SET for the scalar field is \cite[Eq.~(9)]{Decanini:2005eg}
\begin{align}
    T_{\alpha \beta } & = \left( 1 -2\xi \right)  \Phi _{;\alpha } \Phi _{;\beta }
    + \left( 2\xi - \frac{1}{2} \right) g_{\alpha \beta } g^{\rho \sigma } \Phi _{;\rho }\Phi _{;\sigma }
    \nonumber \\ & \qquad 
    -2\xi \Phi \Phi _{;\alpha \beta }+2\xi g_{\alpha \beta } \Phi \Box \Phi 
    \nonumber \\ & \qquad
    + \xi \left(  R_{\alpha \beta } -\frac{1}{2}Rg_{\alpha \beta } \right) \Phi ^{2} 
    - \frac{1}{2} g_{\alpha \beta } \mu ^{2} \Phi ^{2}
    \nonumber \\ & = 
    \left( 1 -2\xi \right)  \Phi _{;\alpha } \Phi _{;\beta }
    -2\xi \Phi \Phi _{;\alpha \beta }
    \nonumber \\ & \qquad
    + \left( 2\xi - \frac{1}{2} \right) g_{\alpha \beta } g^{\rho \sigma } \Phi _{;\rho }\Phi _{;\sigma }
    \nonumber \\ & \qquad
    + \frac{1}{2} \left( 4\xi - 1 \right)  g_{\alpha \beta }  \mu ^{2} \Phi ^{2}
    + \xi  R g_{\alpha \beta } \left( 2 \xi - \frac{3}{10}\right)  \Phi ^{2} ,
    \label{eq:SET}
 \end{align}
 where a semicolon ; denotes a covariant derivative, and in the second equality we have used the scalar field equation (\ref{eq:KG}) and the fact that $R_{\alpha \beta } = Rg_{\alpha \beta }/5$ for the equal-angular-momenta black holes. 
 Therefore, while solutions of the scalar field equation (\ref{eq:KG}) depend on the constants $\mu ^{2}$
and $\xi $ only through the combination $\nu $ (\ref{eq:nu}), the SET (\ref{eq:SET}) depends on $\mu ^{2}$ and $\xi $ separately. 

The scalar field equation (\ref{eq:KG}) is separable \cite{Frolov:2002xf,Kunduri:2005fq,Frolov:2006pe,Kunduri:2006qa,BarraganAmado:2018zpa}, and we seek mode solutions of the form:
\begin{equation}
    \Phi (t,r,\theta, \varphi , \psi ) = e^{-i\omega t}e^{ip \psi }X(r) Y(\theta ,\varphi ),
    \label{eq:mode}
\end{equation}
where $X(r)$ and $Y(\theta ,\varphi )$ are, respectively, the radial and angular functions.
The constant $\omega $ is the frequency of the mode, and $p$ is the azimuthal quantum number.
Since $\psi \in [0,2\pi )$, we have $p\in {\mathbb{Z}}$.

In this section we review the angular and radial equations.
Since we are only considering black holes with $\Omega _{+}L<1$ the classical scalar field modes we consider are linearly stable \cite{Kunduri:2006qa,Koga:2022vun}. 

\subsection{Angular equation}
\label{sec:angular}

Substituting the separable mode ansatz (\ref{eq:mode}) for the scalar field into (\ref{eq:KG}) and separating variables, the resulting equation for the angular function $Y(\theta ,\varphi )$ can be written as
\begin{multline}
    0 = \frac{\partial ^{2}Y}{\partial \theta ^{2}}  + \cot \theta \frac{\partial Y}{\partial \theta }
    - ip \cot \theta \csc \theta \frac{\partial Y}{\partial \varphi }
    + \csc ^{2} \theta \frac{\partial ^{2}Y}{\partial \varphi ^{2}}
    \\
    - \frac{1}{4}p^{2} Y\cot ^{2} \theta  + \lambda Y ,
    \label{eq:ang1}
    \end{multline}
where $\lambda $ is a separation constant. 
We require $Y(\theta, \varphi )$ to be regular at the poles $\theta = 0 , \pi $, and to be periodic in $\varphi $ with period $4\pi $.

There are several different (but equivalent) representations of the angular functions. 
First, Eq.~(\ref{eq:ang1}) is the equation satisfied by a charged scalar field on ${\mathbb{CP}}^{1}$, having charge $p$ and with the gauge potential (\ref{eq:CPgauge})  \cite{Kunduri:2006qa}.
On ${\mathbb{CP}}^{1}$, the angular functions  $Y(\theta, \varphi )$ are monopole harmonics \cite{Wu:1976ge,Wu:1977qk}.

Second, the angular function $Y(\theta, \varphi ) $ can further be separated as follows:
\begin{equation}
    Y(\theta, \varphi ) = e^{im\varphi } {\widetilde {Y}}(\theta ).
    \label{eq:angsep}
\end{equation}
Since $\varphi \in [0,4\pi )$, we have that $2m\in {\mathbb{Z}}$.
The resulting equation for ${\widetilde {Y}}(\theta )$ can be cast into the form of a hypergeometric equation, with the solutions regular at $\theta = 0,  \pi $ given in terms of Jacobi polynomials \cite{Frolov:2002xf,Kunduri:2005fq} (note that \cite{Frolov:2002xf,Kunduri:2005fq} use a different coordinate system than the one employed here, so their angular equation is modified in our case).

We shall find it more convenient to use a third representation, in which the angular functions $Y(\theta, \varphi )$ are given by spin-weighted spherical harmonics (see, for example, \cite{Newman:1966ub,Goldberg:1966uu,Campbell:1971,Breuer:1977,TorresDelCastillo:2007,Michel:2020,Freeden:2022} for a selection of the extensive references on these, and \cite{Michel:2020,Freeden:2022} for  longer bibliographies).
We compare (\ref{eq:ang1}) with the governing equation for spin-weighted spherical harmonics ${}_{s}Y_{\ell }^{n}(\theta, \varphi )$ having spin $s$, total angular momentum quantum number $\ell $ and azimuthal quantum number $n$ \cite{Breuer:1977,Michel:2020,Freeden:2022}:
\begin{multline}
    0 = \frac{\partial ^{2}{}_{s}Y_{\ell }^{n}}{\partial \theta ^{2}}  + \cot \theta \frac{\partial {}_{s}Y_{\ell }^{n}}{\partial \theta }
    +2is \cot \theta \csc \theta \frac{\partial {}_{s}Y_{\ell }^{n}}{\partial \varphi }
    \\
    + \csc ^{2} \theta \frac{\partial ^{2}{}_{s}Y_{\ell }^{n}}{\partial \varphi ^{2}}
    - s^{2}{}_{s}Y_{\ell }^{n}\cot ^{2} \theta  + \left[ \ell ^{2} + \ell - s^{2} \right] {}_{s}Y_{\ell }^{n}.
    \label{eq:ang2}
    \end{multline}
Here, $s$ is a (positive or negative) integer or half-integer, $\ell = |s|, |s|+1, |s|+2, \ldots $ is a positive integer or half-integer and $m = -\ell , - \ell + 1, \ldots, \ell - 1, \ell $ is also an integer or half-integer (taking both positive and negative values).
Comparing (\ref{eq:ang1}, \ref{eq:ang2}), we see that the spin  $s$ is related to the quantum number $p$ by 
\begin{equation}
s=-\frac{p}{2}, 
\label{eq:spin}
\end{equation}
and the separation constant $\lambda $ is given by 
\begin{equation}
    \lambda = \ell ^{2} + \ell - s^{2} = \ell ^{2}+\ell - \frac{1}{4}p^{2}.
    \label{eq:lambda}
\end{equation}
Using the further separation (\ref{eq:angsep}), we may identify $m=n$. 
The advantage of working with spin-weighted spherical harmonics is the existence of addition theorems \cite{Monteverdi:2024xyp} for sums over the azimuthal quantum number $m$, which will enable us, in Sec.~\ref{sec:observables}, to simplify the expressions for the square of the scalar field and SET for the quantum scalar field. 
The relevant addition theorems are summarized in App.~\ref{sec:addition}.

The standard normalization for the spin-weighted spherical harmonics having integer spin $s$ is
\begin{multline}
\int _{\varphi =0}^{2\pi } \int _{\theta = 0}^{\pi }  {}_{s}Y_{\ell }^{m*}(\theta , \varphi ) {}_{s'}Y_{\ell '}^{m'} (\theta , \varphi ) \, \sin \theta \, d\theta \, d\varphi 
\\ = \delta _{ss'}\delta _{\ell \ell'}\delta _{mm'}.
\end{multline}
We use this normalization (so that the addition theorems in App.~\ref{sec:addition} apply), but since we have $\varphi \in [0,4\pi ) $ rather than $\varphi \in [0,2\pi )$, our normalization is
\begin{multline}
\int _{\varphi =0}^{4\pi } \int _{\theta = 0}^{\pi }  {}_{s}Y_{\ell }^{m*}(\theta , \varphi ) {}_{s'}Y_{\ell '}^{m'} (\theta , \varphi ) \, \sin \theta \, d\theta \, d\varphi 
\\ = 2\delta _{ss'}\delta _{\ell \ell'}\delta _{mm'}.
\label{eq:swshnorm}
\end{multline}
This normalization condition is valid for half-integer as well as integer spin.
The angular equation (\ref{eq:ang2}) is invariant under the mapping $(s,m)\rightarrow (-s,-m)$ (or, equivalently $(p,m)\rightarrow (-p,-m)$ from (\ref{eq:spin})), and hence the normalized spin-weighted spherical harmonics ${}_{s}Y_{\ell}^{m}(\theta, \varphi )$ and ${}_{-s}Y_{\ell }^{-m}(\theta, \varphi )$ are equal up to an overall phase, so that $|{}_{s}Y_{\ell }^{m}(\theta, \varphi )|^{2}=|{}_{-s}Y_{\ell }^{-m}(\theta, \varphi )|^{2}$.

\subsection{Radial equation}
\label{sec:radial}

We next describe the radial equation in two forms: the first is a Schr\"odinger-like equation \cite{Kunduri:2006qa} and is useful for defining an orthonormal basis of field modes for the canonical quantization of the scalar field in Sec.~\ref{sec:quantum}; while the second involves a Heun equation \cite{Aliev:2008yk,BarraganAmado:2018zpa,Noda:2022zgk,Koga:2022vun} and is useful for our numerical computations in Sec.~\ref{sec:observables}.

\subsubsection{Potential form}
\label{sec:potential}

The radial equation resulting from substituting the separable mode ansatz (\ref{eq:mode}) into (\ref{eq:KG}) is
\begin{multline}
    0  = X''(r) + \left[ \frac{3}{r} - \frac{2g'(r)}{g(r)} \right] X'(r)
    \\ 
    + g(r)^{2} \left[ \frac{\left\{ \omega - p \Omega (r)\right\} ^{2}}{f(r)^{2}} - \frac{p^{2}}{h(r)^{2}} - \frac{4\lambda }{r^{2}} - \nu ^{2} \right] X(r), 
    \label{eq:radial}
\end{multline}
where $\lambda $ is the separation constant (\ref{eq:lambda}) and $\nu ^{2}$ is given in (\ref{eq:nu}).
Following \cite{Kunduri:2006qa}, we use the ``tortoise'' coordinate $r_{\star }$ given by (\ref{eq:tortoise}).
Near the event horizon, as $r\rightarrow r_{+}$, we have $g(r)/f(r) \sim {\mathcal {O}}(r-r_{+})^{-1}$ and hence $r_{\star } \rightarrow - \infty $, while, as $r\rightarrow \infty $, since $g(r)/f(r) \sim {\mathcal {O}}(r^{-2})$, by a suitable choice of integration constant, we may take $r_{\star } \rightarrow 0$.
Defining a new radial function ${\widetilde {X}}(r)$ by 
\begin{equation}
    X(r) = \frac{{\widetilde{X}}(r_{\star })}{r{\sqrt {h(r)}}},
    \label{eq:tildeX}
\end{equation}
the radial equation (\ref{eq:radial}) becomes
\begin{equation}
\frac{d^{2}{\widetilde{X}}}{dr_{\star }^{2}} + V(r) {\widetilde{X}} = 0, 
\label{eq:radialpot}
\end{equation}
where the potential $V(r)$ is given by \cite[Eq.~(15)]{Kunduri:2006qa} 
\begin{multline}
V(r) = 
\left[ \omega - p\Omega (r) \right] ^{2} -f(r)^{2}\left[ \nu ^{2} + \frac{4\lambda }{r^{2}} + \frac{p^{2}}{h(r)^{2}} \right]
\\ - \frac{f(r)^{2}{\sqrt{h(r)}}}{r^{2}} \frac{d}{dr} \left[ \frac{f(r)^{2}h(r)}{r} \frac{d}{dr} \left\{ r{\sqrt {h(r)}} \right\}  \right] ,
\end{multline}
and we have used the form of $f(r)$ (\ref{eq:metricfunctionsend}) to simplify $V(r)$.

In the asymptotic regions $r_{\star} \rightarrow -\infty $, $r_{\star }\rightarrow 0$, the potential takes the asymptotic forms
\begin{equation}
    V(r) \sim 
    \begin{cases}
        {\widetilde {\omega }}^{2}, & r_{\star} \rightarrow -\infty , \quad r\rightarrow r_{+},
        \\
         - \dfrac{r^{2}}{L^{2}} \left[ \nu ^{2} + \dfrac{15}{4L^{2}} \right] , & r_{\star }\rightarrow 0, \quad r\rightarrow \infty ,
    \end{cases}
\end{equation}
where $\nu ^{2}$ is given in (\ref{eq:nu}) and we have defined a new, shifted, frequency ${\widetilde {\omega }}$ by
\begin{equation}
    {\widetilde {\omega }} = \omega - p \Omega _{+} ,
    \label{eq:tildeomega}
\end{equation}
where $\Omega _{+}$ is given in (\ref{eq:OmegaH}).
Near the horizon, as $r\rightarrow r_{+}$ and $r_{\star} \rightarrow -\infty $, the solutions of the radial equation (\ref{eq:radialpot}) therefore take the form
\begin{equation}
    {\widetilde {X}}(r_{\star }) \sim {\mathcal {C}}_{+} e^{i{\widetilde {\omega }}r_{\star }} + {\mathcal {C}}_{-} e^{-i{\widetilde {\omega }}r_{\star }},
    \label{eq:Xhor}
\end{equation}
where ${\mathcal {C}}_{\pm }$ are complex constants, giving ingoing and outgoing plane waves.
As $r\rightarrow \infty $ and $r_{\star} \rightarrow 0$, the solutions of (\ref{eq:radialpot}) are \cite[Eq.~(43)]{Kunduri:2006qa}
\begin{equation}
    {\widetilde {X}}(r_{\star }) \sim {\mathcal {D}}_{+} r_{\star }^{\frac{1}{2}+{\widetilde {\nu }}}  + {\mathcal {D}}_{-} r_{\star }^{\frac{1}{2}- {\widetilde {\nu }}},
    \label{eq:Xinf}
\end{equation}
where ${\mathcal {D}}_{\pm }$ are complex constants, and 
\begin{subequations}
\label{eq:bfbound}
\begin{equation}
    {\widetilde {\nu }}^{2} = 4 +  \nu ^{2} L^{2} .
    \label{eq:tildenu}
\end{equation}
In order  that the scalar field has no classical mode instabilities, it must be the case that the Breitenl\"ohner-Freedman bound \cite{Breitenlohner:1982bm,Breitenlohner:1982jf} is satisfied, namely:
\begin{equation}
    {\widetilde {\nu }}^{2}  >0 .
\end{equation}
\end{subequations}
From here on, we shall assume this to be the case. 
We then consider only the regular decaying solution in (\ref{eq:Xinf}), that is, we assume that ${\widetilde {\nu }}>0$ and
\begin{equation}
    {\widetilde {X}}(r_{\star }) \sim {\mathcal {D}}_{+}  r_{\star }^{\frac{1}{2}+{\widetilde {\nu }}},
\end{equation}
as $r\rightarrow \infty $ and $r_{\star} \rightarrow 0$.

Absorbing the constant ${\mathcal {C}}_{+}$ into an overall normalization constant, we can, without loss of generality, take the radial function ${\widetilde {X}}(r_{\star })$ to have the form
\begin{equation}
    {\widetilde {X}}_{\omega p \ell }(r_{\star }) \sim 
    \begin{cases}
        e^{i{\widetilde {\omega }}r_{\star}} + {\mathcal {R}}_{\omega p \ell }e^{-i{\widetilde {\omega }}r_{\star}}, & r_{\star} \rightarrow -\infty , 
        \\
        {\mathcal {T}}_{\omega p \ell } r_{\star }^{\frac{1}{2}+{\widetilde {\nu }}}, & r_{\star} \rightarrow 0,
    \end{cases}
    \label{eq:Xtildefinal}
\end{equation}
where ${\mathcal {R}}_{\omega p \ell }$ and ${\mathcal {T}}_{\omega p \ell }$ are complex constants, and we have introduced the subscripts to indicate that the radial function depends only on the frequency $\omega $ and the quantum numbers $p$ and $\ell $.
Since the radial equation (\ref{eq:radialpot}) is invariant under the mapping $(\omega, p)\rightarrow (-\omega , -p)$, we have that ${\widetilde {X}}_{-\omega, - p, \ell }(r_{\star})={\widetilde {X}}_{\omega p \ell }^{*}(r_{\star })$.

Since (\ref{eq:radialpot}) takes a Schr\"odinger-like form, the Wronskian of any two linearly independent solutions is a nonzero constant.
In particular, considering the Wronskian of ${\widetilde {X}}_{\omega p \ell }$ and its complex conjugate, we find that
\begin{equation}
    \left| {\mathcal {R}}_{\omega p \ell }\right| ^{2} = 1.
\end{equation}
Due to our choice of boundary conditions, the scalar field flux through the boundary at $r\rightarrow \infty $ is zero, and the field is reflected at the boundary.  
In particular, this means that, unlike the situation for rotating asymptotically flat black holes, there is no classical superradiance in this case, and hence no superradiant instability \cite{Hawking:1999dp,Koga:2022vun}. 

\subsubsection{Heun form}
\label{sec:Heun}

To transform the radial equation (\ref{eq:radial}) into the form of a Heun differential equation \cite{Ronveaux:1995}, we follow the method of \cite{Aliev:2008yk,BarraganAmado:2018zpa,Noda:2022zgk,Koga:2022vun}.
In \cite{Aliev:2008yk,BarraganAmado:2018zpa,Noda:2022zgk,Koga:2022vun}, the authors use a different coordinate system from that employed here; this means that their expressions for the various constants introduced below differ from ours.
We first note that the radial equation (\ref{eq:radial}) has four regular singular points, at $r^{2}=r_{+}^{2}$, $r^{2}=r_{-}^{2}$, $r^{2}=r_{0}^{2}$ and $r^{2}=\infty $. 
Since the Heun differential equation also has four regular singular points, we anticipate that it will be possible to cast the radial equation (\ref{eq:radial}) in the Heun form using an appropriate transformation.

To this end, we define a new independent variable $z$ by \cite[Eq.~(40)]{Noda:2022zgk} (see also \cite{BarraganAmado:2018zpa,Koga:2022vun})
\begin{equation}
    z = \frac{r^{2}-r_{+}^{2}}{r^{2}-r_{0}^{2}}.
    \label{eq:zdef}
\end{equation}
The regular singular points are then at $z=0$ ($r=r_{+}$), $z=1$ ($r=\infty $), $z=z_{-}$ ($r=r_{-}$) and $z=\infty $ ($r=r_{0}$), where
\begin{equation}
    z_{-} = \frac{r_{-}^{2}-r_{+}^{2}}{r_{-}^{2}-r_{0}^{2}}.
\end{equation}
Next we define a new dependent variable ${\mathcal {X}}(z)$ by
\begin{equation}
    X(r) = z^{\theta _{+}}\left( z-1 \right)^{\theta _{\infty }} \left( z - z_{-} \right) ^{\theta _{-}} {\mathcal {X}}(z) ,
    \label{eq:calX1}
\end{equation}
where $\theta _{+}$, $\theta _{-}$ and $\theta _{\infty }$ are (possibly complex) constants to be determined.
We transform the radial equation (\ref{eq:radial}) to the new independent variable $z$ and substitute in for $X(r)$ from (\ref{eq:calX1}). 
The resulting differential equation takes the form
\begin{multline}
    0 = {\mathcal {X}}''(z)  + \left[ \frac{\gamma }{z} + \frac{\delta }{z-1} + \frac{\epsilon }{z-z_{-}} \right] {\mathcal {X}}'(z) \\ + {\mathfrak {V}}(z) {\mathcal {X}}(z),
    \label{eq:Heun1}
\end{multline}
where the constants $\gamma $, $\delta $ and $\epsilon $ are given by
\begin{equation}
    \gamma = 1 + 2 \theta _{+}, \quad \delta = -1 + 2 \theta _{\infty }, \quad \epsilon = 1 + 2 \theta _{-},
    \label{eq:Heunconstants}
\end{equation}
and ${\mathfrak {V}}(z)$ is a function of $z$ which is too lengthy to display here.
For (\ref{eq:Heun1}) to have the Heun form, we require ${\mathfrak {V}}(z)$ to take the form
\begin{equation}
    {\mathfrak {V}}(z) = \frac{\sigma \tau z - q}{z(z-1)(z-z_{-})} ,
    \label{eq:VHeun}
\end{equation}
where $\sigma $ and $\tau  $ are constants such that
\begin{equation}
    \sigma + \tau + 1 = \gamma + \delta + \epsilon  .
    \label{eq:alphabeta}
\end{equation}
The constraint (\ref{eq:alphabeta}) is satisfied for $\sigma $, $\tau $ given by
\begin{align}
    \sigma & = \theta _{+} + \theta _{-} + \theta _{\infty } + \theta _{0},
    \nonumber \\
    \tau  &  = \theta _{+} + \theta _{-} + \theta _{\infty } - \theta _{0},
\end{align}
where $\theta _{0}$ is another constant to be determined.
The constants $\theta _{+}$, $\theta _{-}$, $\theta _{\infty }$ and $\theta _{0}$ are found by requiring ${\mathfrak {V}}(z)$ to have the required form (\ref{eq:VHeun}). 
After a considerable amount of lengthy algebra, we find
\begin{subequations}
\label{eq:thetas}
\begin{align}
    \theta _{+} & = 
    \frac{i}{2\kappa _{+}}\left( \omega - p \Omega _{+} \right) ,
    \label{eq:theta+}
    \\
    \theta _{-} & = 
    \frac{i}{2\kappa _{-}}\left( \omega - p \Omega _{-} \right) ,
    \label{eq:theta-}
    \\
    \theta _{0} & =
    \frac{i}{2\kappa _{0}}\left( \omega - p \Omega _{0} \right) ,
    \label{eq:theta0}
    \\
    \theta _{\infty } & = 1 + {\sqrt {1+\frac{1}{4}\nu ^{2}L^{2} }},
    \label{eq:thetainf}
\end{align}
where $\nu $  is given in (\ref{eq:nu}), the constants $\kappa _{+}$ (\ref{eq:kappa+}) and $\Omega _{+}$ (\ref{eq:OmegaH}) can be written in the alternative form
\begin{align}
    \kappa _{+} & =  \frac{\left( r_{+}^{2}-r_{-}^{2}\right) \left( r_{+}^{2}-r_{0}^{2}\right) }{Lr_{+}^{2}{\sqrt {\left( r_{+}^{2}+r_{-}^{2}\right) \left( r_{-}^{2}+L^{2}\right) }} },
    \\
    \Omega _{+} & = \frac{ir_{-}r_{0}{\sqrt {r_{+}^{2}+L^{2}}}}{Lr_{+}{\sqrt {\left( r_{+}^{2}+r_{-}^{2}\right) \left( r_{-}^{2}+L^{2}\right)}}},
\end{align}
and we have defined quantities $\kappa _{-}$, $\Omega _{-}$, $\kappa _{0}$ and $\Omega _{0}$ in a similar fashion:
\begin{align}
    \kappa _{-} & =  \frac{\left( r_{-}^{2}-r_{0}^{2}\right) \left( r_{-}^{2}-r_{+}^{2}\right) }{Lr_{-}^{2}{\sqrt {\left( r_{+}^{2}+L^{2}\right) \left( r_{-}^{2}+r_{+}^{2}\right) }} },
    \\
    \Omega _{-} & = \frac{ir_{+}r_{0}{\sqrt {r_{-}^{2}+L^{2}}}}{Lr_{-}{\sqrt {\left( r_{+}^{2}+L^{2}\right) \left( r_{-}^{2}+r_{+}^{2}\right)}} } ,
    \\
    \kappa _{0} & =  \frac{i\left( r_{0}^{2}-r_{+}^{2}\right) \left( r_{0}^{2}-r_{-}^{2}\right) }{Lr_{0}^{2}{\sqrt {\left( r_{+}^{2}+L^{2}\right) \left( r_{-}^{2}+L^{2}\right) }} },
    \\
    \Omega _{0} & = \frac{ir_{+}r_{-}{\sqrt {r_{+}^{2}+r_{-}^{2}}}}{Lr_{0}{\sqrt {\left( r_{+}^{2}+L^{2}\right) \left( r_{-}^{2}+L^{2}\right)}} }.
\end{align}
All quantities under a square root sign in (\ref{eq:thetas}) are positive. 
Since $r_{0}^{2}<0$, we have that $r_{0}$ is purely imaginary. 
Therefore $\kappa _{\pm}$ and $\Omega _{\pm }$ are real, while $\kappa _{0}$ and $\Omega _{0}$ are purely imaginary.
This means that $\theta _{\pm }$ are purely imaginary but $\theta _{0}$ is real.
Since $\nu ^{2}L^{2}>-4$ (\ref{eq:bfbound}), we also have that $\theta _{\infty }$ is real. 
Finally, the constant $q$ is given by 
\begin{multline}
    q = 
    \frac{L^{2}}{4\left( r_{-}^{2}-r_{0}^{2}\right)}\left[ \omega ^{2}L^{2} -p^{2} -4\lambda -r_{+}^{2}\nu ^{2}\right] 
    \\ +z_{-}\left( 2 \theta _{+}\theta _{\infty } +\theta _{\infty }- \theta _{+} \right) 
    \\
    + \left( \theta _{+} + \theta _{-} \right) \left( \theta _{+}+\theta _{-}+1  \right) 
    -  \theta _{0}^{2}, 
\end{multline}
where $\nu $ can be found in (\ref{eq:nu}) and $\lambda $ in (\ref{eq:lambda}).
\end{subequations}

Let ${\mathsf {H}}l(z_{-},q;\sigma , \tau , \gamma, \delta ; z)$ denote the solution of (\ref{eq:Heun1}), with ${\mathfrak {V}}(z)$ given by (\ref{eq:VHeun}),  which is regular at $z=0$.
Then, linearly independent solutions of the radial equation (\ref{eq:Heun1}) near $z=1$ (that is, as $r\rightarrow \infty $) take the form \cite[Eq.~(46--47)]{Noda:2022zgk} (see also \cite{Koga:2022vun})
\begin{subequations}
    \begin{align}
       {\mathcal {X}}_{1}(z) & = {\mathsf {H}}l(1-z_{-},\sigma \tau - q; \sigma , \tau , \delta, \gamma; 1-z ),
       \label{eq:XHeun1}
       \\
       {\mathcal {X}}_{2}(z) & = (1-z)^{1-\delta } 
       \nonumber \\ 
       & \!\!\! \times {\mathsf{H}}l (1-z_{-}, {\widetilde{q}}; \sigma - \delta + 1, \tau -\delta + 1,  2-\delta , \gamma; 1-z) ,
       \label{eq:XHeun2}
    \end{align}
    where
    \begin{equation}
        {\widetilde {q}}=\left[ (1-z_{-})\gamma + \epsilon \right] (1-\delta )+\sigma \tau - q .
    \end{equation}
\end{subequations}
Since ${\mathsf{H}}l(z_{-},q;\sigma, \tau , \gamma, \delta ; 0)=1$ for any parameters $z_{-}$, $q$, $\sigma $, $\tau $, $\gamma $, $\delta$, we see that ${\mathcal {X}}_{1}(z)$ is regular as $z\rightarrow 1$, but ${\mathcal {X}}_{2}(z) \sim (1-z)^{2(1-\theta _{\infty })}$, where  $\theta _{\infty }$ is given by (\ref{eq:thetainf}). 
Since $\theta _{\infty }>1$, we see that ${\mathcal {X}}_{2}(z)$ diverges as $z\rightarrow 1$.
We therefore choose the solution ${\mathcal {X}}_{1}(z)$ to be the appropriate radial function,  so that
\begin{multline}
    X_{\omega p \ell}(r) = {\mathfrak {X}}_{\omega p \ell }z^{\theta _{+}}\left( z-1 \right)^{\theta _{\infty }} \left( z - z_{-} \right) ^{\theta _{-}}
    \\ \times {\mathsf {H}}l(1-z_{-},\sigma \tau - q; \sigma , \tau , \delta, \gamma; 1-z ) ,
    \label{eq:XHeunfinal}
\end{multline}
where ${\mathfrak {X}}_{\omega p \ell }$ is a constant to be determined in the next subsection.
In making the choice (\ref{eq:XHeunfinal}), we assume that the radial functions tend to zero as quickly as possible as $r\rightarrow \infty $.
The form of the radial function given in (\ref{eq:XHeunfinal}) is that which we shall use for our numerical computations in Sec.~\ref{sec:observables}, since the Heun functions ${\mathsf{H}}l(z_{-},q;\sigma , \tau , \gamma, \delta ; z)$ are built-in to {\tt {Mathematica}}.

\subsubsection{Matching the two forms of the radial function}
\label{sec:matching}

In the previous two subsections, we have derived two forms of the radial function $X_{\omega p \ell }(r)$, namely
\begin{subequations}
\label{eq:Xforms}
\begin{align} 
X_{\omega p \ell } (r) & = \dfrac{{\widetilde {X}}_{\omega p \ell }(r_{\star })}{r{\sqrt {h(r)}}},
\label{eq:X1}
\\
X_{\omega p \ell } (r) & = {\mathfrak {X}}_{\omega p \ell } z^{\theta _{+}}\left( z-1 \right)^{\theta _{\infty }} \left( z - z_{-} \right) ^{\theta _{-}} {\mathcal {X}}_{1}(z),
\label{eq:X2}
\end{align}
\end{subequations}
where ${\widetilde {X}}(r_{\star })$ has the asymptotic forms given in (\ref{eq:Xtildefinal}), the variable $z$ can be found in (\ref{eq:zdef}), and ${\mathcal {X}}_{1}(z)$ is the Heun function (\ref{eq:XHeun1}).
In Sec.~\ref{sec:normalization}, we will use the form (\ref{eq:X1}) near the past event horizon ${\mathcal {H}}^{-}$ to find the overall normalization of the modes. 
In this subsection we therefore seek the constant ${\mathfrak {X}}_{\omega p \ell }$, so that, near the past horizon, the two asymptotic forms of (\ref{eq:Xforms}) match.
Our analysis follows that in \cite{Noda:2022zgk}, although we use different coordinates, in particular our definition of the tortoise coordinate (\ref{eq:tortoise}) differs from that in \cite{Noda:2022zgk}.

From (\ref{eq:Xtildefinal}), near the past horizon ${\mathcal {H}}^{-}$ (where $r=r_{+}$) we have
\begin{equation}
    X_{\omega p \ell }(r) \sim \dfrac{e^{i{\widetilde {\omega }}r_{\star }}}{r_{+}{\sqrt {h(r_{+})}}} ,
    \label{eq:X1hor}
\end{equation}
where the tortoise coordinate $r_{\star }$ is determined by the differential equation (\ref{eq:tortoise}).
As $r\sim r_{+}$, we have, for $r> r_{+}$
\begin{equation}
    r_{\star } \sim \frac{1}{2\kappa _{+}}\log \left( \frac{r-r_{+}}{r_{+}} \right)  + \ldots  
    \label{eq:rstarhor}
\end{equation}
where $\kappa _{+}$ is the surface gravity (\ref{eq:kappa+}).  
The next term in (\ref{eq:rstarhor}) is a constant which leads to an irrelevant phase in (\ref{eq:X1hor}).
Substituting (\ref{eq:rstarhor}) into (\ref{eq:X1hor}), we have
\begin{equation}
    X_{\omega p \ell }(r) \sim \frac{1}{r_{+}{\sqrt {h(r_{+})}}} \left( \frac{r-r_{+}}{r_{+}}
    \right)  ^{\frac{i{\widetilde {\omega }}}{2\kappa _{+}}}.
    \label{eq:X1compare}
\end{equation}
This is the form we wish to match to the expression (\ref{eq:XHeunfinal}) for $X_{\omega p \ell }(r)$ involving a Heun function.

The expression (\ref{eq:XHeunfinal}) involves a Heun function whose asymptotics as $z\rightarrow 0 $ are not readily obtained. 
However, near $z=0$, the two linearly independent solutions of the radial equation (\ref{eq:Heun1}) are
\cite[Eq.~(44--45)]{Noda:2022zgk}
\begin{subequations}
\label{eq:Heun0}
\begin{align}
    {\mathcal {X}}_{3}(z) & = {\mathsf{H}}l(z_{-},q;\sigma, \tau, \gamma, \delta ; z) ,
    \\
    {\mathcal {X}}_{4}(z) & = z^{1-\gamma }  {\mathsf{H}}l(z_{-},{\overline {q}};\sigma - \gamma + 1, \tau - \gamma + 1,2- \gamma, \delta ; z) ,
\end{align}
where
\begin{equation}
    {\overline {q}} = \left( z_{-}\delta + \epsilon  \right) \left( 1 - \gamma \right) + q . 
\end{equation}
\end{subequations}
Therefore we can write the Heun function ${\mathcal {X}}_{1}(z)$ (\ref{eq:XHeun1}) as a linear combination of those in (\ref{eq:Heun0}), as follows:
\begin{equation}
    {\mathcal {X}}_{1}(z) = {\mathcal {W}}_{3}{\mathcal {X}}_{3}(z) + {\mathcal {W}}_{4} {\mathcal {X}}_{4}(z) ,
    \label{eq:linearcombHeun}
\end{equation}
where ${\mathcal {W}}_{3,4}$ are constants which can be expressed as the ratios of Wronskians $W\{ \, , \}$ of Heun functions:
\begin{equation}
    {\mathcal {W}}_{3} = \frac{ W\{ {\mathcal {X}}_{1}, {\mathcal {X}}_{4} \} }{W\{ {\mathcal {X}}_{3}, {\mathcal {X}}_{4} \} }, \qquad
    {\mathcal {W}}_{4}  =  - \frac{ W\{ {\mathcal {X}}_{1}, {\mathcal {X}}_{3} \} }{W\{ {\mathcal {X}}_{3}, {\mathcal {X}}_{4} \} } .
    \label{eq:Wronskians}
\end{equation}
We note that while the Wronskians depend on $z$, their ratios are constants \cite{Noda:2022zgk}.
Alternatively, the constants ${\mathcal {W}}_{3,4}$ can be written in terms of the Nekrasov-Shatashvili partition function and derived using Liouville conformal field theory \cite{Bonelli:2022ten,Aminov:2023jve,BarraganAmado:2024tfu}. 
However, the ratios of Wronskians (\ref{eq:Wronskians}) are convenient for our numerical computations in Sec.~\ref{sec:observables}.

Using (\ref{eq:X2}, \ref{eq:linearcombHeun}), we have, for $z\sim 0$, since $\theta _{+}$ (\ref{eq:theta+}) is purely imaginary, 
\begin{equation}
    X_{\omega p \ell }(r) \sim {\mathfrak {X}}_{\omega p \ell } {\mathcal {W}}_{3} z^{\theta _{+}}(-1)^{\theta _{\infty }}\left( - z_{-} \right) ^{\theta _{-}}  .
\end{equation}
Next, since, for $r\sim r_{+}$ 
\begin{equation}
    z \sim \frac{2r_{+}(r-r_{+})}{r_{+}^{2}-r_{0}^{2}}, 
\end{equation}
we have
\begin{equation}
    X_{\omega p \ell }(r) \sim {\mathfrak {X}}_{\omega p \ell } {\mathcal {W}}_{3} (-1)^{\theta _{\infty }}\left( - z_{-} \right) ^{\theta _{-}} \left( \frac{2r_{+}[ r-r_{+}]}{r_{+}^{2}-r_{0}^{2}} \right) ^{\theta _{+}}.
    \label{eq:X2compare}
\end{equation}
Comparing (\ref{eq:X1compare}, \ref{eq:X2compare}), the constant ${\mathfrak {X}}_{\omega p \ell }$ is determined to be
\begin{multline}
    {\mathfrak {X}}_{\omega p \ell } =\frac{1}{{\mathcal {W}}_{3}r_{+}{\sqrt {h(r_{+})}}} (-1)^{-\theta _{\infty }}\left( - z_{-} \right) ^{-\theta _{-}} 
    \\ \times \left( \frac{1}{2r_{+}^{2}} \left[ r_{+}^{2} - r_{0}^{2} \right] \right) ^{\theta _{+}}.
    \label{eq:frakX}
\end{multline}
We have already chosen the overall phase in our derivation of the form (\ref{eq:X1compare}) of the radial function near the past horizon.
As a result, we need to keep all the phases in the constant ${\mathfrak {X}}_{\omega p \ell }$.

\subsection{Normalization of the scalar field modes}
\label{sec:normalization}

The final step in the construction of an orthonormal basis of scalar field modes is to ensure that the modes are normalized.
The scalar field modes take the form
\begin{multline}
    \phi _{\omega p \ell m }(t,r,\theta ,\phi , \psi ) \\ =  
    {\mathcal {N}}_{\omega p \ell} e^{-i\omega t}e^{ip \psi } X_{\omega p \ell }(r) {}_{-p/2}Y_{\ell }^{m}(\theta ,\varphi ) ,
    \label{eq:modeN}
\end{multline}
where the quantum numbers are the frequency $\omega \in {\mathbb{R}}$, the azimuthal quantum number $p\in {\mathbb{Z}}$, the total angular momentum quantum number $\ell = |p/2|, |p/2|+1, |p/2|+2,\ldots $ and the further angular quantum number $m=-\ell , -\ell + 1, \ldots, \ell -1, \ell $.
In (\ref{eq:modeN}), the constant ${\mathcal {N}}_{\omega p \ell }$ is a normalization constant, and we have anticipated our result (derived below) that this depends only on $\omega $, $p$ and $\ell $.
The only dependence on the quantum number $m$ is in the spin-weighted spherical harmonics ${}_{-p/2}Y_{\ell }^{m}(\theta ,\varphi ) $, which will prove to be useful for simplifying the expectation values of observables in Sec.~\ref{sec:observables}.

The modes (\ref{eq:modeN}) are normalized using the inner product $\langle \Phi _{1},\Phi _{2}\rangle $ of any two solutions of (\ref{eq:KG}), which is defined by
\begin{equation}
    \langle \Phi _{1},\Phi _{2}\rangle  = i \int _{\Sigma } \left[  \left( \nabla _{\mu }\Phi _{1}^{*} \right)  \Phi _{2}  - \Phi _{1}^{*} \nabla _{\mu } \Phi _{2}  \right] \, d\Sigma ^{\mu } , 
    \label{eq:inner}
\end{equation}
where a star denotes complex conjugate. 
The space-like hypersurface $\Sigma $ extends from the bifurcation two-sphere to the space-time boundary.
Since the black hole is asymptotically AdS, the surface $\Sigma $ is not a Cauchy surface. 
The boundary conditions we have imposed on the radial function $X_{\omega p \ell }(r)$ ensure that the scalar field modes vanish at the space-time boundary where $r\rightarrow \infty $.
As a result, the inner product (\ref{eq:inner}) is independent of the choice of the surface $\Sigma $.

We take $\Sigma $ to be a surface close to the past horizon ${\mathcal {H}}^{-}$ (where the Kruskal coordinate $V=0$ (\ref{eq:Kruskal})), parameterized by the Kruskal coordinate $U$ (\ref{eq:Kruskal}). 
On this surface, using the form (\ref{eq:X1}) of the radial function, and the asymptotic form (\ref{eq:Xtildefinal}), we have
\medskip
\begin{multline}
     \phi _{\omega p \ell m }(t,r,\theta ,\phi , \psi ) \\ \sim  
   \frac{1}{r{\sqrt {h(r)}}} {\mathcal {N}}_{\omega p \ell} e^{-i\omega t}e^{ip \psi }e^{i{\widetilde {\omega }}r_{\star }}  {}_{-p/2}Y_{\ell }^{m}(\theta ,\varphi ) 
   \\ = 
   \frac{1}{r{\sqrt {h(r)}}} {\mathcal {N}}_{\omega p \ell} e^{-i{\widetilde {\omega }}u}e^{ip \psi _{+}} {}_{-p/2}Y_{\ell }^{m}(\theta ,\varphi ) , \,\,
   \label{eq:modeH-}
\end{multline}
where the corotating angle $\psi _{+}$ is given by (\ref{eq:psi+}). 
The surface element is $d\Sigma ^{\mu }=n^{\mu} \,  d\Sigma $, where $n^{\mu }$ is the normal
\begin{equation}
    n^{\mu } = -\frac{2\kappa _{+}^{2}UV}{f(r)^{2}} \delta ^{\mu }_{U} ,
\end{equation}
with $\kappa _{+}$ the surface gravity (\ref{eq:kappa+}), and 
\begin{equation}
    d\Sigma = {\sqrt {-g_{\rm {K}}}} \, dU \, d\theta \, d\varphi \, d\psi _{+}. 
\end{equation}
Using the following result for the square root of minus the determinant of the metric (\ref{eq:metricKruskal}): 
\begin{multline}
    {\sqrt { -g_{\rm {K}}}} \\ = \frac{r^{2}h(r)}{8\kappa _{+}^{2}UV}\left[ 2\kappa _{+}f(r)^{2} + \left\{ \Omega (r) - \Omega _{+} \right\} ^{2} h(r)^{2} \right] \sin \theta ,
\end{multline}
we have
\begin{multline}
    d\Sigma ^{\mu } = -\frac{r^{2}h(r)}{4\kappa _{+}f(r)^{2}} \left[ 2\kappa _{+}f(r)^{2} + \left\{ \Omega (r) - \Omega _{+} \right\} ^{2} h(r)^{2} \right] 
    \\ \times \, \delta ^{\mu }_{U} \sin \theta  \, dU \, d\theta \, d\varphi \, d\psi _{+}  
    \\ \sim -\frac{1}{2} r^{2}h(r)  \, \delta ^{\mu }_{U} \, \sin \theta \,  dU \, d\theta \, d\varphi \, d\psi _{+}   ,
\end{multline}
where in the last line we give the leading-order expression near ${\mathcal {H}}^{-}$.
Changing integration variable to $u$ (\ref{eq:null}) rather than $U$ (\ref{eq:Kruskal}), the inner product of two scalar field modes is
\begin{widetext}
\begin{multline}
    \langle \phi _{\omega p \ell m }(t,r,\theta ,\phi , \psi ) , \phi _{\omega 'p' \ell ' m' }(t,r,\theta ,\phi , \psi )\rangle  \\ = 
    \frac{1}{2} \int _{u=-\infty }^{\infty } \int _{\theta =0}^{\pi }\int_{\varphi =0}^{4\pi }
    \int _{\psi _{+}=0}^{2\pi }
    {\mathcal {N}}_{\omega p \ell}^{*} {\mathcal {N}}_{\omega 'p' \ell'}\left[{\widetilde {\omega }} + {\widetilde {\omega }}' \right] e^{i\left( {\widetilde {\omega}} - {\widetilde {\omega }}' \right) u}e^{-i\left( p - p'\right) \psi _{+}} X_{\omega p \ell }^{*}(r) X_{\omega 'p' \ell '}(r)
    \\ \times 
    {}_{-p/2}Y_{\ell }^{m*}(\theta ,\varphi ) {}_{-p'/2}Y_{\ell '}^{m'}(\theta ,\varphi ) 
    \, du \, d\theta \, d\varphi \, d\psi .
    \label{eq:normintermediate}
    \end{multline} 
 \end{widetext}
Using the normalization (\ref{eq:swshnorm}) of the spin-weighted spherical harmonics and the results
\begin{subequations}
\begin{align}
    \int _{0}^{2\pi } e^{-i\left( p - p'\right) \psi _{+}} \, d\psi _{+} & =2 \pi  \delta _{pp'},
    \\
    \int _{-\infty }^{\infty } e^{i\left( {\widetilde {\omega}} - {\widetilde {\omega }}' \right) u} \, du 
    & =2\pi  \delta \left( {\widetilde {\omega}} - {\widetilde {\omega }}' \right) ,
\end{align}
\end{subequations}
the inner product (\ref{eq:normintermediate}) becomes
\begin{multline}
    \langle \phi _{\omega p \ell m }(t,r,\theta ,\phi , \psi ) , \phi _{\omega 'p' \ell ' m' }(t,r,\theta ,\phi , \psi )\rangle  \\ = 
    8\pi ^{2} {\widetilde {\omega }} \left| {\mathcal {N}}_{\omega p \ell} \right| ^{2}  \, \delta \left( {\widetilde {\omega}} - {\widetilde {\omega }}' \right) \delta _{pp'} \delta _{\ell \ell'} \delta _{mm'} ,
    \label{eq:innerfinal}
\end{multline}
and we therefore take the normalization constant to be,  making a choice of phase,
\begin{equation}
    {\mathcal {N}}_{\omega p \ell}  = \frac{1}{2{\sqrt {2}}\pi \left| {\widetilde {\omega }} \right|}.
    \label{eq:normN}
\end{equation}
Note that, from (\ref{eq:innerfinal}), modes having ${\widetilde {\omega }}>0$ have positive ``norm'', whereas those with ${\widetilde {\omega }}<0$ have negative ``norm''. 
This will be important when we come to perform the canonical quantization of the field in Sec.~\ref{sec:quantum}.

\section{Canonical quantization of the scalar field}
\label{sec:quantum}

In the previous section we constructed an orthonormal set of scalar field modes given by (\ref{eq:modeN}), with normalization constant (\ref{eq:normN}).  
We expect this set to form a basis of mode solutions of the scalar field equation (\ref{eq:KG}) regular at the event horizon and vanishing as quickly as possible as $r\rightarrow \infty $. 
In this section we use these modes to perform the canonical quantization of the scalar field. 
We construct two quantum states: a Boulware state $|{\rm {B}}\rangle $ and a Hartle-Hawking state $|{\rm {H}}\rangle $.

\subsection{Boulware state}
\label{sec:Boulware}

In the canonical quantization of the scalar field, we need to make a choice of positive and negative frequency modes. 
It is useful to write the modes (\ref{eq:modeN}) in terms of the corotating coordinate $\psi _{+}$ (\ref{eq:psi+}):
\begin{multline}
    \phi _{\omega p \ell m}(t,r,\theta ,\varphi,  \psi ) 
\\
    = \frac{1}{2{\sqrt {2}}\pi |{\widetilde {\omega }}|} e^{-i{\widetilde {\omega }}t}e^{ip \psi _{+}} X_{\omega p \ell }(r) {}_{-p/2}Y_{\ell }^{m}(\theta ,\varphi )  .
    \label{eq:modetilde}
\end{multline}
Hence a natural definition of positive frequency with respect to time $t$ is to set ${\widetilde {\omega }}>0$. 
From (\ref{eq:innerfinal}), modes having ${\widetilde {\omega }}>0$ have positive ``norm'' and hence this is an appropriate choice of positive frequency, which will lead to a consistent quantization (see, for example, \cite{Letaw:1979wy,Duffy:2002ss,Ambrus:2014uqa} for more detailed discussion of this requirement in the context of rotating quantum states in Minkowski space-time, and also \cite[Sec.~IIIA]{Balakumar:2022yvx}).   
Therefore we define positive frequency modes $\phi ^{+}_{\omega p\ell m }$ to take the form (\ref{eq:modeN}) with ${\widetilde {\omega }}>0$.
Since we are considering a real scalar field, we may take the negative frequency modes $\phi ^{-}_{\omega p\ell m }$ to simply be the complex conjugates of the positive frequency modes. 
We therefore expand the classical scalar field $\Phi $ as follows:
\begin{multline}
    \Phi (t,r, \theta, \varphi,  \psi ) =  \sum _{p=-\infty }^{\infty } \sum _{\ell = |p|/2}^{\infty } \sum _{m=-\ell }^{\ell } \int _{{\widetilde {\omega }}=0}^{\infty } \, d{\widetilde {\omega }}
    \\ \times \left[
    a_{\omega p \ell m} \phi _{\omega p\ell m } + a^{\dagger }_{\omega p \ell m}\phi _{\omega p\ell m } ^{*} 
    \right] ,
\end{multline}
where $a_{\omega p \ell m}$ and $a^{\dagger }_{\omega p \ell m}$ are constants.
To quantize the field, the expansion coefficients $a_{\omega p \ell m}$ and $a^{\dagger }_{\omega p \ell m}$ are promoted to operators ${\hat {a}}_{\omega p \ell m}$ and ${\hat {a}}^{\dagger }_{\omega p \ell m}$ respectively:
\begin{multline}
    {\hat {\Phi }} (t,r, \theta, \varphi,  \psi ) =  \sum _{p=-\infty }^{\infty } \sum _{\ell = |p|/2}^{\infty } \sum _{m=-\ell }^{\ell } \int _{{\widetilde {\omega }}=0}^{\infty } \, d{\widetilde {\omega }}
    \\ \times \left[
    {\hat {a}}_{\omega p \ell m} \phi _{\omega p\ell m } + {\hat {a}}^{\dagger }_{\omega p \ell m}\phi _{\omega p\ell m } ^{*} 
    \right] .
    \label{eq:PhiB}
\end{multline}
The operators ${\hat {a}}_{\omega p \ell m}$ and ${\hat {a}}^{\dagger }_{\omega p \ell m}$ satisfy the usual bosonic commutation relations (for ${\widetilde {\omega }}>0$, and all $p$, $\ell $ and $m$)
\begin{subequations}
\begin{align}
    \left[ {\hat {a}}_{\omega p \ell m}, {\hat {a}}_{\omega ' p' \ell' m'} \right] & = 0,
    \\
    \left[ {\hat {a}}_{\omega p \ell m}^{\dagger }, {\hat {a}}_{\omega ' p' \ell' m'}^{\dagger } \right] & = 0,
    \\
    \left[ {\hat {a}}_{\omega p \ell m}, {\hat {a}}_{\omega ' p' \ell' m'}^{\dagger } \right] & =
    \delta ({\widetilde {\omega }} - {\widetilde {\omega }}') \delta _{pp'}\delta _{\ell \ell'} \delta _{mm'} .
\end{align}
\end{subequations}
Hence we interpret the operators ${\hat {a}}_{\omega p \ell m}$ as particle annihilation operators and the operators ${\hat {a}}_{\omega p \ell m}^{\dagger }$ as particle creation operators.
The Boulware state $|{\rm {B}}\rangle $ is then defined as the state annihilated by the annihilation operators ${\hat {a}}_{\omega p \ell m}$:
\begin{equation}
    {\hat {a}}_{\omega p \ell m}|{\rm {B}}\rangle  = 0.
\end{equation}
The Boulware state $|{\rm {B}}\rangle $ is the natural definition of a ground state for an observer at constant $(r,\theta, \varphi, \psi _{+})$, in other words an observer corotating with the event horizon of the black hole.

\subsection{Hartle-Hawking state}
\label{sec:HH}

The Boulware state $|{\rm {B}}\rangle $ was defined using a notion of positive frequency with respect to the time coordinate $t$. 
This coordinate is not regular across the event horizon of the black hole.
We now construct the Hartle-Hawking state $|{\rm {H}}\rangle $, by using an alternative definition of positive frequency.
In particular, we shall use a set of modes which are positive frequency with respect to $U$ (\ref{eq:Kruskal}), the Kruskal coordinate which parameterizes the past event horizon ${\mathcal {H}}^{-}$. 
We follow the method of \cite{Unruh:1976db,Novikov:1989sz}.

We start with the form of the modes (\ref{eq:modeH-}) near the past event horizon ${\mathcal {H}}^{-}$ (where $V=0$), written in terms of Kruskal coordinates:
\begin{multline}
    \phi _{\omega p\ell m} \sim  \frac{1}{2{\sqrt {2}}\pi \left| {\widetilde {\omega }} \right|  r{\sqrt {h(r)}}} \exp \left[ i\frac{{\widetilde {\omega }}}{\kappa _{+}}
    \ln \left(-\kappa _{+}U\right) \right]  \\ \times  \Theta \left( - \kappa _{+}U \right)  e^{ip\psi _{+}}{}_{-p/2}Y_{\ell }^{m}(\theta, \varphi ).
    \label{eq:modesH-K}
\end{multline}
Here we have used the definition (\ref{eq:Kruskal}) of the Kruskal coordinates, and included the Heaviside step function $\Theta ({\mathsf {X}})$
\begin{equation}
    \Theta ({\mathsf {X}}) = \begin{cases}
        1, & {\mathsf {X}} \ge 0 ,\\
        0, &  {\mathsf {X}}<0, 
    \end{cases}
\end{equation}
so that the argument of the logarithm is positive.
The quantity $\kappa _{+}$ is the surface gravity (\ref{eq:kappa+}), and ${\widetilde {\omega }}$ the shifted frequency (\ref{eq:tildeomega}). 
From (\ref{eq:innerfinal}), the modes (\ref{eq:modesH-K}) have positive ``norm'' when ${\widetilde {\omega }}>0$ and negative ``norm'' when ${\widetilde {\omega }}<0$.

The modes constructed in Sec.~\ref{sec:scalar} are nonzero in region I of the extended space-time (see Fig.~\ref{fig:penrose}), but vanish in region IV.
A set of modes which are nonzero in region IV (but vanish in region I) can be constructed by making the transformation $U\rightarrow - U$, $V\rightarrow - V$ in the modes (\ref{eq:modeN}).
We denote the resulting modes by ${\widetilde {\phi }} _{\omega p \ell m}$.
Near the surface $V=0$, these take the form
\begin{multline}
    {\widetilde {\phi }}_{\omega p\ell m} \sim  \frac{1}{2{\sqrt {2}}\pi \left| {\widetilde {\omega }} \right|  r{\sqrt {h(r)}}} \exp \left[ i\frac{{\widetilde {\omega }}}{\kappa _{+}}
    \ln \left(\kappa _{+}U\right) \right]  \\ \times  \Theta \left( \kappa _{+}U \right)  e^{ip\psi _{+}}{}_{-p/2}Y_{\ell }^{m}(\theta, \varphi ).
\end{multline}
These modes have positive ``norm'' when ${\widetilde {\omega }}<0$ and negative ``norm'' when ${\widetilde {\omega }}>0$. 

We now make use of the following Lemma (from \cite[App.~H]{Novikov:1989sz}), valid for all real ${\mathfrak{q}}$ and arbitrary positive ${\mathfrak {p}}$:
\begin{multline}
    \int _{-\infty }^{\infty } d{\mathsf{X}} \, e^{-i{\mathfrak{p}}{\mathsf {X}}} \left[ 
    e^{-i{\mathfrak{q}} \ln ({\mathsf{X}})} \Theta \left( {\mathsf{X}} \right) 
    \right. \\ \left. 
    + e^{-\pi {\mathfrak{q}}} e^{i{\mathfrak{q}} \ln (-{\mathsf{X}})} \Theta \left( -{\mathsf{X}} \right) 
    \right]
     = 0 .
\end{multline}
Setting ${\mathfrak {q}}=-{\widetilde {\omega }}/\kappa _{+}$ we find that, for ${\mathfrak {p}}>0$,
\begin{equation}
    \int _{-\infty }^{\infty } d{U} \, e^{-i{\mathfrak{p}}{U}} \left[ 
    {\widetilde {\phi }}_{\omega p\ell m} + e^{\frac{\pi {\widetilde {\omega }}}{\kappa _{+}} } \phi _{\omega p\ell m} 
    \right] 
    =0. 
\end{equation}
We deduce that the modes ${\widetilde {\phi }} _{\omega p\ell m} + e^{\frac{\pi {\widetilde {\omega }}}{\kappa } } \phi _{\omega p\ell m} $ have positive frequency with respect to $U$ for {\em {all}} values of ${\widetilde {\omega }}$.
Since the modes $\phi _{\omega p\ell m}$ (and hence also the modes ${\widetilde {\phi }} _{\omega p\ell m}$) are already normalized, a suitable set of normalized modes having positive frequency with respect to $U$ is, for all ${\widetilde {\omega}}$,
\begin{subequations}
    \label{eq:chimodes}
\begin{equation}
    \chi _{\omega p \ell m}^{+} =
    \frac{1}{{\sqrt { 2 \sinh \left| \frac{\pi {\widetilde {\omega }}}{\kappa _{+}} \right| }}}
    \left[ e^{\frac{\pi {\widetilde {\omega }}}{2\kappa _{+}} } \phi _{\omega p\ell m} +
    e^{-\frac{\pi {\widetilde {\omega }}}{2\kappa _{+}} } {\widetilde {\phi }} _{\omega p\ell m} 
    \right]  .
    \label{eq:chimodes+}
\end{equation}
Using a similar argument, a set of normalized modes having negative frequency with respect to $U$ is, again for all ${\widetilde {\omega }}$:
\begin{equation}
    \chi _{\omega p \ell m}^{-} =
    \frac{1}{{\sqrt { 2 \sinh \left| \frac{\pi {\widetilde {\omega }}}{\kappa _{+}} \right| }}}
    \left[ e^{-\frac{\pi {\widetilde {\omega }}}{2\kappa _{+}} } \phi _{\omega p\ell m} +
    e^{\frac{\pi {\widetilde {\omega }}}{2\kappa _{+}} } {\widetilde {\phi }} _{\omega p\ell m} 
    \right]  .
\end{equation}
\end{subequations}
In order to obtain a manifestly real expansion for the scalar field, we shall use the fact that the complex conjugates of the positive frequency modes (\ref{eq:chimodes+}) also have negative frequency for all values of ${\widetilde {\omega }}$.
Since the angular (\ref{eq:ang2}) and radial (\ref{eq:radialpot}) equations are invariant under the mapping $(\omega, p, m)\rightarrow (-\omega , -p, -m)$, the complex conjugates of the scalar field modes $\phi ^{*}_{\omega p \ell m }$ are equal, up to an irrelevant phase, to the mode functions $\phi _{-\omega , -p, \ell, -m}$.
The same is true for the mode functions ${\widetilde {\phi }}_{\omega p \ell m}$.
Therefore the complex conjugates $(\chi _{\omega , p, \ell m }^{+})^{*}$ are equal, again up to an irrelevant phase, to the mode functions $\chi ^{-}_{-\omega , -p, \ell ,-m}$.

Expanding the quantum scalar field in terms of the modes (\ref{eq:chimodes}) gives

\begin{multline}
    {\hat {\Phi }} = \sum _{p=-\infty }^{\infty }
    \sum _{\ell = |p|/2}^{\infty } \sum _{m=-\ell }^{\ell } \int _{{\widetilde {\omega }}=-\infty }^{\infty }  \, d{\widetilde {\omega }}
    \\ \times 
    \left[ 
    {\hat {b}}_{\omega p \ell m }\chi _{\omega p \ell m}^{+} 
    + {\hat {b}}^{\dagger} _{\omega p \ell m} \left( \chi _{\omega p \ell m}^{+}\right) ^{*}
    \right] ,
    \label{eq:PhiH}
\end{multline}
where the particle annihilation operators ${\hat {b}}_{\omega p \ell m }$ and particle creation operators ${\hat {b}}_{\omega p \ell m }^{\dagger}$ satisfy the usual commutation relations (for all ${\widetilde {\omega }}$, $p$, $\ell $, $m$):
\begin{subequations}
\begin{align}
    \left[ {\hat {b}}_{\omega p \ell m}, {\hat {b}}_{\omega ' p' \ell' m'} \right] & = 0,
    \\
    \left[ {\hat {b}}_{\omega p \ell m}^{\dagger }, {\hat {b}}_{\omega ' p' \ell' m'}^{\dagger } \right] & = 0,
    \\
    \left[ {\hat {b}}_{\omega p \ell m}, {\hat {b}}_{\omega ' p' \ell' m'}^{\dagger } \right] & =
    \delta ({\widetilde {\omega }} - {\widetilde {\omega }}') \delta _{pp'}\delta _{\ell \ell'} \delta _{mm'} .
\end{align}
\end{subequations}
If we consider only region I of the space-time (see Fig.~\ref{fig:penrose}), the modes ${\widetilde {\phi }}_{\omega p \ell m }$ vanish, and (\ref{eq:PhiH}) reduces to
\begin{multline}
        {\hat {\Phi }} = \sum _{p=-\infty }^{\infty }
    \sum _{\ell = |p|/2}^{\infty } \sum _{m=-\ell }^{\ell } \int _{{\widetilde {\omega }}=-\infty }^{\infty }   \, d{\widetilde {\omega }} \,
    \frac{1}{{\sqrt { 2 \sinh \left| \frac{\pi {\widetilde {\omega }}}{\kappa _{+}} \right| }}}
    \\ \times 
    e^{\frac{\pi {\widetilde {\omega }}}{2\kappa _{+}} } 
    \left[ 
     {\hat {b}}_{\omega p \ell m } \phi _{\omega p\ell m} 
    + {\hat {b}}^{\dagger} _{\omega p \ell m}   \phi _{\omega p\ell m} ^{*}
    \right] .
    \label{eq:PhiH1}
\end{multline}
The Hartle-Hawking state $|{\rm {H}}\rangle $ is defined as the state annihilated by all the operators ${\hat {b}}_{\omega p \ell m} $:
\begin{equation}
    {\hat {b}}_{\omega p \ell m}|{\rm {H}}\rangle  = 0.
\end{equation}
The Hartle-Hawking state $|{\rm {H}}\rangle $ is the natural definition of a ``vacuum'' state for an observer freely-falling towards the event horizon of the black hole.

\section{Quantum scalar field observables}
\label{sec:observables}

Having defined the Boulware $|{\rm {B}}\rangle $ and Hartle-Hawking $|{\rm {H}}\rangle $ states in the previous section, we now study the expectation values of the square of the quantum scalar field operator (also known as the ``vacuum polarization'' [the terminology we shall adopt here] or the ``scalar condensate''), and the SET operator in these two states.
Since a practical method for computing renormalized expectation values has yet to be developed for a higher-dimensional rotating black hole, we focus on the differences in expectation values between these two states, since these differences do not require renormalization.

\subsection{Vacuum polarization}
\label{sec:vacpol}

The simplest nontrivial expectation value is the square of the scalar field $\langle {\hat {\Phi }}^{2} \rangle $, which we term the ``vacuum polarization''.
Using the expansion (\ref{eq:PhiB}) for the scalar field operator, we find that the unrenormalized vacuum polarization in the Boulware state $|{\rm {B}}\rangle $ is
\begin{align}
    \langle {\rm {B}} | {\hat {\Phi }}^{2} | {\rm {B}} \rangle &  = 
    \sum _{p=-\infty }^{\infty } \sum _{\ell = |p|/2}^{\infty } \sum _{m=-\ell }^{\ell }
    \int _{{\widetilde {\omega }}=0}^{\infty } d{\widetilde {\omega }}
    \left| \phi _{\omega p \ell m } \right| ^{2} 
    \nonumber \\ & 
    = \frac{1}{2} \sum _{p=-\infty }^{\infty } \sum _{\ell = |p|/2}^{\infty } \sum _{m=-\ell }^{\ell }
    \int _{{\widetilde {\omega }}=-\infty }^{\infty } d{\widetilde {\omega }}
    \left| \phi _{\omega p \ell m } \right| ^{2} ,
    \label{eq:VPB}
\end{align}
where we have used the result $|\phi _{\omega p \ell m}|^{2}=|\phi _{-\omega , -p, \ell , -m}|^{2}$.
Similarly, using the expansion (\ref{eq:PhiH1}) for the scalar field operator gives the unrenormalized vacuum polarization in region I, when the scalar field is in the Hartle-Hawking state $|{\rm {H}} \rangle $, to be
\begin{align}
    \langle {\rm {H}} | {\hat {\Phi }}^{2} | {\rm {H}} \rangle &  = \frac{1}{2}
    \sum _{p=-\infty }^{\infty } \sum _{\ell = |p|/2}^{\infty } \sum _{m=-\ell }^{\ell }
    \int _{{\widetilde {\omega }}=-\infty }^{\infty } d{\widetilde {\omega }}
    \nonumber \\ & \qquad  \times 
    \left| \phi _{\omega p \ell m } \right| ^{2} \exp \left(  \frac{\pi {\widetilde {\omega }}}{\kappa _{+}} \right) 
    \cosech \left| \frac{\pi {\widetilde {\omega }}}{\kappa _{+}} \right| 
    \nonumber \\ & =
    \frac{1}{2}\sum _{p=-\infty }^{\infty } \sum _{\ell = |p|/2}^{\infty } \sum _{m=-\ell }^{\ell }
    \int _{{\widetilde {\omega }}=-\infty }^{\infty } d{\widetilde {\omega }}
    \nonumber \\ & \qquad \qquad \qquad  \times 
    \left| \phi _{\omega p \ell m } \right| ^{2} \coth \left| \frac{\pi {\widetilde {\omega }}}{\kappa _{+}} \right| ,
    \label{eq:VPH}
\end{align}
again using $|\phi _{\omega p \ell m}|^{2}=|\phi _{-\omega , -p, \ell , -m}|^{2}$. Both the expectation values (\ref{eq:VPB}, \ref{eq:VPH}) can be written as integrals over ${\widetilde {\omega }}>0$ only; however we have found it convenient in our computations to use the forms involving integrals over all ${\widetilde {\omega }}$.
In this paper we do not address the technically challenging question of renormalizing the above expectation values. 
Instead, we consider the difference between the two expectation values (\ref{eq:VPB}, \ref{eq:VPH}):
\begin{multline}
\Delta {\hat {\Phi }}^{2} 
= 
    \langle {\rm {H}} | {\hat {\Phi }}^{2} | {\rm {H}} \rangle  - \langle {\rm {B}} | {\hat {\Phi }}^{2} | {\rm {B}} \rangle
   \\ = \sum _{p=-\infty }^{\infty } \sum _{\ell = |p|/2}^{\infty } \sum _{m=-\ell }^{\ell }
    \int _{{\widetilde {\omega }}=-\infty }^{\infty } d{\widetilde {\omega }}
    \\ \times  \frac{\left| \phi _{\omega p \ell m } \right| ^{2} }{\exp \left( 2\pi \left| {\widetilde {\omega }}\right| /{\kappa _{+}} \right) -1}.
    \label{eq:VPH-B}
\end{multline}
Assuming that both the states $|{\rm {H}}\rangle $ and $|{\rm {B}}\rangle$ are Hadamard states in the region exterior to the event horizon, we expect that this difference in expectation values does not require renormalization, since the 
singular terms in the Green's function for the scalar field are the same for all Hadamard quantum states
\cite{Decanini:2005eg}.
In the rest of this subsection, we first describe the numerical method employed to compute (\ref{eq:VPH-B}), before discussing our numerical results. 

\subsubsection{Numerical method}
\label{sec:numerics}

To evaluate (\ref{eq:VPH-B}), we use the separated form (\ref{eq:modetilde}) of the scalar field modes:
\begin{multline}
    \Delta {\hat {\Phi }}^{2} 
    =\frac{1}{8\pi ^{2}} \sum _{p=-\infty }^{\infty } \sum _{\ell = |p|/2}^{\infty } \sum _{m=-\ell }^{\ell }
    \int _{{\widetilde {\omega }}=-\infty }^{\infty } d{\widetilde {\omega }}
    \\ \times  \frac{\left| X_{\omega p \ell }(r) \right| ^{2} \left| {}_{-p/2}Y_{\ell }^{m}(\theta ,\varphi ) \right|^{2} }{|{\widetilde {\omega }}| \left[ \exp \left( 2\pi |{\widetilde {\omega }}|/{\kappa _{+}} \right) -1  \right] }.
    \label{eq:VPH-B1}
\end{multline}
We see that there is no dependence on the time coordinate $t$ or azimuthal coordinate $\psi $; in addition, the norm of the spin-weighted spherical harmonics $\left| {}_{-p/2}Y_{\ell }^{m}(\theta ,\varphi ) \right|^{2}$ does not depend on $\varphi $, so the final answer (\ref{eq:VPH-B1}) depends only on the radial coordinate $r$ and polar angle $\theta $. 
In our numerical work, we use the coordinate $z$ (\ref{eq:zdef}) instead of $r$ as the radial coordinate. 
This is because the radial function $X_{\omega p \ell }(r)$ (\ref{eq:XHeunfinal}) is  given in terms of $z$ (and the Heun function in (\ref{eq:XHeunfinal}) is built-in to {\tt {Mathematica}}, which aids computation), and has the advantage that the entire region exterior to the event horizon, $r\in [r_{+},\infty )$ is mapped to $z\in [0,1)$, which is convenient for our purposes.  

In (\ref{eq:VPH-B1}), we first find the sum over $m$ as this can be done analytically using the addition theorem 
for the spin-weighted spherical harmonics (see App.~\ref{sec:addition}, specifically (\ref{eq:additionVP})). 
This gives
\begin{multline}
     \Delta {\hat {\Phi }}^{2} 
    =\frac{1}{32\pi ^{3}} \sum _{p=-\infty }^{\infty } \sum _{\ell = |p|/2}^{\infty } 
    \int _{{\widetilde {\omega }}=-\infty }^{\infty }  d{\widetilde {\omega }}
    \\ \times 
    \frac{\left( 2 \ell + 1 \right) \left| X_{\omega p \ell }(r) \right| ^{2} }{|{\widetilde {\omega }}| \left[ \exp \left( 2\pi |{\widetilde {\omega }}|/{\kappa _{+}} \right) -1  \right] }.
    \label{eq:VPH-B2}
\end{multline}
Note that (\ref{eq:VPH-B2}) depends only on the radial coordinate $r$; there is no dependence on the polar angle $\theta $.  

\begin{figure}
    \centering
    \includegraphics[scale=0.7]{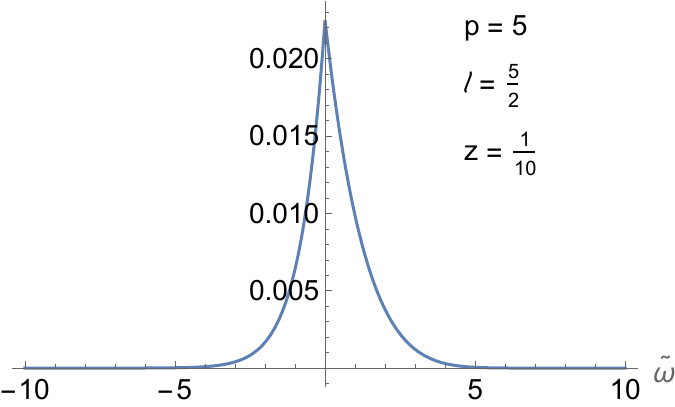}
    \caption{Typical integrand in (\ref{eq:integral}) as a function of the shifted frequency ${\widetilde {\omega }}$ (\ref{eq:tildeomega}). We have fixed the space-time parameters to be $M=10$, $L=1$, $a=1/2$ and the scalar field effective mass (\ref{eq:nu}) to be $\nu = 1/100$. The integrand is shown for radial coordinate $z=1/10$ (\ref{eq:zdef}), for the scalar field mode with $p=5$ and $\ell = 5/2$.}
    \label{fig:integrand}
\end{figure}

Next we compute the integral over the shifted frequency ${\widetilde {\omega }}$, defining
\begin{equation}
    I_{p\ell }(r) = \int _{{\widetilde {\omega }}=-\infty }^{\infty } d{\widetilde {\omega }}
\, \frac{\left| X_{\omega p \ell }(r) \right| ^{2} }{|{\widetilde {\omega }}| \left[ \exp \left( 2\pi |{\widetilde {\omega }}|/{\kappa _{+}} \right)  -1 \right] }.
\label{eq:integral}
\end{equation}
We use the radial function (\ref{eq:XHeunfinal}) with the constant ${\mathfrak{X}}_{\omega p \ell }$ given by (\ref{eq:frakX}).
A typical integrand in (\ref{eq:integral}) is shown in Fig.~\ref{fig:integrand}, where we have fixed the space-time parameters to be $M=10$, $L=1$, $a=1/2$ and the scalar field effective mass (\ref{eq:nu}) to be $\nu = 1/100$; we have set $z=1/10$ and shown the integrand for the scalar field mode with $p=5$ and $\ell = 5/2$.
The integrand in (\ref{eq:integral}) has the following key features.  
First, for $p\neq 0 $ it is not symmetric about ${\widetilde {\omega }}=0$.
We therefore compute the integral over the whole range of positive and negative ${\widetilde {\omega }}$.  
The radial function $X_{\omega p \ell }(r)$ (and hence the integrand in (\ref{eq:integral})) is
unchanged under the transformation $\omega \rightarrow -\omega $, $p\rightarrow -p$ (from (\ref{eq:radial})), and hence the integrand is invariant if we take $p\rightarrow - p$ as well as ${\widetilde {\omega }} \rightarrow - {\widetilde {\omega }}$.
This means that it is sufficient to compute the integrals for $p\ge 0$.
Second, our numerical investigations indicate that the integrand is regular for all ${\widetilde {\omega }}$.
At ${\widetilde {\omega }}=0$, the double zero in the denominator of the integrand does not lead to a divergence, as 
we find numerically that $|X_{\omega p \ell }(r)|$ vanishes sufficiently quickly for ${\widetilde {\omega }}\rightarrow 0$ to give a regular integrand. This may be understood heuristically as follows. As ${\widetilde {\omega }}\rightarrow 0$, we have $\theta _{+}\rightarrow 0$ (\ref{eq:theta+}) and thus $\gamma \rightarrow 1$ (\ref{eq:Heunconstants}). Thus, in this limit, the functions ${\mathcal {X}}_{3}(z)$ and ${\mathcal {X}}_{4}(z)$  (\ref{eq:Heun0}) are no longer linearly independent. Therefore the ratio ${\mathcal {W}}_{3}$ (\ref{eq:Wronskians}) diverges and thus the constant ${\mathfrak {X}}_{\omega p \ell }$ (\ref{eq:frakX}) tends to zero as ${\widetilde {\omega}} \rightarrow 0$.
We do however see a cusp in the integrand when ${\widetilde {\omega }}=0$, due to the presence of the absolute value of ${\widetilde {\omega }}$ in the terms in the denominator of the integrand. 
Third, as expected from the form of the denominator in the integrand in (\ref{eq:integral}), the integrand tends to zero very rapidly for $|{\widetilde {\omega }}| \rightarrow \infty $.

The integrals $I_{p\ell }(r)$ are computed using {\tt {Mathematica}}'s built-in {\tt {NIntegrate}} function. 
We use a working precision of 32 figures and integrate over $|{\widetilde {\omega }}|\le 30$.
The relative errors due to truncating the integrals at this value of $|{\widetilde {\omega }}|$ are extremely small, for example, we estimate this error to be less than $10^{-12}$ for the mode shown in Fig.~\ref{fig:integrand}.
The peak in the integrand typically increases with decreasing $z$, and decreases as either $p$ or $\ell $ increases. 
While it is convenient from the point of view of coding that the radial functions are given in terms of Heun functions, in practice the integrals $I_{p\ell }(r)$ require significant computation time, due to the numerical evaluation of these Heun functions.  
The integrals $I_{p\ell }(r)$ are evaluated on an evenly-spaced grid of 99 values of $z\in (0,1)$,  for values of $p$ and $\ell $ discussed below.

Once we have the integrals $I_{p\ell }(r)$, it remains to perform the sum over the quantum numbers $p$ and $\ell $. 
We write (\ref{eq:VPH-B2}) as
\begin{align}
   \Delta {\hat {\Phi }}^{2} 
    & =\frac{1}{32\pi ^{3}} \sum _{p=-\infty }^{\infty } \sum _{\ell = |p|/2}^{\infty } 
    \left( 2 \ell + 1 \right) I_{p\ell }(r) 
   \nonumber  \\
     & =\frac{1}{32\pi ^{3}} \sum _{2\ell =0}^{\infty } \sum _{p = -2\ell }^{2\ell } 
     \left( 2 \ell + 1 \right) I_{p\ell }(r)  ,
     \label{eq:VPH-B3}
\end{align}
where in the second line we have rewritten the two infinite sums over $\ell $ and $p$ in an equivalent form.
This leaves us with a sum over a finite range of the quantum number $p$, and only a final sum over an infinite range of values of $\ell $.
Recall that $\ell $ is either an integer or a half-integer, therefore we sum over the positive integer values of $2\ell $.

\begin{figure}
    \centering
    \includegraphics[scale=0.7]{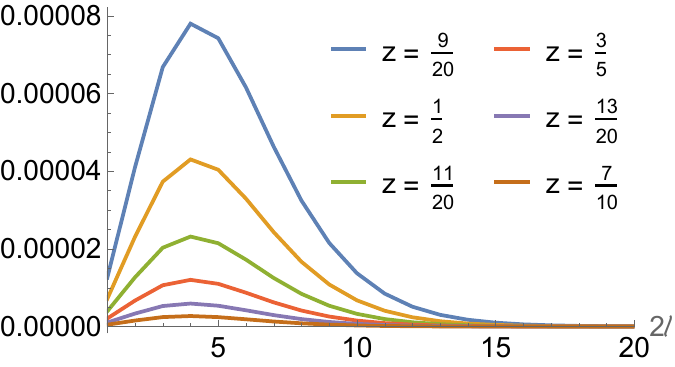}
    \caption{Summand $S_{\ell}(r)$ (\ref{eq:Sigma}) as a function of $2\ell $ for a selection of values of $z$ (\ref{eq:zdef}). The space-time and scalar field parameters are as in Fig.~\ref{fig:integrand}.}
    \label{fig:summand}
\end{figure}

The finite sum over $p$ is readily computed for each value of $\ell $ and $z$.
Defining
\begin{equation}
    S _{\ell }(r) = \frac{1}{32\pi ^{3}} \sum _{p = -2\ell }^{2\ell }\left( 2\ell + 1 \right) 
     I_{p\ell }(r) ,
     \label{eq:Sigma}
\end{equation}
the final step in our computation of the vacuum polarization is to evaluate 
\begin{equation}
     \Delta {\hat {\Phi }}^{2} 
   = \sum _{2\ell =0}^{\infty } S _{\ell }(r).
   \label{eq:finalsum}
\end{equation}
Typical summands $S _{\ell }(r)$ are shown in Fig.~\ref{fig:summand} for a selection of values of the radial coordinate $z$ (\ref{eq:zdef}) and the same space-time and scalar field parameters as in Fig.~\ref{fig:integrand}.
The profiles of $S_{\ell}$ as a function of $2\ell $ have similar shapes for all values of $z$ in Fig.~\ref{fig:summand}. 
In particular, there is a peak in the value of $S _{\ell }$ at $\ell \sim 4$ for each value of $z$, and $S_{\ell }$ then decreases rapidly as $\ell $ increases.
The value of $S _{\ell }$ at the peak increases as $z$ decreases, and $S _{\ell }$ is significantly greater than zero for larger values of $\ell $ as $z$ decreases.

\begin{figure}
    \centering
    \includegraphics[scale=0.46]{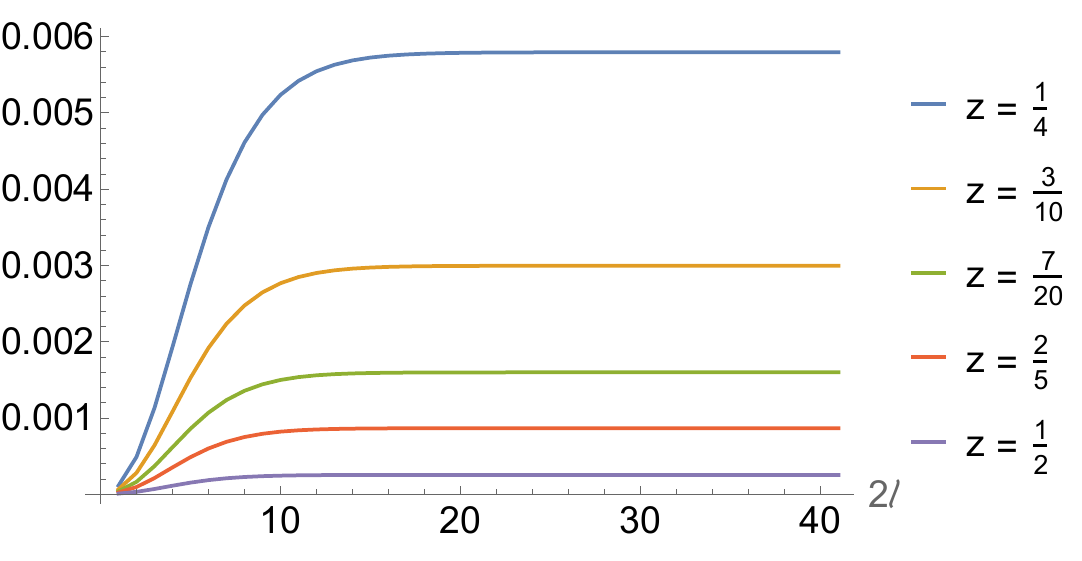}
    \caption{Partial sums ${\widetilde {S}}_{\ell }$ (\ref{eq:tildeSigma}) as a function of $2\ell $ for a selection of values of $z$ (\ref{eq:zdef}). 
    The space-time and scalar field parameters are as in Fig.~\ref{fig:integrand}.}
    \label{fig:partialsums}
\end{figure}

\begin{figure}
    \centering
    \includegraphics[scale=0.65]{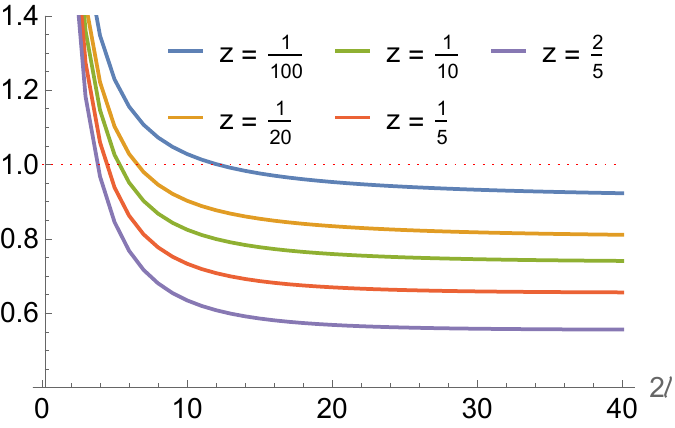}
    \caption{Ratio of the summands $S_{\ell + 1}/S _{\ell }$ (\ref{eq:Sigma}) as a function of $2\ell $ for a selection of values of $z$ (\ref{eq:zdef}).  
    The space-time and scalar field parameters are as in Fig.~\ref{fig:integrand}.}
    \label{fig:ratio}
\end{figure}

The partial sums
\begin{equation}
    {\widetilde {S}} _{\ell } (r) = \sum _{2{{\ell '}}=0}^{2\ell } S_{{\ell }'}(r) 
    \label{eq:tildeSigma}
\end{equation}
are shown in Fig.~\ref{fig:partialsums} for a selection of values of $z$.
For each value of $z$, the partial sums converge for large $2\ell $, with the limit increasing as $z$ decreases.
To check the convergence of the sums over $\ell $, in Fig.~\ref{fig:ratio} we plot the ratio of the summands $S _{\ell + 1}/S_{\ell }$ as a function of $2\ell $, again for a selection of values of $z$. 
We find that the ratio is less than unity for $2\ell  $ sufficiently large, demonstrating the convergence of the sums. 
However, the sums over $2\ell $ are not uniformly convergent as $z$ varies, with the rate of convergence decreasing as $z$ decreases.  
We also check our final answers for the sum (\ref{eq:finalsum}) obtained by direct summation with those obtained using sequence acceleration methods such as the Shanks transformation \cite{Schmidt:1941,Shanks:1955,Howard:1984ttx} (see also \cite{Corless:2023} for a review of sequence acceleration methods). 
We truncate the sum (\ref{eq:finalsum}) at $2\ell = 40$, which gives small relative errors, for example, at $z=\frac{1}{2}$ we estimate the error to be of the order of $10^{-8}$.
 
\subsubsection{Numerical results}
\label{sec:VPres}

\begin{figure}
    \centering
    \includegraphics[width=\columnwidth]{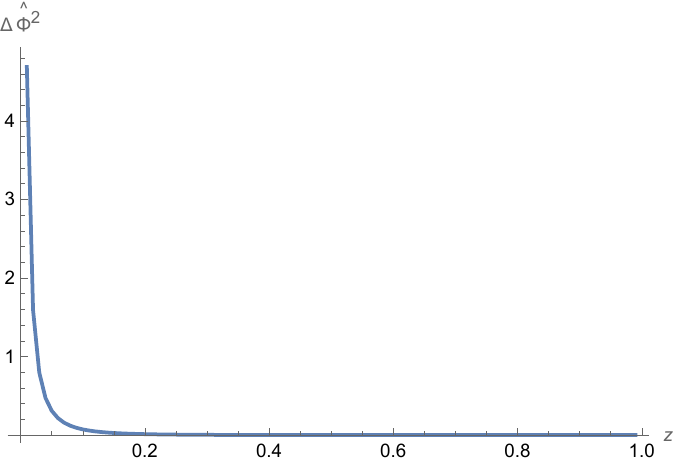}
    \includegraphics[width=\columnwidth]{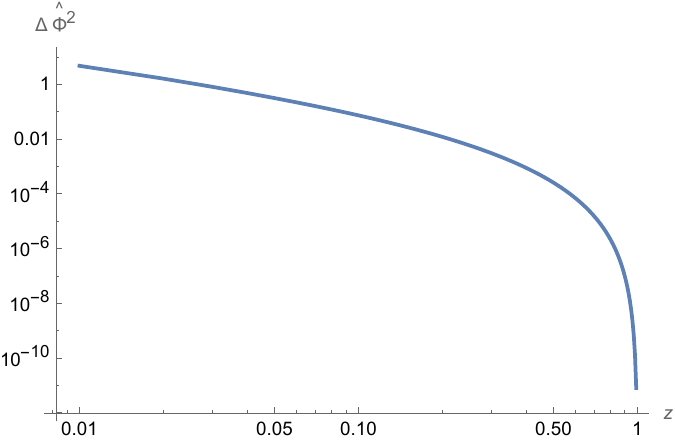}
    \caption{Difference in expectation values of the vacuum polarization in the Hartle-Hawking and Boulware states (\ref{eq:VPH-B}) as a function of the radial coordinate $z$ (\ref{eq:zdef}), using a linear scale (upper plot) or a log-log scale (lower plot). 
    The space-time and scalar field parameters are as in Fig.~\ref{fig:integrand}.}
    \label{fig:VP}
\end{figure}

We are now in a position to present, in Fig.~\ref{fig:VP}, our results for the vacuum polarization (\ref{eq:VPH-B}).
We fix the space-time parameters to be $M=10$, $L=1$ and $a=1/2$, with a scalar field effective mass (\ref{eq:nu}) of $\nu =1/100$.
The upper plot in Fig.~\ref{fig:VP} shows the difference in the vacuum polarization between the Hartle-Hawking and Boulware states, using a linear scale.
It can be seen that this difference tends to zero as $z\rightarrow 1 $ (and we approach the anti-de Sitter boundary at $r\rightarrow \infty )$ but diverges as $z\rightarrow 0 $ (and we approach the event horizon). 
In the upper plot in Fig.~\ref{fig:VP}, the difference in vacuum polarization is indistinguishable  from zero for $z\gtrsim 0.2$.
Therefore, in the lower plot in Fig.~\ref{fig:VP}, we show the same results but on a log-log scale,
which makes the range of scales involved clearer.

The Boulware state $|{\mathrm {B}}\rangle $ is expected to be a ground state far from the black hole, that is, to be as empty of particles as possible. 
We would therefore expect that $\langle {\rm {B}} | {\hat {\Phi }}^{2} | {\rm {B}} \rangle $ tends to zero as $r\rightarrow \infty $ and $z\rightarrow 1$.
If this is the case, then the results in Fig.~\ref{fig:VP} imply that $\langle {\rm {H}} | {\hat {\Phi }}^{2} | {\rm {H}} \rangle $ also tends to zero as $z\rightarrow 1$.
The Hartle-Hawking state $|{\rm {H}}\rangle $ represents a black hole in thermal equilibrium with a heat bath at the Hawking temperature $\kappa _{+}/2\pi $.
The vacuum polarization for a quantum scalar field in a thermal equilibrium state in pure anti-de Sitter space-time tends to its vacuum expectation value at the boundary \cite{Allen:1986ty,Ambrus:2018olh,Barroso:2019cwp,Morley:2020ayr,Namasivayam:2022bky}.
We would therefore expect that the vacuum polarization in the Hartle-Hawking state (a thermal state) approaches that in the Boulware state (a ground state) far from the black hole.

At the event horizon, we anticipate that the Hartle-Hawking state $|{\rm {H}}\rangle $ is regular, in which case our results imply that the vacuum polarization in the Boulware state $|{\rm {B}}\rangle $ is divergent.
This is in agreement with the divergence of $\langle {\rm {B}} | {\hat {\Phi }}^{2} | {\rm {B}} \rangle $ at the event horizon of, for example, a Schwarzschild black hole \cite{Candelas:1980zt,Levi:2015eea}.

\subsection{Stress-energy tensor}
\label{sec:SETexp}

We now turn to the expectation value of the quantum SET operator ${\hat {T}}_{\mu \nu }$.
Analogously to the vacuum polarization (\ref{eq:VPH-B}), we find
\begin{multline}
    \Delta {\hat {T}}_{\mu \nu } 
= 
    \langle {\rm {H}} | {\hat {T}}_{\mu \nu }  | {\rm {H}} \rangle  - \langle {\rm {B}} | {\hat {T}}_{\mu \nu }  | {\rm {B}} \rangle
   \\ = \sum _{p=-\infty }^{\infty } \sum _{\ell = |p|/2}^{\infty } \sum _{m=-\ell }^{\ell }
    \int _{{\widetilde {\omega }}=-\infty }^{\infty } d{\widetilde {\omega }}
    \\ \times  \frac{{}^{\aleph}T_{\mu \nu } }{\exp \left( 2\pi \left|{\widetilde {\omega }}\right| /{\kappa _{+}} \right) -1},
    \label{eq:SETH-B}
\end{multline}
where ${}^{\aleph}T_{\mu \nu } $ is the classical SET for a scalar field mode (\ref{eq:mode}) and we use the notation $\aleph = \{ \omega, p, \ell, m \}$ to denote the quantum numbers on which this mode contribution to the SET depends. 
In five space-time dimensions, there are fifteen independent components of the SET, since it is a symmetric tensor. 
Using the mode solutions (\ref{eq:mode}) of the scalar field equation, the form of these components ${}^{\aleph}T_{\mu \nu } $ can be derived in terms of the radial and angular functions $X_{\omega p \ell }(r)$ (\ref{eq:XHeunfinal}) and ${}_{s}{\widetilde {Y}}_{\ell }^{m}(\theta )$ (\ref{eq:angsep}). 
The resulting expressions are rather lengthy, so are presented in App.~\ref{sec:SETcomp}.

In this section, before we proceed to the numerical computation of $\Delta {\hat {T}}_{\mu \nu }$ (\ref{eq:SETH-B}) and a discussion of our numerical results, we first simplify the components of $\Delta {\hat {T}}_{\mu \nu }$ using the general principles of conservation and the underlying symmetries of the background space-time (\ref{eq:metric}). 

\subsubsection{General properties of the stress-energy tensor}
\label{sec:SETgen}

In this subsection, we use the symmetries of the space-time (\ref{eq:metric}) to constrain the form of the SET (\ref{eq:SETH-B}), assuming that this shares these symmetries of the underlying black hole geometry.

In particular, we assume that the SET preserves the space-time symmetries resulting from the Killing vectors \eqref{eq:Killing}, so that the Lie derivatives of the SET along each of these Killing vectors vanish:
\begin{equation}
    {\mathcal {L}}_{{\mathcal {V}} _{i}} \langle {\hat {T}}^{\mu \nu } \rangle  =0,  \qquad i = 0,1,\ldots ,5.
    \label{eq:Lie}
\end{equation}
Applying this to the first three Killing vectors (\ref{eq:xi123}), we conclude that the components of $\langle {\hat {T}}^{\mu \nu } \rangle $ are independent of the coordinates $t$, $\varphi $ and $\psi $.
The remaining independent Killing vectors, ${\mathcal {V}} _{4}$ \eqref{eq:xi4} and ${\mathcal {V}} _{5}$ \eqref{eq:xi5}, give more complicated constraints.
Writing the Lie derivative ${\mathcal {L}}_{{\mathcal {V}} _{i}}\langle {\hat {T}}_{\mu \nu } \rangle $ as
\begin{align}
    0 = & ~ {\mathcal {L}}_{{\mathcal {V}} _{i}}\langle {\hat {T}}_{\mu \nu } \rangle 
    \nonumber \\
    =  & ~{\mathcal {V}} _{i}^{\alpha } \partial _{\alpha } \langle {\hat {T}}^{\mu \nu } \rangle 
    - \left( \partial _{\alpha } {\mathcal {V}} _{i}^{\mu } \right) \langle {\hat {T}}^{\alpha  \nu } \rangle 
    - \left( \partial _{\alpha } {\mathcal {V}} _{i}^{\nu } \right) \langle {\hat {T}}^{\mu  \alpha  } \rangle ,
\end{align}
gives fifteen equations for each of the two remaining Killing vectors.
Considering the combination ${\mathcal {L}}_{{\mathcal {V}} _{4}}\langle {\hat {T}}^{\mu \nu } \rangle  \sin \varphi  + {\mathcal {L}}_{{\mathcal {V}} _{5}}\langle {\hat {T}}^{\mu \nu } \rangle  \cos \varphi$ immediately gives that the following SET components vanish identically:
\begin{equation}
 \langle {\hat {T}}^{t\theta } \rangle =\langle {\hat {T}}^{t\varphi } \rangle
   = \langle {\hat {T}}^{r\theta } \rangle =\langle {\hat {T}}^{r\varphi } \rangle
   = \langle {\hat {T}}^{\theta  \varphi } \rangle =\langle {\hat {T}}^{\theta \psi } \rangle
   = 0,
\end{equation}
as well as the following relations:
\begin{subequations}
\begin{align}
    \langle {\hat {T}}^{\varphi \varphi } \rangle & = \frac{\langle {\hat {T}}^{\theta \theta } \rangle}{\sin ^{2} \theta } ,
    \\
    \langle {\hat {T}}^{\varphi \psi } \rangle & = -\frac{\langle {\hat {T}}^{\theta \theta } \rangle}{2\tan \theta \sin \theta } .
\end{align}
\end{subequations}
Next we consider ${\mathcal {L}}_{{\mathcal {V}} _{4}}\langle {\hat {T}}^{\mu \nu } \rangle  \cos \varphi  - {\mathcal {L}}_{{\mathcal {V}} _{5}}\langle {\hat {T}}^{\mu \nu } \rangle  \sin \varphi$, from which we deduce that the following components of the SET do not depend on the angle $\theta $: $\langle {\hat {T}}^{tt} \rangle $, $\langle {\hat {T}}^{tr} \rangle $, $\langle {\hat {T}}^{t\psi } \rangle $, $\langle {\hat {T}}^{rr} \rangle $, $\langle {\hat {T}}^{r\psi } \rangle $ and $\langle {\hat {T}}^{\theta \theta } \rangle $.
There is one further relation arising from this combination of Lie derivatives, which takes the form
\begin{subequations}
\begin{equation}
    \partial _{\theta  } \langle {\hat {T}}^{\psi \psi } \rangle =  -\frac{\langle {\hat {T}}^{\theta \theta } \rangle}{2\tan \theta \sin ^{2} \theta } ,
\end{equation}
and which is readily integrated to give
\begin{equation}
\langle {\hat {T}}^{\psi \psi } \rangle =  \frac{\langle {\hat {T}}^{\theta \theta } \rangle}{4\sin ^{2} \theta }  + {\mathcal {F}}^{\psi \psi }(r),
\end{equation}
\end{subequations}
where ${\mathcal {F}}^{\psi \psi }(r)$ is an arbitrary function of $r$. 
In summary, we can write the SET in matrix form as follows:
\begin{widetext}
\begin{equation}
    \langle {\hat {T}}^{\mu \nu } \rangle  = \left(
    \begin{array}{ccccc}
    {\mathcal {F}}^{tt}(r) & {\mathcal {F}}^{tr}(r) & 0 & 0 & {\mathcal {F}}^{t\psi }(r) \\
    {\mathcal {F}}^{tr}(r)  & {\mathcal {F}}^{rr}(r) & 0 & 0 & {\mathcal {F}}^{r\psi }(r) \\
    0 & 0 & {\mathcal {F}}^{\theta \theta } (r) & 0 & 0 \\
    0 & 0 & 0 & \dfrac{{\mathcal {F}}^{\theta \theta}(r)}{\sin ^{2} \theta } & 
    -\dfrac{{\mathcal {F}}^{\theta \theta}(r)}{2\tan \theta \sin  \theta } \\[0.3cm]
    {\mathcal {F}}^{t\psi }(r) & {\mathcal {F}}^{r\psi }(r) & 0 & -\dfrac{{\mathcal {F}}^{\theta \theta}(r)}{2\tan \theta \sin  \theta } & \dfrac{{\mathcal {F}}^{\theta \theta }(r)}{4\sin ^{2} \theta }  + {\mathcal {F}}^{\psi \psi }(r)
\end{array}
    \right) ,
    \label{eq:SETsimp1}
\end{equation}
\end{widetext}
where the ${\mathcal {F}}^{\bullet \bullet}(r)$ are arbitrary functions of $r$.
We note that the underlying symmetries of the black hole geometry have completely fixed the dependence of the SET components on the angle $\theta $ (as is the case for static, spherically symmetric black holes), and we are left with seven arbitrary functions of $r$ which are to be determined. 

We can further constrain these seven arbitrary functions of $r$ by imposing the requirement that the SET is conserved:
\begin{equation}
    \nabla _{\nu } \langle {\hat {T}}_{\mu }{}^{\nu } \rangle   = 0,
\end{equation}
which we write in the alternative form
\begin{equation}
\partial _{\nu } \left( {\mathsf {g}} \langle {\hat {T}}_{\mu }{}^{\nu } \rangle   \right) 
= \frac{1}{2} {\mathsf {g}} \left(  \partial _{\mu } g_{\alpha \beta } \right) 
\langle {\hat {T}}^{\alpha \beta } \rangle  ,
\label{eq:conservation}
\end{equation}
where ${\mathsf {g}}$ is given by (\ref{eq:metricdet}). 
Comparing the form of the metric (\ref{eq:metric}) and the SET (\ref{eq:SETsimp1}), we have immediately that the $\mu = \theta $ equation is trivial. 
Since the metric (\ref{eq:metric}) does not depend on $t$, $\varphi $ or $\psi $, there are three simple conservation equations arising from (\ref{eq:conservation}). 
The $\varphi $ and $\psi $ equations are identical, and give
\begin{subequations}
\label{eq:consf1}
\begin{equation}
    \frac{d}{dr} \left[ r^{3}h(r)^{2} {\mathcal {F}}_{1}(r) \right] =  0,
    \label{eq:cons1}
\end{equation}
where we have defined
\begin{equation}
    {\mathcal {F}}_{1} (r) = {\mathcal {F}}^{r\psi }(r) - \Omega (r) {\mathcal {F}}^{tr}(r) 
    \label{eq:f1def}
\end{equation}
\end{subequations}
and $f(r)$, $g(r)$, $h(r)$ and $\Omega (r) $ are the metric functions given in (\ref{eq:metricfunctionsstart}--\ref{eq:metricfunctionsend}).
Integrating (\ref{eq:consf1}) yields
\begin{equation}
  {\mathcal {F}}_{1}(r) 
    = \frac{{\mathcal {Y}}}{r^{3}h(r)^2}, 
    \label{eq:calX}
\end{equation}
where ${\mathcal {Y}}$ is an arbitrary constant.
The $\mu = t$ equation arising from (\ref{eq:conservation}) then takes the form
\begin{equation}
\frac{d}{dr} \left\{ r^{3} \left[   f(r)^{2} {\mathcal {F}} ^{tr}(r)  + h(r)^{2}\Omega (r) {\mathcal {F}}_{1}(r) \right] \right\}   = 0. 
\end{equation}
This is also readily integrated to give
\begin{equation}
{\mathcal {F}}^{tr}(r) 
 =  \frac{{\mathcal {Z}}- \Omega (r) {\mathcal {Y}}}{r^{3}f(r)^{2}}, 
    \label{eq:calZ}
\end{equation}
where ${\mathcal {Z}}$ is an arbitrary constant. 
Hence, using (\ref{eq:f1def}), we have
\begin{multline}
    {\mathcal {F}}^{r\psi }(r) = \frac{1}{r^{3}f(r)^{2}h(r)^{2}} 
    \\ \times \left( {\mathcal {Y}} f(r)^2 +  h(r)^2 \Omega (r) \left[ {\mathcal {Z}} - {\mathcal {Y}}\Omega (r) \right]  \right)  .
\end{multline}
The remaining conservation equation (\ref{eq:conservation}) has $\mu= r$ and is more complicated:
\begin{widetext}
    \begin{align}
     \frac{1}{r^{3}}\frac{d}{dr} \left[ r^{3}g(r)^{2}{\mathcal {F}}^{rr} \right] = & ~  
        g(r)g'(r) {\mathcal {F}}^{rr}(r) 
        + \frac{1}{4} \left[ 2r + h(r)h'(r) \right]  {\mathcal {F}}^{\theta \theta} (r) 
        + h(r)h'(r) {\mathcal {F}}^{\psi \psi }(r)
        \nonumber \\ & ~
        +  \left[  -f(r)f'(r)+ h(r)h'(r) \Omega (r)^2 + h(r)^{2}\Omega (r) \Omega '(r) \right]  {\mathcal {F}}^{tt}(r)
        \nonumber \\ & ~
        - \left[ 2h(r)h'(r) \Omega (r) + h(r)^{2} \Omega '(r) \right] {\mathcal {F}}^{t\psi } (r).
        \label{eq:cons2}
    \end{align} 
Nonetheless, (\ref{eq:cons2}) can be integrated directly to give ${\mathcal {F}}^{rr}(r)$:
\begin{equation}
    {\mathcal {F}}^{rr}(r) = \frac{1}{r^{3}g(r)} 
    \left[ {\mathcal {K}}  + \int _{r'=r_{+}}^{r} \Upsilon  (r') \, dr' \right]  ,
    \label{eq:Frr}
\end{equation}
where ${\mathcal {K}}$ is an arbitrary constant and we have defined
    \begin{multline}
    \Upsilon (r) = 
    \frac{r^{3}}{g(r)} \bigg\{
    \left[ - f(r)f'(r)+ h(r)h'(r) \Omega (r)^2 + h(r)^{2}\Omega (r) \Omega '(r) \right]  {\mathcal {F}}^{tt}(r)
        - \left[ 2h(r)h'(r) \Omega (r) + h(r)^{2} \Omega '(r) \right] {\mathcal {F}}^{t\psi } (r)
          \\  
         + h(r)h'(r) {\mathcal {F}}^{\psi \psi }(r)
         + \frac{1}{4} \left[ 2r + h(r)h'(r) \right]  {\mathcal {F}}^{\theta \theta} (r)
    \bigg\} .
\end{multline} 
\end{widetext}

In summary, using the symmetries of the underlying space-time and the conservation of the SET, we have found that the SET is determined by three arbitrary constants (${\mathcal {K}}$, ${\mathcal {Y}}$ and ${\mathcal {Z}}$) and four arbitrary functions of the radial coordinate $r$ only, namely ${\mathcal {F}}^{\theta \theta }(r)$, ${\mathcal {F}}^{tt}(r)$, ${\mathcal {F}}^{t\psi }(r)$ and ${\mathcal {F}}^{\psi \psi }(r)$. 
The enhanced symmetry of the space-time compared to the four-dimensional Kerr metric has played a significant role here, enabling us to constrain the SET much more than in the Kerr case \cite{Ottewill:2000qh}. 
For Kerr, using the Killing vectors and the conservation equations gives the four-dimensional SET in terms of two arbitrary functions of the latitudinal angle $\theta $ and six functions of both $\theta $ and $r$, which are constrained by two coupled equations \cite[Eq.~(4.4)]{Ottewill:2000qh}. 
In our situation, the enhanced symmetry has completely determined the angular dependence of the SET, as well as reducing the number of unknown components. 
At the same time, the SET structure in our scenario is more complicated than that on a five-dimensional static, spherically symmetric space-time \cite{Morgan:2007hp}, which is determined by just two arbitrary constants and two arbitrary functions of the radial coordinate.

There is one further constraint on the SET components, namely the trace $\langle {\hat {T}}_{\alpha }{}^{\alpha } \rangle $. 
For a massless, conformally-coupled scalar field, this is given by the trace anomaly, which vanishes in five space-time dimensions \cite{Decanini:2005eg}.
When the scalar field has general mass $\mu $ and coupling $\xi $ to the scalar curvature, the trace $\langle {\hat {T}}_{\alpha }{}^{\alpha } \rangle $ is given by \cite{Decanini:2005eg}
\begin{equation}
    \langle {\hat {T}}_{\alpha }{}^{\alpha } \rangle = 
    -\mu ^{2} \langle {\hat {\Phi }}^{2} \rangle + 4 \left( \xi - \frac{3}{16} \right) \Box \langle {\hat {\Phi }}^{2} \rangle ,
    \label{eq:trace}
\end{equation}
which depends on the vacuum polarization $\langle {\hat {\Phi }}^{2} \rangle $ and its derivatives and clearly vanishes when the field is massless ($\mu =0$) and conformally coupled ($\xi = 3/16$).
Since the vacuum polarization is not known {\em {a priori}} and can only be computed numerically,
(\ref{eq:trace}) does not reduce the number of unknown functions, but it does potentially provide a useful check of our numerical results. 
In particular, using the SET form (\ref{eq:SETsimp1}) and metric (\ref{eq:metric}), we have 
\begin{align}
     \langle {\hat {T}}_{\alpha }{}^{\alpha } \rangle 
     = &  \left[ -f(r)^{2}+ h(r)^{2}\Omega (r)^{2} \right] {\mathcal {F}}^{tt}(r)
     \nonumber \\ & - 2h(r)^{2} \Omega (r) {\mathcal {F}}^{t\psi }(r)
     +g(r)^{2}{\mathcal {F}}^{rr}(r)
     \nonumber \\ & 
     + \frac{1}{4} \left[ 2r^{2} +  h(r)^{2} \right] {\mathcal {F}}^{\theta \theta } (r)
     +h(r)^{2}{\mathcal {F}}^{\psi \psi }(r) .
     \label{eq:trace1}
\end{align}
At least in principle, we could use (\ref{eq:trace1}) to eliminate say ${\mathcal {F}}^{\psi \psi }(r)$ from (\ref{eq:cons2}) and then integrate to give an alternative expression for ${\mathcal {F}}^{rr}(r)$, which would involve the trace $\langle {\hat {T}}_{\alpha }{}^{\alpha } \rangle $.
However, there is no great advantage in doing so, and we shall instead use (\ref{eq:trace1}) as a check of our numerical results, which are discussed in the next subsection.

\subsubsection{Numerical method}
\label{sec:SETnum}

To find the difference in expectation values of the SET between the Hartle-Hawking and Boulware states, $\Delta {\hat {T}}_{\mu \nu }$ (\ref{eq:SETH-B}), from the analysis of the previous subsection we require the 
determination of three arbitrary constants (${\mathcal {K}}$, ${\mathcal {X}}$ and ${\mathcal {Z}}$) and the numerical computation of four functions of the radial coordinate $r$ (${\mathcal {F}}^{\theta \theta}(r)$, ${\mathcal {F}}^{tt}(r)$, ${\mathcal {F}}^{t\psi }(r)$, and ${\mathcal {F}}^{\psi \psi }(r)$, where the functions ${\mathcal {F}}^{\bullet \bullet }(r)$ now refer to those pertinent to this difference in SET expectation values).

Our overall strategy for the numerical computation of the functions ${\mathcal {F}}^{\bullet \bullet}(r)$ and hence the SET components (\ref{eq:SETH-B}) follows that for the vacuum polarization in Sec.~\ref{sec:numerics}.
First we require mode sum expressions for the functions ${\mathcal {F}}^{\bullet \bullet}(r)$.
To find these, we start with the expressions for ${}^{\aleph}T_{\mu \nu }$, the classical SET components for a scalar field mode (\ref{eq:mode}), which are given in (\ref{eq:SETcomp1}).
Next we perform the sum over the quantum number $m$, using the addition theorems for spin-weighted spherical harmonics in App.~\ref{sec:addition}.
The resulting quantities can be found in (\ref{eq:SETcomp2}). 
From these, we can write
each ${\mathcal {F}}^{\bullet \bullet}(r)$ as a mode sum over the shifted frequency ${\widetilde {\omega }}$ and quantum numbers $p$ and $\ell $:
\begin{equation}
    {\mathcal {F}}^{\bullet \bullet }(r) = \frac{1}{4\pi }\sum _{2\ell = 0}^{\infty } \sum _{p=-2\ell }^{2\ell } 
    \int _{{\widetilde {\omega }}=-\infty }^{\infty } d{\widetilde {\omega }} \,
    \frac{\left( 2\ell + 1\right) {\mathfrak {F}}^{\bullet \bullet }(r)}{\exp \left( 2\pi |{\widetilde {\omega }}|/{\kappa _{+}} \right) -1  },
    \label{eq:Fmode}
\end{equation}
where expressions for the individual ${\mathfrak {F}}^{\bullet \bullet }(r)$ can be found in (\ref{eq:frakF}).
In (\ref{eq:Fmode}), following (\ref{eq:VPH-B3}), we have rewritten the sums over $p=-\infty ,\ldots , \infty $ and $2\ell =|p|, \ldots \infty $ as a sum over a finite number of values of $p$ and a sum over $2\ell =0,\ldots \infty $.

From (\ref{eq:frakF}), we see that ${\mathfrak {F}}^{tr}(r)=0={\mathfrak {F}}^{r\psi }(r)$ and hence 
${\mathcal {F}}^{tr}(r)=0={\mathcal {F}}^{r\psi }(r)$.
Hence, using (\ref{eq:calX}, \ref{eq:calZ}), we can immediately fix two of our constants: ${\mathcal {Y}}=0={\mathcal {Z}}$.
Instead of finding the constant ${\mathcal {K}}$ and then ${\mathcal {F}}^{rr}(r)$ using (\ref{eq:Frr}), we found it more straightforward to calculate ${\mathcal {F}}^{rr}(r)$ directly. 
This means that we will compute five functions of $r$, namely ${\mathcal {F}}^{\theta \theta}(r)$, ${\mathcal {F}}^{tt}(r)$, ${\mathcal {F}}^{t\psi }(r)$, ${\mathcal {F}}^{\psi \psi }(r)$ and ${\mathcal {F}}^{rr}(r)$.

As in Sec.\ref{sec:numerics}, we first perform the integral over the shifted frequency ${\widetilde {\omega }}$ in (\ref{eq:Fmode}). 
Examination of the expressions in (\ref{eq:frakF}) reveal that, for each $p$ and $\ell $, we require the following integrals:
\begin{subequations}
\label{eq:SETintegrals}
\begin{align}
I^{(1)}_{p \ell }(r) & = \int _{{\widetilde {\omega }}=-\infty }^{\infty } d{\widetilde {\omega }}
\, \frac{\omega \left| X_{\omega p \ell }(r) \right| ^{2} }{|{\widetilde {\omega }}| \left[ \exp \left( 2\pi |{\widetilde {\omega }}|/{\kappa _{+}} \right)  -1 \right] },
\\
I^{(2)}_{p \ell }(r)& = \int _{{\widetilde {\omega }}=-\infty }^{\infty } d{\widetilde {\omega }}
\, \frac{\omega ^{2}\left| X_{\omega p \ell }(r) \right| ^{2} }{|{\widetilde {\omega }}| \left[ \exp \left( 2\pi |{\widetilde {\omega }}|/{\kappa _{+}} \right)  -1 \right] },
\\
I^{(3)}_{p \ell }(r) & =  \int _{{\widetilde {\omega }}=-\infty }^{\infty } d{\widetilde {\omega }}
\, \frac{\Re \{ X_{p\ell }^{*}(r) X'_{p\ell }(r) \} }{|{\widetilde {\omega }}| \left[ \exp \left( 2\pi |{\widetilde {\omega }}|/{\kappa _{+}} \right)  -1 \right] } ,
\\
I^{(4)}_{p\ell }(r) & = \int _{{\widetilde {\omega }}=-\infty }^{\infty } d{\widetilde {\omega }}
\, \frac{\left| X_{\omega p \ell }'(r) \right| ^{2} }{|{\widetilde {\omega }}| \left[ \exp \left( 2\pi |{\widetilde {\omega }}|/{\kappa _{+}} \right)  -1 \right] } ,
\end{align}
\end{subequations}
in addition to the integral $I_{p\ell }(r)$ (\ref{eq:integral}) which we have already computed for the vacuum polarization.
As for  the integrand in $I_{p\ell }(r)$ (see Fig.~\ref{fig:integrand}), the integrands in (\ref{eq:SETintegrals}) are regular and rapidly decaying as $|{\widetilde {\omega }}|\rightarrow \infty $.
The integrals (\ref{eq:SETintegrals}) are computed in a similar way to $I_{p\ell }(r)$, using {\tt {Mathematica}}'s built-in {\tt {NIntegrate}} function, although these have a longer computation time than that required for $I_{p\ell }(r)$.

Once we have found the integrals (\ref{eq:SETintegrals}),
we take appropriate combinations of these, using (\ref{eq:frakF}), to give
\begin{equation}
    {\widetilde {\mathfrak {F}}}^{\bullet \bullet }_{p\ell }(r) = \int _{{\widetilde {\omega }}=-\infty }^{\infty } d{\widetilde {\omega }} \,
    \frac{{\mathfrak {F}}^{\bullet \bullet }(r)}{\exp \left( 2\pi |{\widetilde {\omega }}|/{\kappa _{+}} \right) -1  },
\end{equation}
in terms of which we have
\begin{equation}
  {\mathcal {F}}^{\bullet \bullet }(r) = \frac{1}{4\pi }\sum _{2\ell = 0}^{\infty } \sum _{p=-2\ell }^{2\ell } 
  \left( 2\ell + 1 \right) {\widetilde {\mathfrak {F}}}^{\bullet \bullet }_{p\ell }(r) .
 \label{eq:Fmode1}   
\end{equation}
The sums over the quantum number $p$ in (\ref{eq:Fmode1}) are then straightforward to compute, leaving just the sum over $2\ell $.
As for the vacuum polarization, we find that summing over values of $2\ell $ from 0 to 40 gives results which are sufficiently accurate for our purposes. 

We validate our results 
by computing the trace of $\Delta {\hat {T}}_{\mu \nu }$ (\ref{eq:SETH-B}) using (\ref{eq:trace1}) and compare with the result (\ref{eq:trace}) which involves the difference in vacuum polarization between the Hartle-Hawking and Boulware states.
For a conformally coupled field with $\xi = 3/16$, we find agreement between these two expressions to one part in $10^{12}$.
Ideally, one would also check that our functions ${\mathcal {F}}^{\bullet \bullet}(r)$ satisfy the conservation equation (\ref{eq:cons2}) (and that (\ref{eq:trace1}) holds for values of $\xi $ other than $3/16$).
Performing either of these checks requires derivatives of quantities we have computed numerically. 
These can be found by interpolating our results between the grid points in $z$ and then differentiating the interpolating function.
As might be expected, this introduces additional numerical errors. 
In our situation, these errors are compounded by the fact that both the difference in vacuum polarization $\Delta {\hat {\Phi }}^{2}$ and the functions ${\mathcal {F}}^{\bullet \bullet}(r)$ vary by several orders of magnitude over the range of values of $z$ (see Figs.~\ref{fig:VP}--\ref{fig:SETxi316log}).
Furthermore, different functions ${\mathcal {F}}^{\bullet \bullet}(r)$ have very different orders of magnitude at the same value of $z$ (see Figs.~\ref{fig:SETxi316}--\ref{fig:SETxi316log}).
As a result, neither the conservation equation test, nor the trace test (for nonconformally-coupled fields) is particularly robust. 
However, we do find, at intermediate values of $z$, that the relative error in the evaluation of the conservation equation (\ref{eq:cons2}) is several orders of magnitude smaller than the largest magnitude of the ${\mathcal {F}}^{\bullet \bullet}(r)$ functions, which at least lends credence to our numerical results.

\subsubsection{Numerical results}
\label{sec:SETres}

While the radial functions $X_{\omega p \ell }(r)$ (satisfying the radial equation (\ref{eq:radial})) and hence the vacuum polarization (\ref{eq:VPH-B}) depend on the scalar field mass $\mu $ and coupling $\xi $ only via the combination $\nu $ (\ref{eq:nu}), it can be seen from (\ref{eq:SET}) that the SET components (and the functions ${\mathcal {F}}^{\bullet \bullet }(r)$) depend separately on $\mu ^{2}$ and $\xi $.
Given that our numerical computations are somewhat CPU-intensive, in this paper we present results for a single value of $\nu = 1/100$.
However, with this fixed value of $\nu $, we can vary the coupling constant $\xi $ (and hence also $\mu $) while keeping $\nu $ fixed and thus study how the SET varies depending on the coupling to the scalar curvature.
A similar approach has been employed in \cite{Taylor:2022sly,Arrechea:2023fas}, where the SET on a four-dimensional Schwarzschild or Reissner-Nordstr\"om background was studied.
In those scenarios, the background Ricci scalar curvature vanishes identically, so the coupling constant $\xi $ does not appear in the scalar field equation and the scalar field mass is analogous to our quantity $\nu $.
In \cite{Taylor:2022sly,Arrechea:2023fas}, it is found that varying the mass of the scalar field does not significantly change the qualitative behaviour of the SET components.
In contrast, varying the coupling constant $\xi $ can make a significant difference to the features of the SET components (such as whether they are monotonically increasing or decreasing as functions of the radial coordinate and the existence of maxima or minima).

\begin{figure*}
    \centering
    \begin{minipage}{0.48\textwidth}
        \centering
        \includegraphics[width=\textwidth]{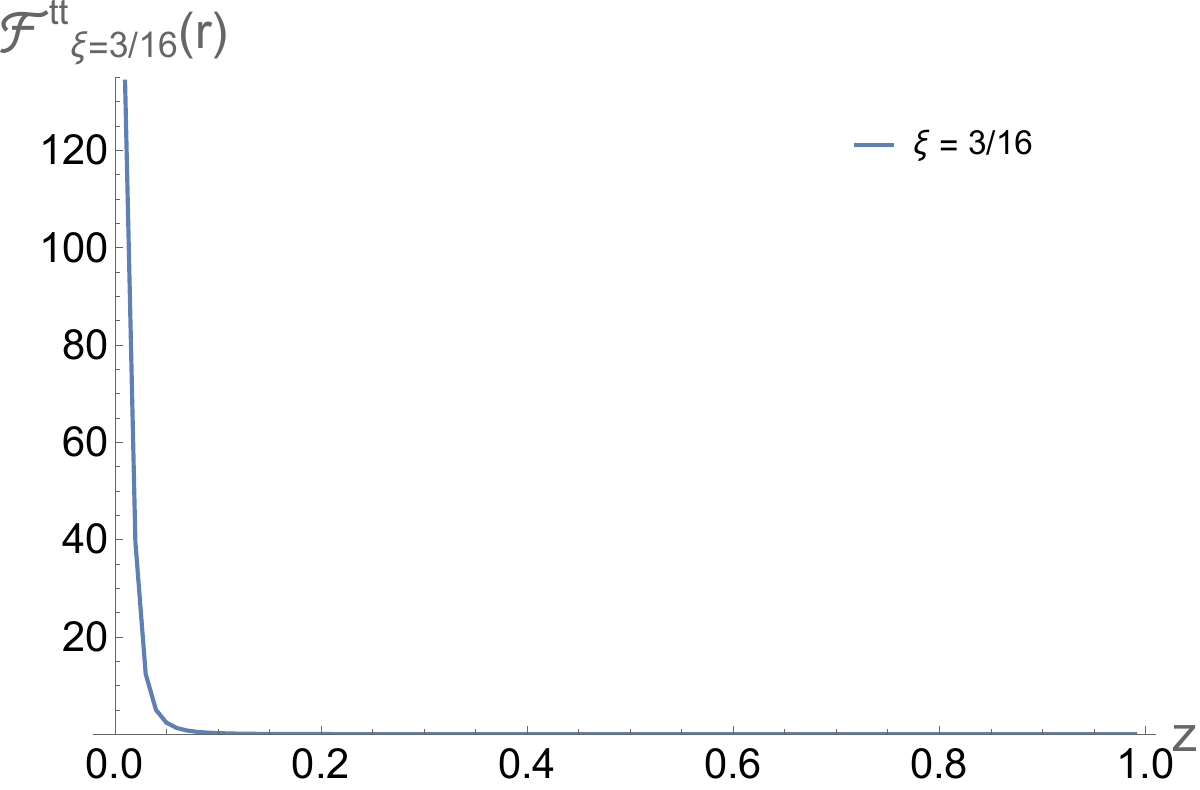}
    \end{minipage}
    \hfill
    \begin{minipage}{0.48\textwidth}
        \centering
        \includegraphics[width=\textwidth]{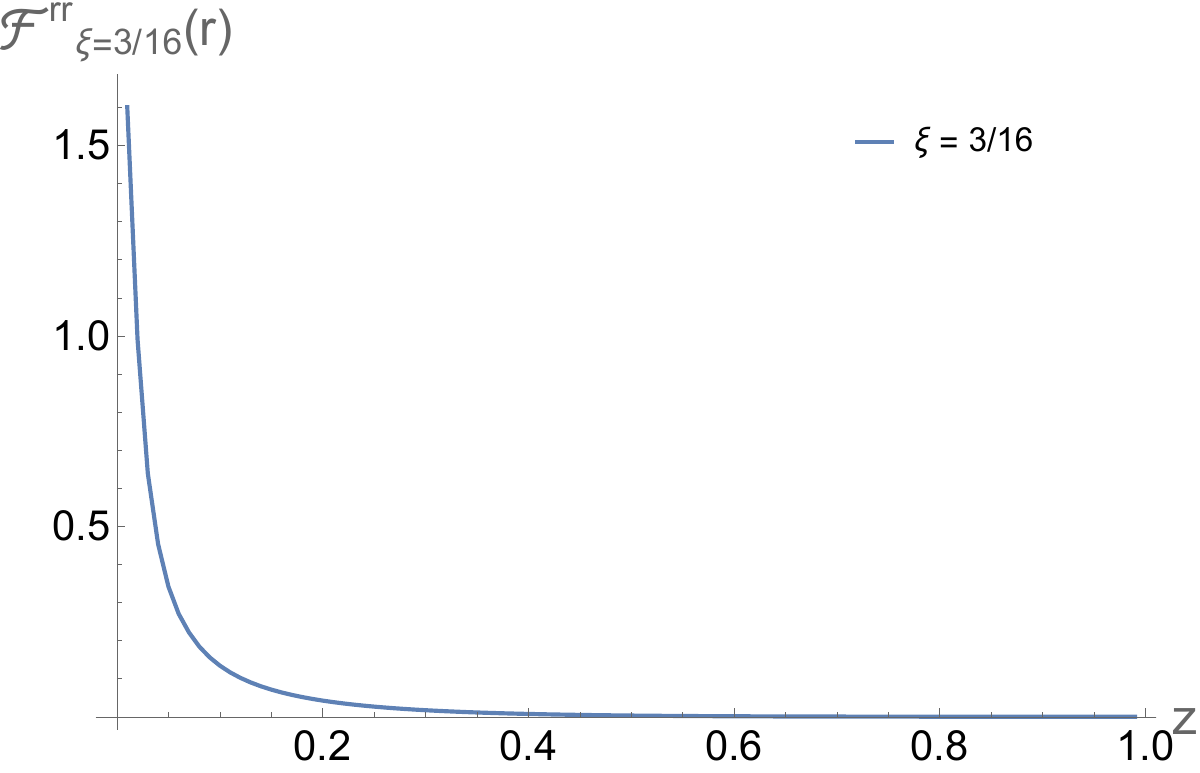}
    \end{minipage}
    \vskip\baselineskip
    \begin{minipage}{0.48\textwidth}
        \centering
        \includegraphics[width=\textwidth]{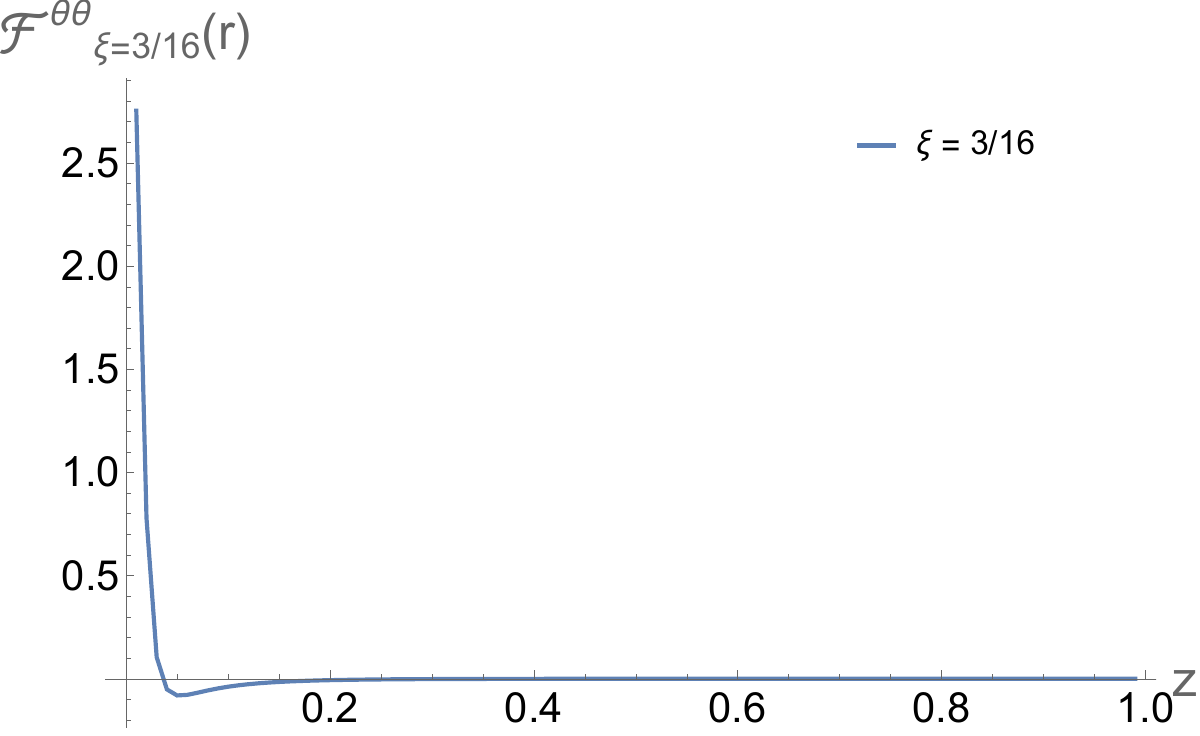}
    \end{minipage}
    \hfill
    \begin{minipage}{0.48\textwidth}
        \centering
        \includegraphics[width=\textwidth]{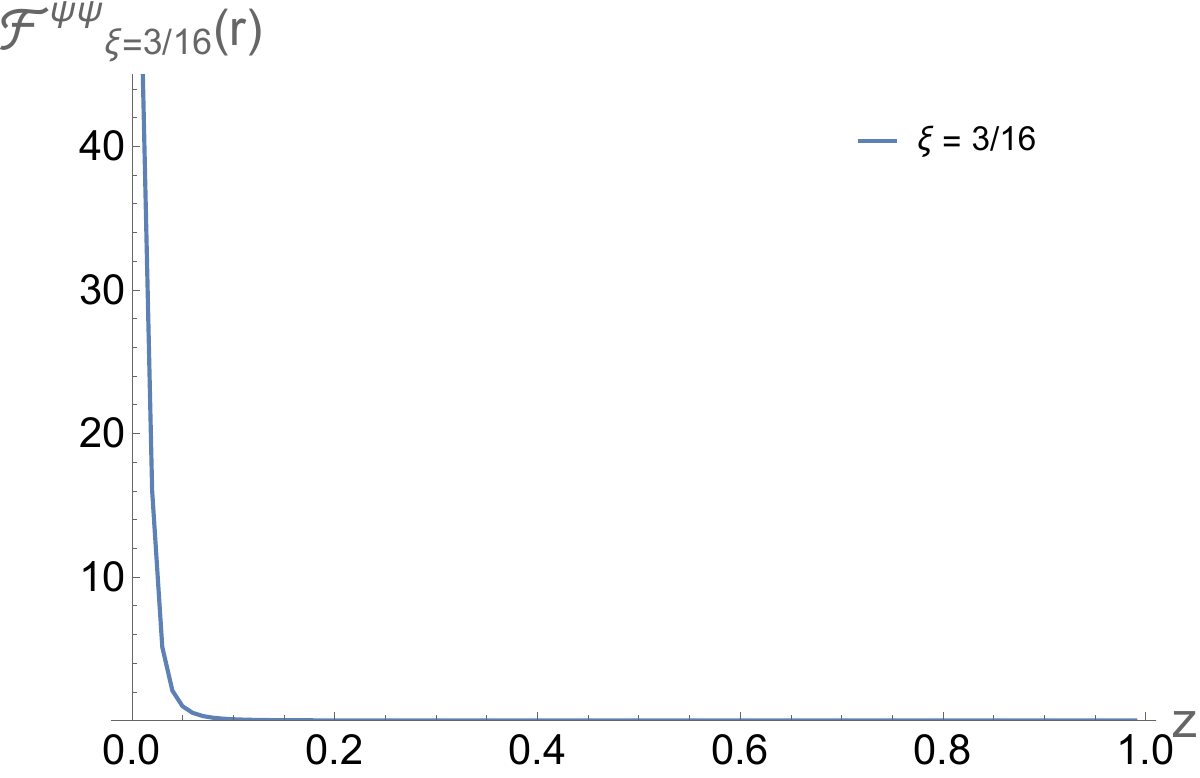}
    \end{minipage}
    \vskip\baselineskip
    \begin{minipage}{0.48\textwidth}
        \centering
        \includegraphics[width=\textwidth]{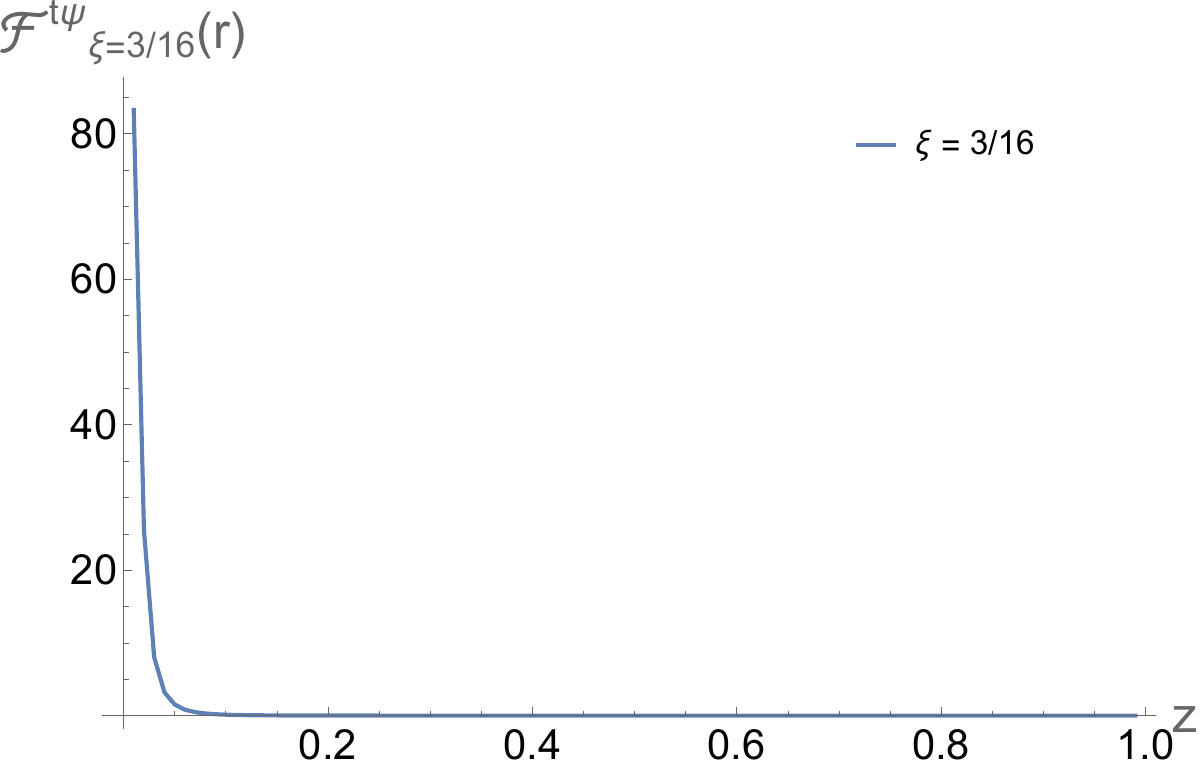}
    \end{minipage}
    \caption{Functions ${\mathcal {F}}^{\bullet \bullet}_{\xi = 3/16}(r)$ in the SET \eqref{eq:SETsimp1}, for a conformally-coupled scalar field with $\xi = 3/16$. The space-time and other scalar field parameters are as in Fig.~\ref{fig:integrand}.}
    \label{fig:SETxi316}
\end{figure*}

\begin{figure*}
    \centering
    \begin{minipage}{0.48\textwidth}
        \centering
        \includegraphics[width=\textwidth]{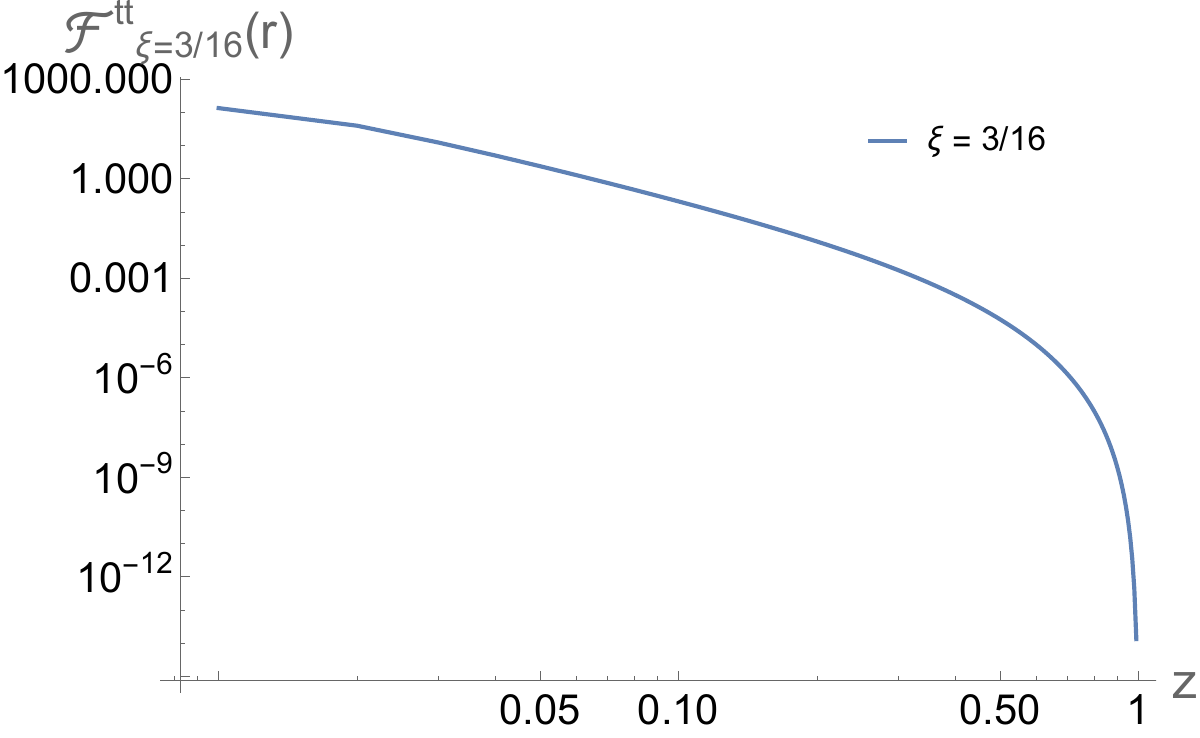}
    \end{minipage}
    \hfill
    \begin{minipage}{0.48\textwidth}
        \centering
        \includegraphics[width=\textwidth]{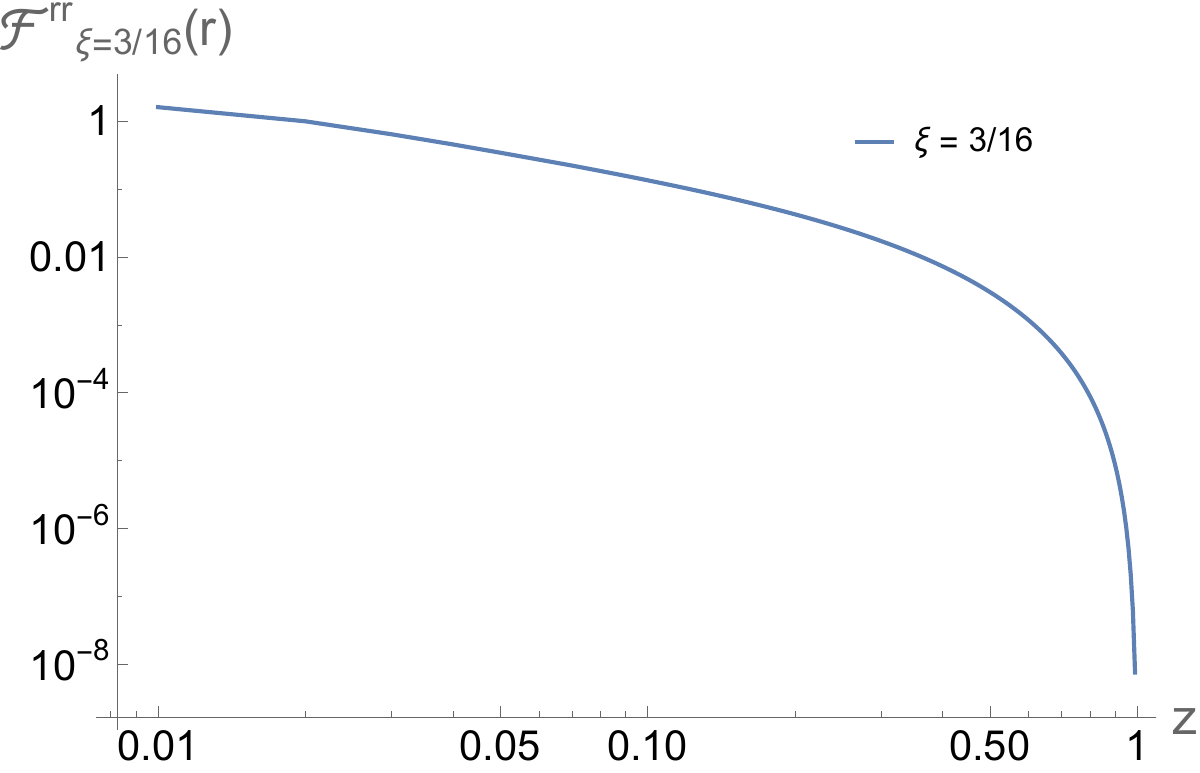}
    \end{minipage}
    \vskip\baselineskip
    \begin{minipage}{0.48\textwidth}
        \centering
        \includegraphics[width=\textwidth]{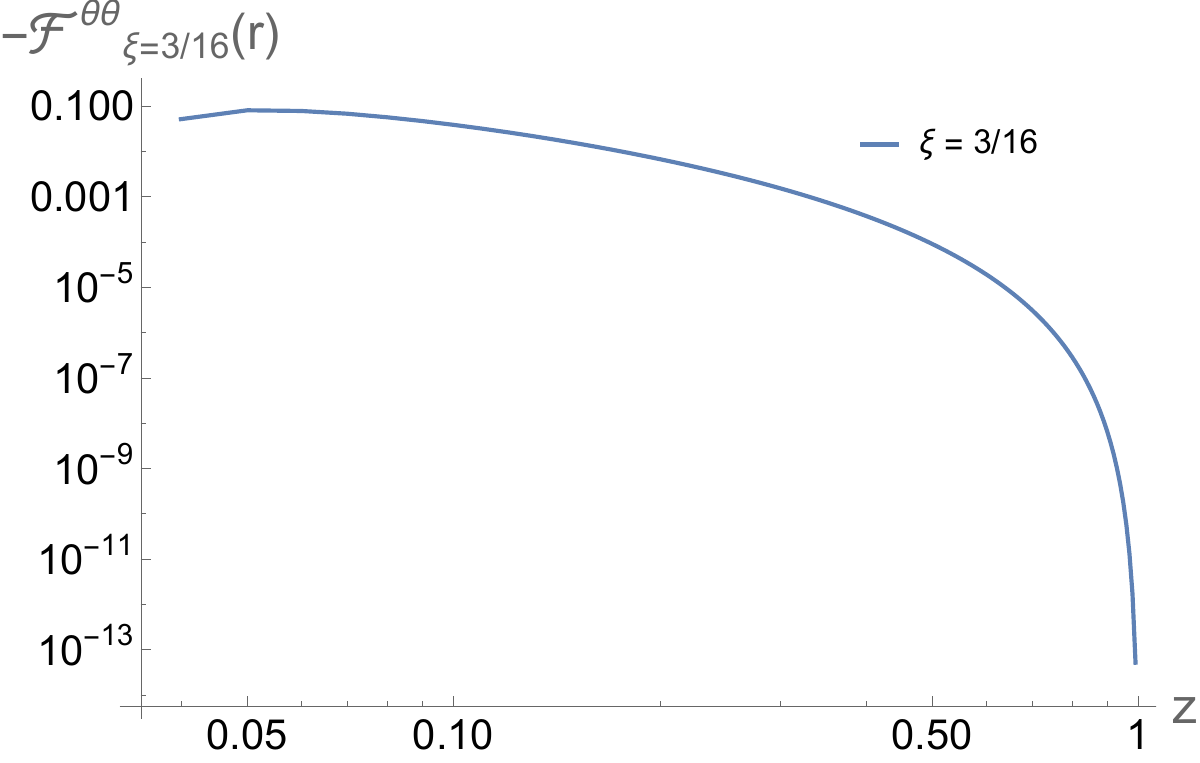}
    \end{minipage}
    \hfill
    \begin{minipage}{0.48\textwidth}
        \centering
        \includegraphics[width=\textwidth]{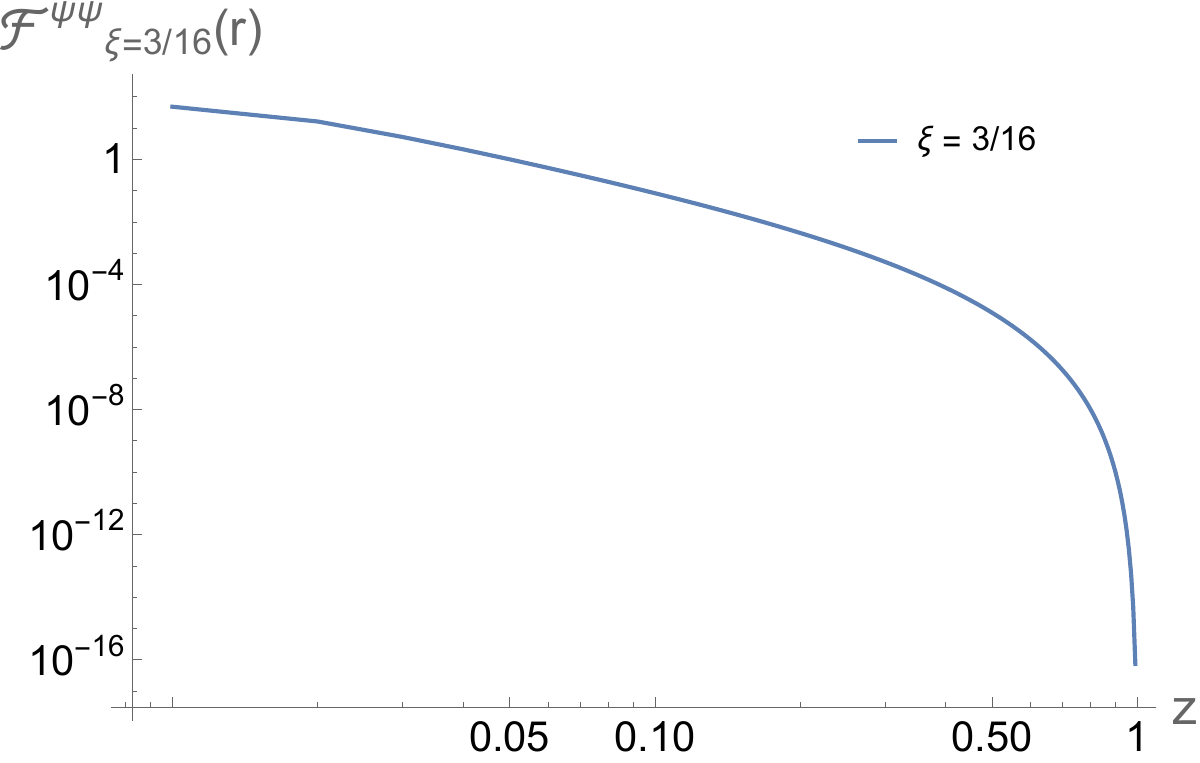}
    \end{minipage}
    \vskip\baselineskip
    \begin{minipage}{0.48\textwidth}
        \centering
        \includegraphics[width=\textwidth]{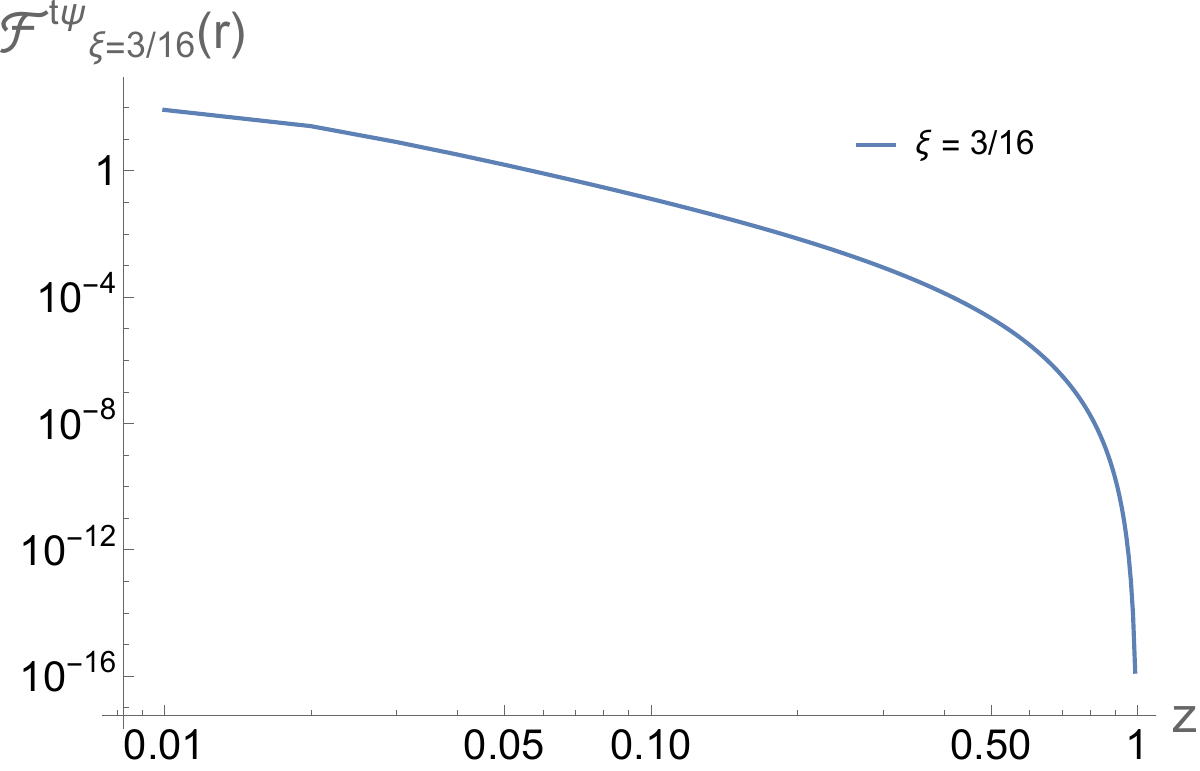}
    \end{minipage}
    \caption{
    Functions ${\mathcal {F}}^{\bullet \bullet}_{\xi = 3/16}(r)$ in the SET \eqref{eq:SETsimp1}, for a conformally-coupled scalar field with $\xi = 3/16$. The space-time and other scalar field parameters are as in Fig.~\ref{fig:integrand}.  The functions are plotted using a log-log scale. 
Since the function ${\mathcal {F}}^{\theta \theta}_{\xi = 3/16}(r)$ is negative for the majority of the data points (see Fig.~\ref{fig:SETxi316}), here we show $-{\mathcal {F}}^{\theta \theta}_{\xi = 3/16}(r)$ only for those values of $z$ for which this quantity is positive, so this function is shown for a different range of values of $z$ than all the other functions.}
    \label{fig:SETxi316log}
\end{figure*}

We present our numerical results for the difference in SET expectation values between the Hartle-Hawking and Boulware states by considering the functions ${\mathcal {F}}^{\bullet \bullet}(r)$ \eqref{eq:SETsimp1}. 
In Figs.~\ref{fig:SETxi316}--\ref{fig:SETxi316log} we consider a conformally-coupled field having $\xi = 3/16$, with the remaining parameters as for the vacuum polarization studied in Sec.~\ref{sec:VPres}.
In Fig.~\ref{fig:SETxi316} these functions are plotted as functions of $z$ using a linear scale, while in Fig.~\ref{fig:SETxi316log} a log-log scale is employed.

Looking first at Fig.~\ref{fig:SETxi316}, we see that all five functions ${\mathcal {F}}^{\bullet \bullet }(r)$ diverge as $z\rightarrow 0$ and the event horizon is approached. 
While the $(t,r, \theta, \varphi , \psi )$ coordinate system is not regular across the event horizon, changing to Kruskal coordinates (\ref{eq:Kruskal}) does not change the angular coordinate $\theta $ and therefore the divergence of ${\mathcal {F}}^{\theta \theta }(r)$ as $z\rightarrow 0 $ indicates that the difference in SET expectation values diverges on the horizon.
As discussed above, this is in accordance with our expectation that the Hartle-Hawking state is regular at the horizon, but the Boulware state diverges there.
All five functions ${\mathcal {F}}^{\bullet \bullet }(r)$ are positive in a neighbourhood of the horizon, but their magnitudes at small fixed $z$ are quite different, with ${\mathcal {F}}^{tt}(r)$ being the largest and ${\mathcal {F}}^{rr}(r)$ the smallest of these functions. 

As $z$ increases, all five functions ${\mathcal {F}}^{\bullet \bullet }(r)$ remain positive, with the exception of ${\mathcal {F}}^{\theta \theta }(r)$ which is negative throughout the space-time except for a small region near the event horizon. 
As was observed for the vacuum polarization, all five functions ${\mathcal {F}}^{\bullet \bullet }(r)$ (and hence the SET itself) tend to zero rapidly as $z\rightarrow 1$ and the space-time boundary is approached.  
This behaviour is more readily seen in the log-log plots in Fig.~\ref{fig:SETxi316log}. 
As discussed above, this is in agreement with our prediction that the Hartle-Hawking and Boulware states are the same far from the black hole. 

\begin{figure*}
    \centering
    \begin{minipage}{0.48\textwidth}
        \centering
        \includegraphics[width=\textwidth]{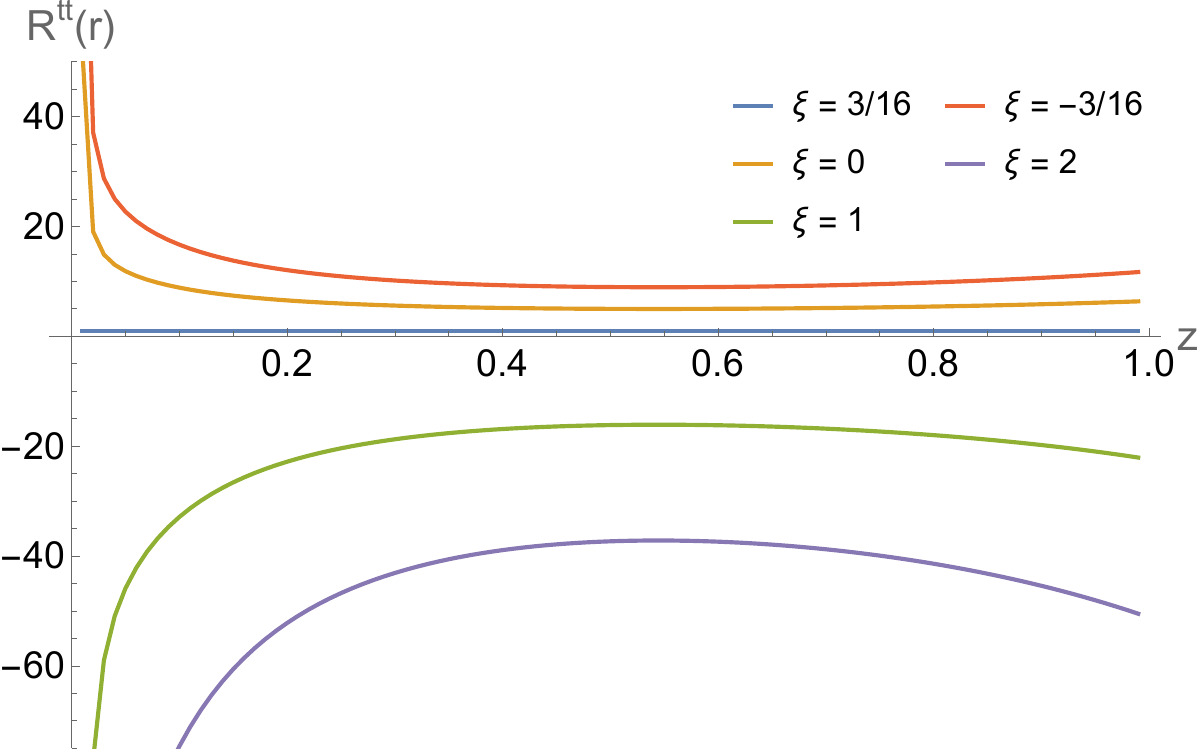}
    \end{minipage}
    \hfill
    \begin{minipage}{0.48\textwidth}
        \centering
        \includegraphics[width=\textwidth]{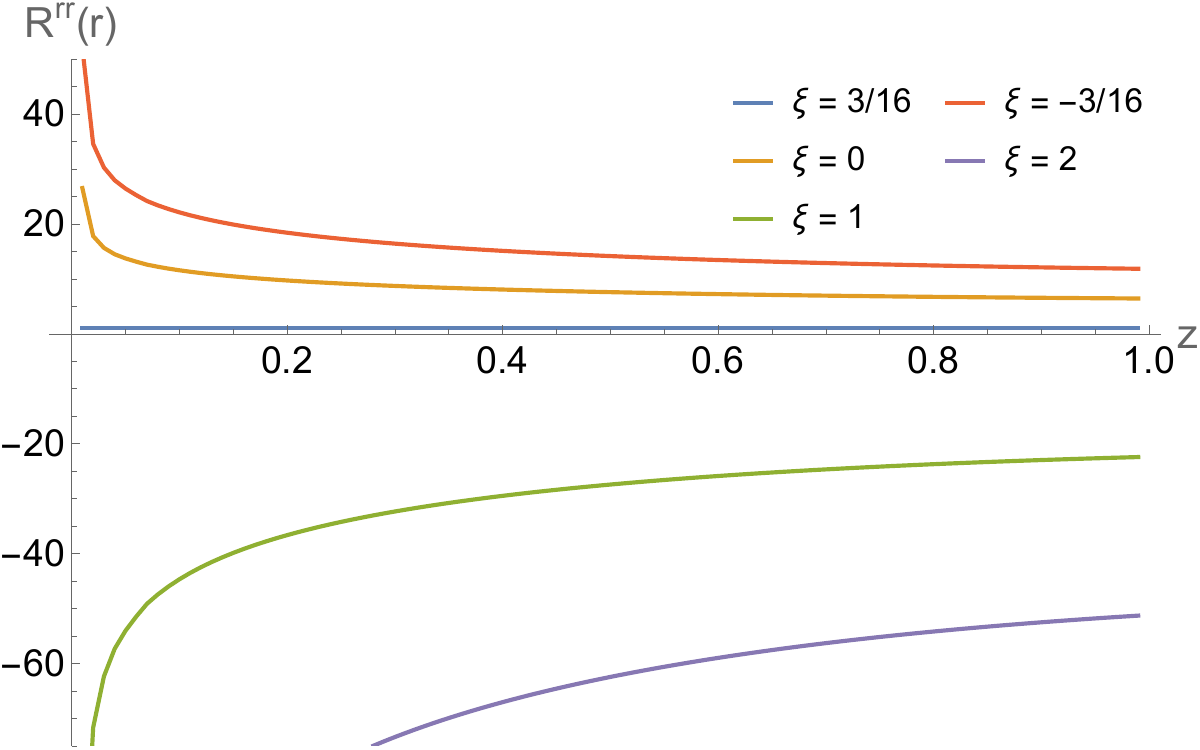}
    \end{minipage}
    \vskip\baselineskip
    \begin{minipage}{0.48\textwidth}
        \centering
        \includegraphics[width=\textwidth]{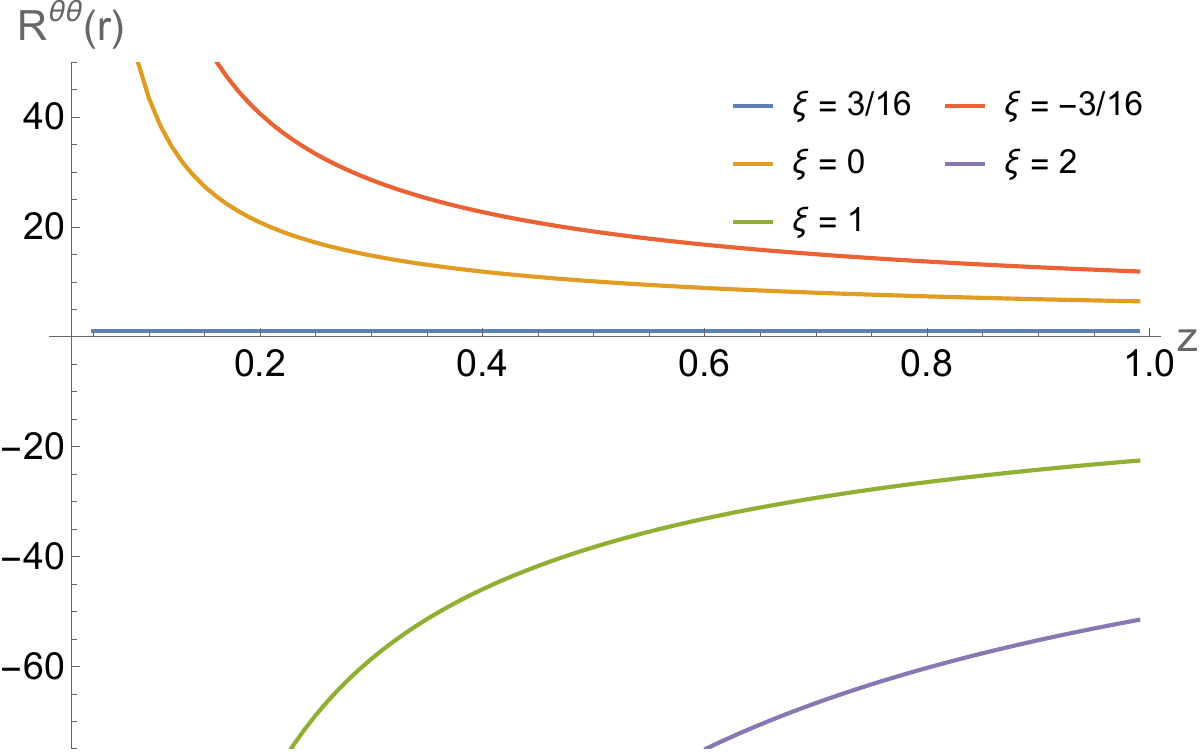}
    \end{minipage}
    \hfill
    \begin{minipage}{0.48\textwidth}
        \centering
        \includegraphics[width=\textwidth]{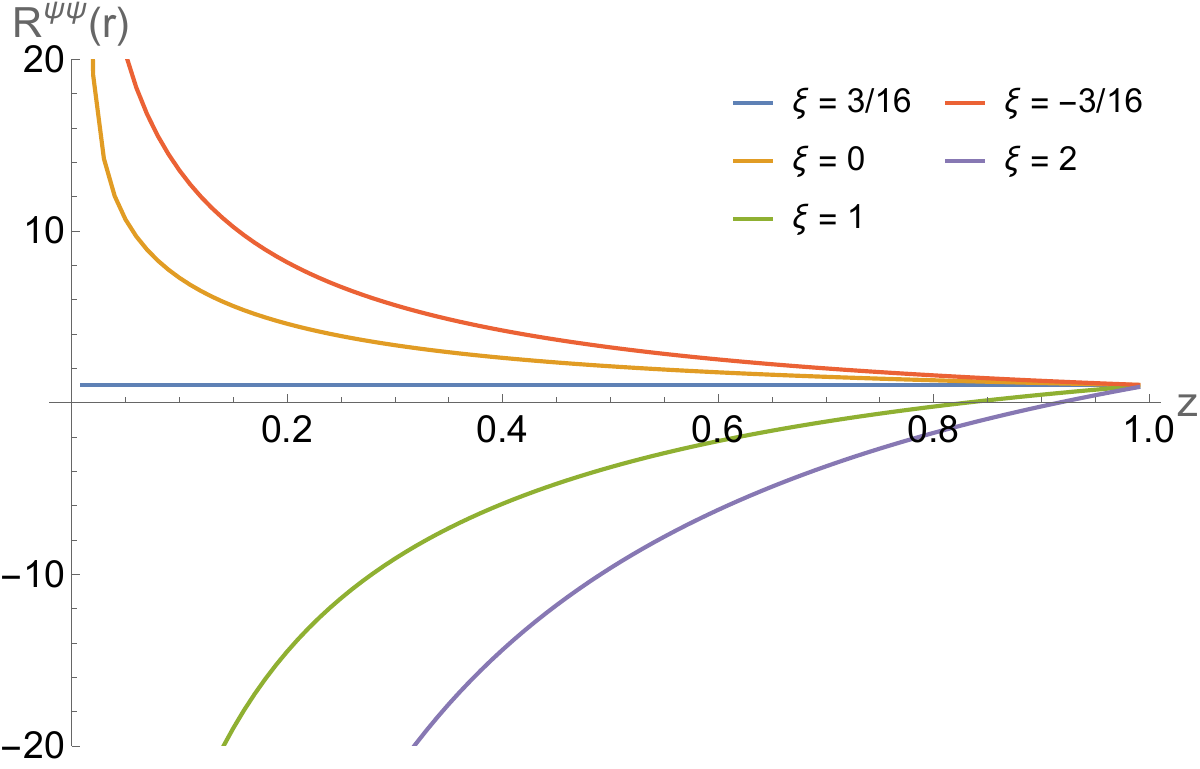}
    \end{minipage}
    \vskip\baselineskip
    \begin{minipage}{0.48\textwidth}
        \centering
        \includegraphics[width=\textwidth]{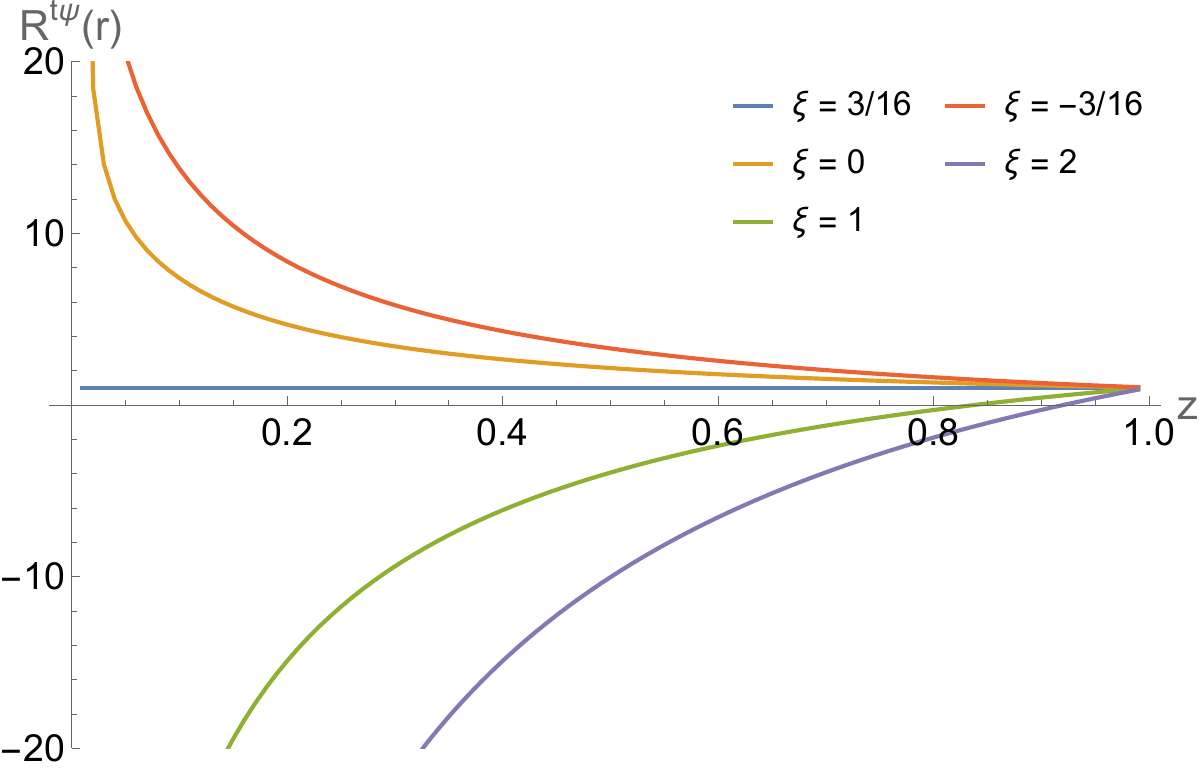}
    \end{minipage}
    \caption{Ratios ${\mathsf {R}}(r)^{\bullet \bullet}={\mathcal {F}}^{\bullet \bullet}(r)/{\mathcal {F}}^{\bullet \bullet}_{\xi = 3/16}(r)$ of the functions ${\mathcal {F}}^{\bullet \bullet}(r)$ in the SET (\ref{eq:SETsimp1}) for a selection of values of the coupling constant $\xi $ with the corresponding functions when $\xi = 3/16$ and the scalar field is conformally coupled. The space-time and other scalar field parameters are as in Fig.~\ref{fig:integrand}.}
    \label{fig:SETFR}
\end{figure*}

In Fig.~\ref{fig:SETFR} we explore the effect of changing the coupling constant $\xi $ (and therefore also the scalar field mass $\mu $) while keeping the effective mass $\nu $ fixed.
We plot the ratios ${\mathsf {R}}^{\bullet \bullet}(r)$ of the functions ${\mathcal {F}}^{\bullet \bullet}(r)$ for various different values of $\xi $ with those functions when $\xi = 3/16$ and the scalar field is conformally coupled.

We see from Fig.~\ref{fig:SETFR} that varying the coupling constant $\xi $ has a significant effect on the difference in expectation values of the SET operator, and can even change the sign of the functions ${\mathcal {F}}^{\bullet \bullet }(r)$.
All the ratios ${\mathsf {R}}^{\bullet \bullet }(r)$ are, trivially, equal to unity when $\xi = 3/16$ and the field is conformally coupled. 
Decreasing the coupling constant $\xi$ below $3/16$ increases all the ratios ${\mathsf {R}}^{\bullet \bullet }(r)$ for every value of the radial coordinate $z$, while increasing $\xi $ above $3/16$ decreases the ratios. 
All five ratios are negative (for nearly all values of $z$) when $\xi = 1$ or $\xi= 2$.

Near the horizon, as $z\rightarrow 0$, the ratios ${\mathsf {R}}^{\bullet \bullet }(r)$ are all diverging, implying that the SET components diverge more quickly on approaching the horizon when the field is not conformally coupled. The rate of divergence increases as $|\xi - 3/16|$ increases. 
The ratios ${\mathsf {R}}^{\bullet \bullet }(r)$ exhibit different behaviour as $z\rightarrow 1 $ and the space-time boundary is approached.
The ratios ${\mathsf {R}}^{tt}(r)$ are slightly increasing in magnitude as $z\rightarrow 1 $, but appear to remain finite. 
In contrast, the ratios ${\mathsf {R}}^{rr}(r)$ and ${\mathsf {R}}^{\theta \theta}(r)$ are slightly decreasing in magnitude on approaching the boundary, but approach nonzero limits. 
Finally, the ratios ${\mathsf {R}}^{t\psi }(r)$ and ${\mathsf {R}}^{\psi \psi }(r)$ tend to unity for all values of $\xi $ as $z\rightarrow 1$.
 The other notable feature is that the ratios ${\mathsf {R}}^{t\psi }(r)$ and ${\mathsf {R}}^{\psi \psi }(r)$ are very similar (they are not quite the same, but are indistinguishable in the plots), although the functions ${\mathcal {F}}^{t\psi }(r)$ and ${\mathcal {F}}^{\psi \psi }(r)$ are not.

On a four-dimensional Reissner-Nordstr\"om black hole, it is found in \cite{Arrechea:2023fas} that, for all nonzero components of the renormalized SET in the Hartle-Hawking, Boulware or Unruh states, changing the value of $\xi $ does not change the expectation value far from the black hole.  
In our situation this appears to happen only for some components of the SET. 
In \cite{Arrechea:2023fas}, changing the coupling constant $\xi $ also does not affect the regularity or rate of divergence (depending on the quantum state under consideration) of the SET components. 
Again, this result appears not to be replicated in our set-up.
A full computation of the renormalized SET (which is beyond the scope of this work) would however be required to address this issue more fully.

We close our study of the SET by considering one further property, namely the rate of rotation of the thermal distribution represented by the difference in expectation values between the Hartle-Hawking and Boulware states.
To find the angular speed with which the thermal radiation is rotating, we use the method of
\cite[Sec.~VII]{Casals:2005kr} (see also \cite[Sec.~IV C 2]{Casals:2012es}).
Suppose that we have an observer on the black-hole space-time (\ref{eq:metric}) at constant $(r,\theta, \varphi)$ and with angular speed $\mho $, given by 
\begin{equation}
    \mho = \frac{d\psi }{dt}. 
\end{equation}
Following \cite{Casals:2005kr,Casals:2012es} an orthonormal {\em {funfbein}} (or {\em {pentrad}}) basis of vectors ${}^{\mho }{\mathsf {e}}_{(a)}$ associated with this observer includes the following:
\begin{subequations}
\label{eq:mhoframe}
\begin{align}
    {}^{\mho }{\mathsf {e}}_{(t)} & = \frac{1}{{\mathfrak {N}}} \left( \partial _{t} + \mho \partial _{\psi } \right), 
    \\
    {}^{\mho }{\mathsf {e}}_{(\psi )} & =  \frac{1}{{\mathfrak {N}}} \left( g_{t\psi} ^{2} - g_{tt}g_{\psi \psi }\right) ^{-\frac{1}{2}} 
    \nonumber \\ & \qquad \times 
    \left[ -\left( g_{t\psi } + \mho g_{\psi \psi } \right) \partial _{t} + \left( g_{tt} + \mho g_{t\psi } \right) \partial _{\psi }  \right] ,
\end{align}
where 
\begin{equation}
    {\mathfrak {N}} = \left| g_{tt} + 2\mho g_{t\psi } + \mho ^{2} g_{\psi \psi }\right| ^{\frac{1}{2}}.
\end{equation}
\end{subequations}
The remaining vectors ${}^{\mho }{\mathsf{e}}_{(r)}={\mathsf{e}}_{(r)}$, ${}^{\mho }{\mathsf{e}}_{(\theta )}={\mathsf{e}}_{(\theta )}$ and ${}^{\mho }{\mathsf {e}}_{(\varphi )}={\mathsf {e}}_{(\varphi )}$ in the funfbein do not depend on the angular speed $\mho $ and are not required for the analysis in this section. 
The metric components in (\ref{eq:mhoframe}) can be found in (\ref{eq:metric}), and using these we have
\begin{equation}
    {\mathfrak {N}} =\left|  -f(r)^{2} + \left[ \mho - \Omega (r) \right]^{2}h(r)^{2} \right| ,
\end{equation}
where $\Omega (r)$ is given in (\ref{eq:Omega}). 
Since we are assuming that there is no speed-of-light surface, ${\mathfrak {N}}$ will be nonvanishing everywhere outside the event horizon for all $\mho \in [0, \Omega _{+}+\epsilon )$ for some $\epsilon >0$, and 
in the following we are concerned only with values of $\mho $ in this interval.
Three natural values of $\mho $ which one might consider correspond to static observers ($\mho = 0$), 
rigidly-rotating observers ($\mho = \Omega _{+}$ (\ref{eq:OmegaH})) and zero angular momentum observers (ZAMOs) \cite{Frolov:1989jh}, whose angular speed is given by 
\begin{equation}
    \mho = \Omega _{{\textrm {\tiny {ZAMO}}}} = - \frac{g_{t\psi }}{g_{\psi \psi }} = \Omega (r) .
    \label{eq:ZAMO}
\end{equation} 
As the name suggests, such observers have vanishing angular momentum about the rotation axis of the black hole.

The observers in which we are interested are, in the nomenclature of \cite{Casals:2005kr,Casals:2012es},
{\em {Zero Energy Flux Observers}} (ZEFOs), who see no angular flux of energy. 
Let $\Omega _{{\textrm {\tiny {ZEFO}}}}$ denote the angular speed of a ZEFO.  
Then, the funfbein component $\Delta {\hat {T}}_{(t)(\psi )}={}^{\mho }{\mathsf {e}}_{(t)}^{\mu }{}^{\mho }{\mathsf {e}}_{(\psi )}^{\nu }\Delta {\hat {T}} _{\mu \nu } $ of the difference in the SET expectation values between the Hartle-Hawking and Boulware states will vanish when evaluated using the funfbein (\ref{eq:mhoframe}) with $\mho = \Omega _{{\textrm {\tiny {ZEFO}}}}$. 
Using (\ref{eq:mhoframe}) and setting $\Delta {\hat {T}}_{(t)(\psi )}=0$ gives $\Omega _{{\textrm {\tiny {ZEFO}}}}$ to be a solution of the quadratic equation \cite{Casals:2005kr}
\begin{subequations}
\label{eq:quadratic}
\begin{equation}
    A_{{\textrm {\tiny {ZEFO}}}}\Omega _{{\textrm {\tiny {ZEFO}}}}^{2}  + B_{{\textrm {\tiny {ZEFO}}}}\Omega _{{\textrm {\tiny {ZEFO}}}} + C_{{\textrm {\tiny {ZEFO}}}} =0,
    \label{eq:quadratic1}
\end{equation}
where the coefficients are given by \cite{Casals:2005kr}
\begin{align}
    A_{{\textrm {\tiny {ZEFO}}}} & = 
    g_{\psi \psi }\Delta {\hat {T}} _{t\psi } - g_{t\psi }\Delta {\hat {T}} _{\psi \psi  },
    \\
    B_{{\textrm {\tiny {ZEFO}}}} & = 
    g_{\psi \psi }\Delta {\hat {T}} _{tt} - g_{tt}\Delta {\hat {T}} _{\psi \psi  },
    \\
    C_{{\textrm {\tiny {ZEFO}}}} & = 
    g_{t\psi }\Delta {\hat {T}} _{tt} - g_{tt}\Delta {\hat {T}} _{t\psi  }.
\end{align}
\end{subequations}
Using the expressions (\ref{eq:SETsimpdown}) for the SET components in terms of the ${\mathcal {F}}^{\bullet \bullet} $ functions, we find
\begin{subequations}
    \begin{align}
     A_{{\textrm {\tiny {ZEFO}}}} & = 
     f(r)^{2}h(r)^{4}\left[ \Omega (r) {\mathcal {F}}^{tt}(r)- {\mathcal {F}}^{t\psi }(r) \right]
     ,
     \label{eq:AZEFO}
     \\
     B_{{\textrm {\tiny {ZEFO}}}} & = f(r)^{2}h(r)^{2} \left[ 
     \left\{ f(r)^{2} - \Omega (r)^{2}h(r)^{2} \right\} {\mathcal {F}}^{tt}(r)
     \right. \nonumber \\ & \qquad \left. 
     +h(r)^{2}{\mathcal {F}}^{\psi \psi }(r) + \frac{1}{4}h(r)^{2} {\mathcal {F}}^{\theta \theta } (r)
     \right]
     ,
     \\
     C_{{\textrm {\tiny {ZEFO}}}} & = -f(r)^{2}h(r)^{2} \left[ 
     \left\{  f(r)^{2} - \Omega (r)^{2}h(r)^{2}  \right\} {\mathcal {F}}^{t\psi}(r)
     \right. \nonumber \\ & \qquad \left. 
     +\Omega (r) h(r)^{2} {\mathcal {F}}^{\psi \psi }(r)
     + \frac{1}{4} \Omega (r) h(r)^{2} {\mathcal {F}}^{\theta \theta }(r)
     \right]
     ,
\end{align}
and the quadratic equation (\ref{eq:quadratic1}) takes the form
\begin{align}
    0 =  & ~ \Omega _{{\textrm {\tiny {ZEFO}}}} \left[ 
    f(r)^{2}+ \Omega (r)h(r)^{2} \left\{ \Omega _{{\textrm {\tiny {ZEFO}}}} - \Omega (r) \right\} 
    \right] {\mathcal {F}}^{tt}(r)
    \nonumber \\ & ~
    -\left[ f(r)^{2} + h(r)^{2} \left\{  \Omega _{{\textrm {\tiny {ZEFO}}}}^{2} - \Omega (r)^{2}\right\} \right] {\mathcal {F}}^{t\psi }(r)
    \nonumber \\ & ~
    + h(r)^{2} \left[ \Omega _{{\textrm {\tiny {ZEFO}}}} - \Omega (r) \right] \left[ {\mathcal{F}}^{\psi \psi }(r)+ \frac{1}{4}{\mathcal {F}}^{\theta \theta }(r) \right] .
    \label{eq:quadratic2}
\end{align}
\end{subequations}
At the event horizon, since the metric function $f(r)^{2}$ (\ref{eq:metricfunctionsend}) vanishes, a solution of (\ref{eq:quadratic2}) is simply
$\Omega _{{\textrm {\tiny {ZEFO}}}} = \Omega (r_{+})=\Omega _{+}$, the angular speed of rotation of the event horizon, irrespective of the details of the SET functions ${\mathcal {F}}^{\bullet \bullet}(r)$.

Away from  the horizon, the algebraic expression for the solution of (\ref{eq:quadratic2}) is not very enlightening, so instead we find $\Omega _{{\textrm {\tiny {ZEFO}}}}$ numerically.
A numerically effective way to compute the solution $\Omega _{{\textrm {\tiny {ZEFO}}}}$ of (\ref{eq:quadratic2}) is to write it as \cite{Casals:2005kr}
\begin{equation}
    \Omega _{{\textrm {\tiny {ZEFO}}}} = -\frac{2C_{{\textrm {\tiny {ZEFO}}}}}{B_{{\textrm {\tiny {ZEFO}}}}\pm {\sqrt {B_{{\textrm {\tiny {ZEFO}}}}^{2}-4A_{{\textrm {\tiny {ZEFO}}}}C_{{\textrm {\tiny {ZEFO}}}}}}},
    \label{eq:OmegaZEFO}
\end{equation}
choosing the sign in the denominator such that $\Omega _{{\textrm {\tiny {ZEFO}}}}$ is regular and positive.

\begin{figure}
    \centering
    \includegraphics[scale=0.4]{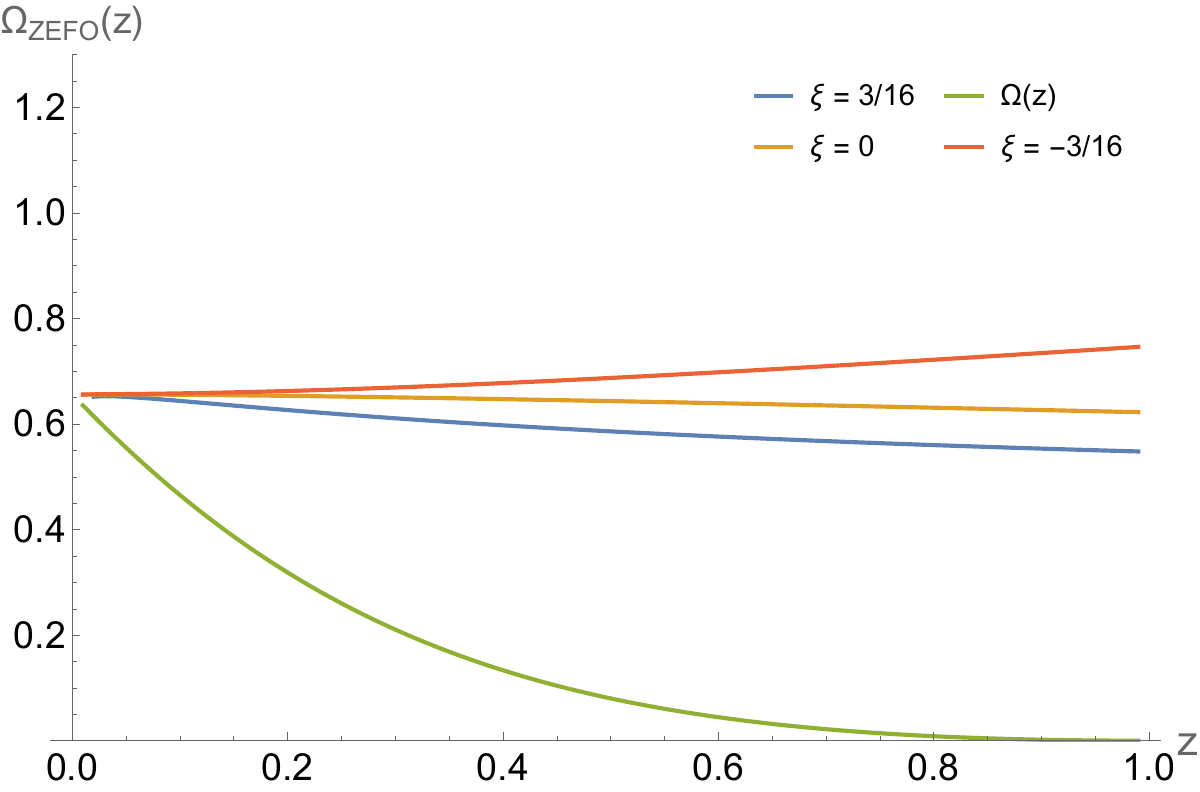}
    \caption{$\Omega _{{\textrm {\tiny {ZEFO}}}}$ (\ref{eq:OmegaZEFO}) computed using the difference in expectation values of the SET between the Hartle-Hawking and Boulware states (blue, red and yellow curves, corresponding to three values of the coupling constant $\xi $), and $\Omega _{{\textrm {\tiny {ZAMO}}}}=\Omega (r)$ (\ref{eq:ZAMO}) (green curve). The space-time and other scalar field parameters are as in Fig.~\ref{fig:integrand}.}
    \label{fig:ZEFO}
\end{figure}

In Fig.~\ref{fig:ZEFO} we plot our results for $\Omega _{{\textrm {\tiny {ZEFO}}}}$ computed using both the difference in SET expectation values between the Hartle-Hawking and Boulware states and $\Omega _{{\textrm {\tiny {ZAMO}}}}=\Omega (r)$ (\ref{eq:ZAMO}) (green curve). 
We see that $\Omega _{{\textrm {\tiny {ZAMO}}}}$ decreases rapidly as we move away from the horizon, and tends to zero at infinity. 
In Fig.~\ref{fig:ZEFO} we show our results for $\Omega _{{\textrm {\tiny {ZEFO}}}}$ for three values of the constant $\xi $ coupling the scalar field to the Ricci scalar curvature. 
For the other values of the coupling constant considered in Fig.~\ref{fig:SETFR}, namely $\xi = 1$ and $\xi = 2$, we are unable to obtain physically reasonable values of $\Omega _{{\textrm {\tiny {ZEFO}}}}$, due to the quantity $A_{{\textrm {\tiny {ZEFO}}}}$ (\ref{eq:AZEFO}) passing through zero between the event horizon and infinity.

For the values of $\xi $ considered in Fig.~\ref{fig:ZEFO}, on the horizon, $\Omega _{{\textrm {\tiny {ZEFO}}}}=\Omega _{{\textrm {\tiny {ZAMO}}}}=\Omega _{+}$ (\ref{eq:OmegaH}) as expected and the thermal distribution is rotating with the same angular speed as the event horizon. 
Away from the horizon, it can be seen in Fig.~\ref{fig:ZEFO} that $\Omega _{{\textrm {\tiny {ZEFO}}}}$ has values close to (but not exactly equal to) $\Omega _{+}$, and, in particular, is significantly larger than the rate of rotation of a ZAMO, $\Omega (r)$. 
For $\xi = 0$ (minimal coupling) and $\xi = 3/16$ (conformal coupling) we find that $\Omega _{{\textrm {\tiny {ZEFO}}}}$ decreases slightly with increasing distance from the horizon, while for $\xi = -3/16$ it can be seen that $\Omega _{{\textrm {\tiny {ZEFO}}}}$ slightly increases as the radial coordinate increases.
We deduce that the difference in SET expectation values between the Hartle-Hawking and Boulware states corresponds to a thermal distribution of particles almost (but not quite) rigidly-rotating with the angular speed of the event horizon.

On a Kerr black hole, a state rigidly-rotating with the same angular speed as the event horizon would be divergent on the speed-of-light surface \cite{Ottewill:2000yr}.
This is not a concern in our situation, as we are assuming that the black hole rotates sufficiently slowly that there is no speed-of-light surface.
Furthermore, it is clear from Fig.~\ref{fig:ZEFO} that the thermal distribution is not exactly rigidly-rotating. 
Similar results were obtained for the corresponding thermal distribution of a quantum scalar \cite{Duffy:2005mz}, fermion \cite{Casals:2012es} and electromagnetic \cite{Casals:2005kr} field on a Kerr space-time. 
For Kerr black holes, it is also found in \cite{Duffy:2005mz,Casals:2005kr,Casals:2012es} that $\Omega _{{\textrm {\tiny {ZEFO}}}}$ is significantly different from $\Omega _{{\textrm {\tiny {ZAMO}}}}$, as is the case in our set-up.

\section{Conclusions}
\label{sec:conc}

In this paper we have studied the canonical quantization of a scalar field on a background Myers-Perry-AdS black hole in five space-time dimensions.
We have set the two angular momentum parameters in the metric to be equal, which results in a geometry with enhanced symmetry compared to the generic case.
We assume that the angular momentum of the black hole is sufficiently small that there is no speed-of-light surface and there exists a Killing vector which is time-like everywhere outside the event horizon.
In this case classical superradiance is absent and there are no unstable scalar field modes.
We thus avoid the complexities of canonical quantization in the presence of classical superradiance \cite{Ottewill:2000qh,Balakumar:2022yvx} and can readily construct a Boulware state $|{\mathrm {B}}\rangle $ and a Hartle-Hawking state $|{\mathrm {H}} \rangle $.

We compute the differences in expectation values of the vacuum polarization (the square of the field operator) and the SET operator between these two states, which have the advantage of not requiring renormalization.
Notwithstanding the simplifications afforded by the enhanced symmetry of the background space-time, our numerical computations are rather time-consuming.
Consequently, we have presented results for one particular choice of the set of parameters of the model, which are the black hole mass parameter $M$, the angular momentum parameter $a$, and the scalar field effective mass $\nu $ (\ref{eq:nu}). 
It would be interesting the explore the effect of varying these parameters on the quantum field expectation values, but this would require the development of a more efficient numerical method. 

Since the black holes considered here are asymptotically-AdS, it is necessary to apply boundary conditions to the scalar field in order to have a well-defined quantum field theory.
In this work, we have applied only the simplest (Dirichlet) boundary conditions to the field, for which the scalar field modes tend to zero as quickly as possible far from the black hole. 
While Dirichlet boundary conditions are the most widely employed in the literature, 
studies of the renormalized vacuum polarization in pure AdS in three and four dimensions \cite{Morley:2020ayr,Namasivayam:2022bky}
has revealed that Neumann rather than Dirichlet boundary conditions give the generic behaviour of the vacuum polarization at the space-time boundary. 
Considering more general Robin boundary conditions changes the behaviour of both the vacuum polarization \cite{Morley:2020ayr,Namasivayam:2022bky} and SET \cite{Morley:2023exv,Namasivayam:2025dub} in pure AdS space-time, and 
the vacuum polarization on asymptotically-AdS black holes in four dimensions \cite{Morley:2020zcd} (the corresponding SET for Robin boundary conditions has yet to be computed).  
It would therefore also be interesting to explore the effect of different boundary conditions on the expectation values on the black hole space-times studied in this paper. 

Our analysis has considered only the region exterior to the event horizon of the black hole, where both the Hartle-Hawking and Boulware states can be defined.
We expect that the Boulware state is divergent at the event horizon, but that it should be possible to extend the construction of the Hartle-Hawking state into the black hole interior.
For a four-dimensional Kerr black hole, there is no Hartle-Hawking state \cite{Kay:1988mu} to extend into the black hole interior, but the Unruh state has been constructed inside the event horizon \cite{Zilberman:2022iij}. 
A number of recent papers (see, for example, \cite{McMaken:2022dqc,Zilberman:2022aum,Klein:2023urp,Klein:2024sdd,McMaken:2024fvq,Zilberman:2024jns}) have studied the properties of expectation values on the interiors of rotating black holes, paying particular attention to the region close to the inner horizon.
A notable exception of a rotating black hole having a Hartle-Hawking state for which expectation values have been computed  in the black hole interior is the three-dimensional BTZ black hole \cite{Steif:1993zv,Casals:2016odj,Casals:2019jfo,Dias:2019ery}, where again the behaviour at the inner horizon has been a particular focus of investigation.

Work to date on a Hartle-Hawking state on a black hole interior in four dimensions is restricted to spherically symmetric black holes (see, for example, \cite{Lanir:2017oia} for the construction of $|{\rm {H}}\rangle $ inside a Reissner-Nordstr\"om black hole). 
The properties of expectation values on the interior of spherically symmetric black holes have attracted a good deal of attention in recent years (a selection of references is  \cite{Lanir:2018rap,Lanir:2018vgb,Hollands:2019whz,Zilberman:2019buh,Hollands:2020qpe,Zilberman:2021vgz,Klein:2023urp,Juarez-Aubry:2021tae,Hintz:2023pak,McMaken:2023uue}).
A particular focus has been the behaviour of these quantities near the inner (Cauchy) horizon.
In general, the expectation value of the SET for a quantum field diverges at the inner horizon, with a rate of divergence which is larger than the corresponding classical SET for classical perturbations \cite{Hollands:2019whz,Zilberman:2019buh,Hollands:2020qpe,Hintz:2023pak}. 
Thus quantum perturbations can restore cosmic censorship in the case of those Reissner-Nordstr\"om-de Sitter black holes for which classical perturbations do not blow up at the Cauchy horizon \cite{Hollands:2019whz,Hollands:2020qpe}.

One motivation for our work was the proof \cite{Gerard:2021yoi} of the existence of a Hartle-Hawking state for a four-dimensional stationary black hole possessing a Killing vector time-like everywhere outside the event horizon.  
We conjecture that a corresponding result holds for the black holes considered in this paper, although it should be emphasized that our construction of the state $|{\rm {H}}\rangle $ in this paper is not rigorous. 
To lend support to this hypothesis, it would be interesting to compute the renormalized expectation value of the SET (and the vacuum polarization) of the quantum scalar field in the state $|{\rm {H}}\rangle $, and in particular to ascertain whether this is regular at the event horizon.
Such work would be a necessary precursor to any detailed investigation of the properties of the state $|{\rm {H}}\rangle $ on the black hole interior. 
We anticipate that significant technical challenges would arise in attempting to perform such a calculation. 
Of the recent methodologies for computing renormalized expectation values in the region exterior to a black hole event horizon, the ``pragmatic mode sum'' approach has been successful on the four-dimensional Kerr black hole \cite{Levi:2016exv}.
However, this approach has, to date, only been employed on four-dimensional black holes, and it remains to be seen whether this would be practical to implement in higher dimensions.
In contrast, the ``extended coordinates'' approach
has been successfully applied to find the vacuum polarization on spherically symmetric black holes in more than four dimensions \cite{Taylor:2016edd,Taylor:2017sux}.
In our case, we would be interested in developing this methodology to both the renormalized SET and rotating black holes, either of which is likely to be rather nontrivial.
We therefore leave this for future work.

\begin{acknowledgments}
We thank the anonymous referee whose detailed feedback has signficantly improved the paper.
We acknowledge IT Services at The University of Sheffield for the provision of services for High Performance Computing.
A.M.~thanks Harkan Kirk-Karakaya for helpful discussions on sequence acceleration methods, in particular for drawing \cite{Corless:2023} to our attention.
E.W.~thanks Christian G\'erard, Dietrich H\"afner, Micha\l{} Wrochna, and participants in the programme ``Quantum and classical fields interacting with geometry''  at the Institut Henri Poincar\'{e} for stimulating discussions. 
The work of A.M.~is supported by an EPSRC studentship.
The work of E.W.~is supported by STFC grant number ST/X000621/1.
E.W.~also acknowledges support of the Institut Henri Poincar\'{e} (UAR 839 CNRS-Sorbonne Universit\'{e}), and LabEx CARMIN (ANR-10-LABX-59-01). 
\end{acknowledgments}

\section*{Data availability}
The data that support the findings of this article are openly available \cite{Figshare}.

\appendix

\section{Derivation of Eq.~(\ref{eq:soleqn})}
\label{sec:solderiv}
Here we give more details of the derivation of Eq.~(\ref{eq:soleqn}), which gives the location of the speed-of-light surface. 
We start with Eq.~(\ref{eq:soleqnfirst}):
\begin{align}
    0 &  = g_{tt} + 2\Omega _{+}g_{t\psi } + \Omega _{+}^{2} g_{\psi \psi } \nonumber \\
    & = -f(r)^{2} + h(r)^{2}\left[ \Omega (r) - \Omega _{+} \right]^{2},
    \label{eq:soleqn1}
\end{align}
where the metric functions $f(r)$, $h(r)$ and $\Omega (r)$ are given in (\ref{eq:metric}).
Multiplying throughout by $h(r)^{2}>0$, the equation for the speed-of-light surface at $r=r_{\mathrm {L}}$  is
\begin{equation}
    0 = -r_{\mathrm {L}}^{2}g(r_{\mathrm {L}})^{-2} + h(r_{\mathrm {L}})^{4}\left[ \Omega (r_{\mathrm {L}}) - \Omega _{+} \right] ^{2}.
    \label{eq:soleqn2}
\end{equation}
To simplify this equation, we proceed as follows.
From the definitions of $\Omega (r)$ (\ref{eq:Omega}) and $\Omega _{+}$ (\ref{eq:OmegaH}), we have
\begin{subequations}
\label{eq:intermediates}
\begin{equation}
    \Omega (r) - \Omega _{+}   = - \frac{\Omega (r) \Omega _{+}}{2Ma} \left(  r^{2}+ r_{+}^{2}  \right) \left( r^{2}-r_{+}^{2} \right) ,
\end{equation}
and making use of the relation (\ref{eq:OmegaHalt}), we can derive the following result:
\begin{equation}
    r^{2}g(r)^{-2} = -\frac{2M}{\Omega (r)} \left[ \Omega (r)- \Omega _{+}\right] -\frac{2Ma}{ r_{+}^{2}r^{2} \Omega (r)} \left( r^{2}-r_{+}^{2} \right) .
\end{equation}
\end{subequations}
Combining  the expressions (\ref{eq:intermediates}), we deduce that
\begin{equation}
    r^{2}g(r)^{-2}  = \left( r^{2} - r_{+}^{2} \right) \left[ \frac{\Omega _{+}}{a}\left( r^{2}+r_{+}^{2}\right)  -\frac{2Ma}{r_{+}^{2}r^{2}\Omega (r) } \right] . 
\end{equation}
Substituting in (\ref{eq:soleqn2}) then gives
\begin{multline}
    0 = \left( r_{\mathrm {L}}^{2}-r_{+}^{2}\right) \left\{ 
    \frac{\Omega _{+}}{a}\left( r_{\mathrm {L}}^{2}+r_{+}^{2}\right)  -\frac{2Ma}{r_{+}^{2}r_{\mathrm {L}}^{2}\Omega (r_{\mathrm {L}}) } 
    \right. \\
 \left.     - 
    \frac{h(r_{\mathrm {L}})^{4}\Omega (r_{\mathrm {L}})^{2} \Omega _{+}^{2}}{4M^{2}a^{2}} \left(  r_{\mathrm {L}}^{2}+ r_{+}^{2}  \right) ^{2} \left( r_{\mathrm {L}}^{2}-r_{+}^{2} \right)
    \right\} .
\end{multline}
We further simplify using $h(r)^{4}\Omega (r)^{2} = 4M^{2}a^{2}/r^{4}$, and (\ref{eq:Omega}, \ref{eq:OmegaHalt}), then multiplying throughout by $r_{{\mathrm {L}}}^{4}$, we obtain the equation
\begin{align}
    0 &  = \left( r_{\mathrm {L}}^{2}-r_{+}^{2}\right) \left\{  
    r_{\mathrm {L}}^{6} \left( \frac{1}{L^{2}}- \Omega _{+}^{2} \right)
     + r_{\mathrm {L}}^{4} \frac{\Omega _{+}r_{+}^{2}}{a} \left( 1-a \Omega _{+} \right) 
    \right. \nonumber 
    \\  & \qquad  \left. 
    + r_{\mathrm {L}}^{2}\left( \Omega _{+}^{2}r_{+}^{4} - \frac{2Ma^{2}}{r_{+}^{2}} \right) + \Omega _{+}^{2} r_{+}^{6} 
   \right\}
    \nonumber \\ & = 
    \left( r_{\mathrm {L}}^{2}-r_{+}^{2}\right) \left\{ 
     r_{\mathrm {L}}^{6} \left( \frac{1}{L^{2}}- \Omega _{+}^{2} \right)
     + r_{\mathrm {L}}^{4} \frac{\Omega _{+}^{2}r_{+}^{6}}{2Ma^{2}}
     \right. \nonumber 
    \\  & \qquad  \left. 
     + 2Ma^{2}r_{\mathrm {L}}^{2} \left( \frac{1}{L^{2}} -\Omega _{+}^{2} \right) + \Omega _{+}^{2} r_{+}^{6} 
    \right\}  .
    \label{eq:intermediate1}
\end{align}
In the final step in (\ref{eq:intermediate1}), 
we have used the result $\Omega _{+}r_{+}^{4} = 2Ma \left( 1 - a\Omega _{+}\right)$, which follows from the definition (\ref{eq:OmegaH}), and also (\ref{eq:OmegaHalt}) again.
Multiplying (\ref{eq:intermediate1}) throughout by $L^{2}$, the cubic in $r_{\mathrm {L}}^{2}$ can be further factorized to give:
\begin{multline}
\label{eq:soleqnfinal}
\left( r_{\rm {L}}^{2}-r_{+}^{2}\right) \left( r_{\rm {L}}^{4} + 2Ma^{2}\right) \left[ \left( 1- \Omega _{+}^{2}L^{2} \right) r_{\rm {L}}^{2} + \frac{\Omega _{+}^{2}r_{+}^{6}L^{2}}{2Ma^{2}}\right]  \\ = 0,
\end{multline}
which is precisely Eq.~(\ref{eq:soleqn}).

\section{Addition theorems for spin-weighted spherical harmonics}
\label{sec:addition}
Below we state the addition theorems for spin-weighted spherical harmonics which enable us to simplify the SET components in App.~\ref{sec:SETcomp}.  
The results below are proven elsewhere \cite{Monteverdi:2024xyp}.
All the results below are valid for $\ell $ a positive integer or half-integer, and all $s$ an integer or half-integer, such that $\ell =|s|, |s|+1, |s|+2, \ldots $.

The addition theorems we require are:
\begin{subequations}
\label{eq:addition}
    \begin{align}
        \sum _{m=-\ell }^{\ell } {}_{s}Y_{\ell }^{m*}(\theta, \varphi ) {}_{s}Y_{\ell }^{m}(\theta, \varphi ) 
        & = \frac{2\ell + 1}{4\pi },
        \label{eq:additionVP}
        \\
        \sum _{m=-\ell }^{\ell }  m \left[ {}_{s}Y_{\ell }^{m*}(\theta, \varphi ) {}_{s}Y_{\ell }^{m}(\theta, \varphi ) \right]  
        & =- \frac{\left(2\ell + 1\right)s}{4\pi }\cos \theta ,
        \\
        \sum _{m=-\ell }^{\ell }  m^{2}\left[ {}_{s}Y_{\ell }^{m*}(\theta, \varphi ) {}_{s}Y_{\ell }^{m}(\theta, \varphi )  \right]  
        & = \frac{2\ell + 1}{8\pi }{\mathfrak {C}} ,
        \\
        \sum _{m=-\ell }^{\ell } {}_{s}Y_{\ell }^{m*}(\theta, \varphi ) \frac{\partial }{\partial \theta }{}_{s}Y_{\ell }^{m}(\theta, \varphi ) 
        & = 0,
        \\
         \sum _{m=-\ell }^{\ell } \frac{\partial }{\partial \theta }{}_{s}Y_{\ell }^{m*}(\theta, \varphi ) \frac{\partial }{\partial \theta }{}_{s}Y_{\ell }^{m}(\theta, \varphi ) 
        & =  \frac{2\ell + 1}{8\pi } \left( \ell ^{2} + \ell - s^{2} \right) ,
    \end{align}
    where
    \begin{equation}
    \label{eq:frakC}
        {\mathfrak {C}} =  \left(  \ell ^{2} +\ell -s^{2} \right) \sin ^{2} \theta + 2s^{2} \cos ^{2} \theta  .  
    \end{equation}
\end{subequations}
It is straightforward to check that in the case $s=0$ (ordinary spherical harmonics), the results (\ref{eq:addition}) reduce to those in, for example,  App.~C of Ref.~\cite{Balakumar:2022yvx}.

Since we have ${}_{s}Y_{\ell }^{m}(\theta, \varphi ) = e^{im\varphi }{}_{s}{\widetilde {Y}}_{\ell }^{m}(\theta )$ (\ref{eq:angsep}), the addition theorems in (\ref{eq:addition}) give 
\begin{subequations}
\label{eq:addition1}
    \begin{align}
        \sum _{m=-\ell }^{\ell } \left| {}_{s}{\widetilde {Y}}_{\ell }^{m}(\theta)\right| ^{2} 
        & = \frac{2\ell + 1}{4\pi },
        \\
        \sum _{m=-\ell }^{\ell } m \left| {}_{s}{\widetilde {Y}}_{\ell }^{m}(\theta)\right| ^{2} 
        & = \frac{\left(2\ell + 1\right)p}{8\pi }\cos \theta ,
        \\
        \sum _{m=-\ell }^{\ell } m^{2} \left| {}_{s}{\widetilde {Y}}_{\ell }^{m}(\theta)\right| ^{2}  
        & = \frac{2\ell + 1}{8\pi }{\mathfrak {C}} ,
        \\
        \sum _{m=-\ell }^{\ell }  {}_{s}{\widetilde {Y}}_{\ell }^{m*}(\theta)  {}_{s}{\widetilde {Y}}_{\ell }^{'m}(\theta) 
        & = 0,
        \\
         \sum _{m=-\ell }^{\ell } \left|  {}_{s}{\widetilde {Y}}_{\ell }^{'m}(\theta) \right| ^{2}
        & =  \frac{2\ell + 1}{32\pi } \left( 4\ell ^{2} + 4\ell - p^{2}\right) ,
    \end{align}
    where we have used the fact that $s=-p/2$ (\ref{eq:spin}), and where ${\mathfrak{C}}$ (\ref{eq:frakC}) is now
    \begin{equation}
          {\mathfrak {C}} = \frac{1}{4} \left(  4\ell ^{2} +4\ell -p^{2} \right) \sin ^{2} \theta + \frac{p^{2}}{2} \cos ^{2} \theta  .
        \label{eq:frakC1}
    \end{equation}
    \end{subequations}
    As a corollary of the results (\ref{eq:addition1}), we have
    \begin{multline}
        \sum _{m=-\ell }^{\ell }\left[ p \cot \theta - 2m  \csc \theta \right] ^{2}   {}_{s}{\widetilde {Y}}_{\ell }^{m}(\theta)^{2}  
        \\
        =
        \frac{2\ell + 1}{8\pi } \left(4\ell ^{2}+4\ell -p^{2}\right) .
    \end{multline}

\begin{widetext}
    
\section{Stress-energy tensor components}
\label{sec:SETcomp}
In this appendix we give explicit expressions for the components of the classical SET (\ref{eq:SET}) evaluated for a scalar field mode (\ref{eq:mode}) with angular function (\ref{eq:angsep}). 
Since these expressions are fairly lengthy, we omit the indices on the radial and angular functions.
In terms of the metric functions $f(r)$, $g(r)$, $h(r)$ and $\Omega (r)$ (\ref{eq:metricfunctionsstart}--\ref{eq:metricfunctionsend}), the SET components are:
\begin{subequations}
\label{eq:SETcomp1}
\begin{align}
    {}^{\aleph}T_{tt} & =  
    \omega ^{2} \left| X(r) \right| ^{2} {\widetilde {Y}}(\theta ) ^{2} 
    +\frac{2\xi}{g(r)^{2}} \left[ f(r)f'(r) - h(r) \Omega (r) \frac{d}{dr} \left\{ h(r)\Omega (r) \right\}  \right] \Re \left\{  X^{*}(r)X'(r) \right\} 
    {\widetilde {Y}}(\theta ) ^{2}
    \nonumber \\ & \qquad
    -\left[ f(r)^{2}-h(r)^{2}\Omega (r)^{2}\right]{\mathfrak {Z}}(r,\theta)  
    ,
    \\
    {}^{\aleph}T_{tr} & = -\omega \Im \left\{  X^{*}(r)X'(r) \right\} {\widetilde {Y}}(\theta ) ^{2}
    ,
    \\
    {}^{\aleph}T_{t\theta } & = 0,
    \\
    {}^{\aleph}T_{t\varphi } & = 
    - m\omega \left| X(r) \right| ^{2} {\widetilde {Y}}(\theta ) ^{2} 
    - \frac{2\xi }{r^{2}}h(r)^{2}\Omega (r) \left| X(r) \right| ^{2} {\widetilde {Y}}'(\theta ) {\widetilde {Y}}(\theta ) \sin \theta 
    \nonumber \\ & \qquad 
   + \frac{\xi h(r)}{2g(r)^{2}} \left[h(r) \Omega '(r) + 2h'(r) \Omega (r) \right] \Re \left\{  X^{*}(r)X'(r) \right\} {\widetilde {Y}}(\theta ) ^{2}  \cos \theta  
    -\frac{1}{2} h(r)^{2} \Omega (r) {\mathfrak {Z}}(r,\theta)  \cos \theta 
    ,
    \\
    {}^{\aleph}T_{t\psi } & = 
    -p\omega \left| X(r) \right| ^{2} {\widetilde {Y}}(\theta ) ^{2} 
    +\frac{\xi h(r)}{g(r)^{2}} \left[h(r) \Omega '(r) + 2h'(r) \Omega (r) \right] \Re \left\{  X^{*}(r)X'(r) \right\} {\widetilde {Y}}(\theta ) ^{2}    
    - h(r)^{2} \Omega (r) {\mathfrak {Z}}(r,\theta)  
    ,
    \\
    {}^{\aleph}T_{rr} & = 
    \left( 1 - 2\xi \right) \left| X'(r) \right| ^{2} {\widetilde {Y}}(\theta ) ^{2} 
    -2\xi  g(r)^{2}
    \left[ 
    \frac{1}{r^{2}h(r)^{2}}\left\{ p^{2}r^{2}  + \left( 4\lambda + r^{2}\nu ^{2}\right) h(r)^{2} \right\} 
    -  \frac{1}{f(r)^{2}}\left\{  \omega - p\Omega (r) \right\}  ^{2}
    \right] \left| X(r) \right| ^{2} {\widetilde {Y}}(\theta ) ^{2} 
    \nonumber \\ & \qquad
    +2\xi \left[ \frac{f'(r)}{f(r)}+ \frac{2h(r) +rh'(r)}{rh(r)} \right] \Re \left\{  X^{*}(r)X'(r) \right\}{\widetilde {Y}}(\theta ) ^{2} 
    + g(r)^{2}{\mathfrak {Z}}(r,\theta) 
    ,
    \\
    {}^{\aleph}T_{r\theta } & = 
   \left[  \left( 1 - 4\xi \right) \Re \left\{  X^{*}(r)X'(r) \right\}  + \frac{2\xi }{r} \left| X(r) \right| ^{2}
   \right] {\widetilde {Y}}'(\theta ) {\widetilde {Y}}(\theta ) 
    ,
    \\
    {}^{\aleph}T_{r\varphi } & =m \Im \left\{  X^{*}(r)X'(r) \right\} {\widetilde {Y}}(\theta ) ^{2}
    ,
    \\
    {}^{\aleph}T_{r\psi } & =p \Im \left\{  X^{*}(r)X'(r) \right\}  {\widetilde {Y}}(\theta ) ^{2}
    ,
    \\
    {}^{\aleph}T_{\theta \theta } & = 
    \left( 1 - 2\xi \right) \left| X(r) \right| ^{2} {\widetilde {Y}}'(\theta ) ^{2} 
    -\frac{\xi r}{2g(r)^{2}}\Re \left\{  X^{*}(r)X'(r) \right\}{\widetilde {Y}}(\theta ) ^{2}  
    +2\xi \left[  \lambda - \frac{1}{4}\left( p \cot \theta - 2m \csc \theta \right)^{2} \right] \left| X(r) \right| ^{2} {\widetilde {Y}}(\theta ) ^{2} 
    \nonumber \\ & \qquad 
   +2\xi \cot \theta \left| X(r) \right| ^{2} {\widetilde {Y}}'(\theta ) {\widetilde {Y}}(\theta ) 
    + \frac{1}{4}r^{2} {\mathfrak {Z}}(r,\theta) 
    ,
    \\
    {}^{\aleph}T_{\theta \varphi } & = 0,
    \\
    {}^{\aleph}T_{\theta \psi } & = 0,
    \\
    {}^{\aleph}T_{\varphi \varphi } & = 
     m^{2} \left| X(r) \right| ^{2} {\widetilde {Y}}(\theta ) ^{2}  
    -\frac{\xi }{2g(r)^{2}}\left[ r\sin ^{2}\theta + h(r)h'(r) \cos^{2}\theta  \right] \Re \left\{  X^{*}(r)X'(r) \right\}{\widetilde {Y}}(\theta ) ^{2} 
    \nonumber \\ & \qquad 
     +\frac{\xi }{r^{2}}\left[ h(r)^{2}-r^{2}\right] \left| X(r) \right| ^{2} {\widetilde {Y}}'(\theta ) {\widetilde {Y}}(\theta ) \sin (2\theta )
    +\frac{1}{4}\left[ r^{2}\sin ^{2} \theta + h(r)^{2}\cos ^{2} \theta  \right] {\mathfrak {Z}}(r,\theta) 
    ,
    \\
    {}^{\aleph}T_{\varphi \psi } & = 
    mp \left| X(r) \right| ^{2} {\widetilde {Y}}(\theta ) ^{2}  
   -\frac{\xi h(r)h'(r)}{g(r)^{2}} \Re \left\{  X^{*}(r)X'(r) \right\}{\widetilde {Y}}(\theta ) ^{2}  \cos \theta 
   +\frac{2\xi h(r)^{2}}{r^{2}}\left| X(r) \right| ^{2} {\widetilde {Y}}'(\theta ) {\widetilde {Y}}(\theta )  \sin \theta 
   \nonumber \\ & \qquad 
    +\frac{1}{2}h(r)^{2}{\mathfrak {Z}}(r,\theta) \cos \theta  
    ,
    \\
    {}^{\aleph}T_{\psi \psi } & = 
     p^{2} \left| X(r) \right| ^{2} {\widetilde {Y}}(\theta ) ^{2}  
    -\frac{2\xi h(r)h'(r)}{g(r)^{2}} \Re \left\{  X^{*}(r)X'(r) \right\}{\widetilde {Y}}(\theta ) ^{2} 
    + h(r)^{2}{\mathfrak {Z}}(r,\theta) 
    ,
\end{align}
where
\begin{align}
{\mathfrak {Z}}(r,\theta)   
= &~ \left( 2 \xi - \frac{1}{2}\right) g^{\rho \sigma }\Phi _{;\rho } \Phi _{;\sigma }
+ \frac{1}{2} \left( 4\xi - 1 \right)  \mu ^{2} \Phi ^{2}
    + \xi  R \left( 2 \xi - \frac{3}{10}\right)  \Phi ^{2} 
\nonumber \\  
 = & ~\left( 2\xi - \frac{1}{2} \right) \left[  \left\{ 
\dfrac{p^{2}}{h(r)^{2}} 
+ \dfrac{1}{r^{2}} \left[ p \cot \theta - 2m  \csc \theta \right] ^{2} 
-  \dfrac{1}{f(r)^{2}}\left[ \omega - p\Omega (r) \right] ^{2}
\right\} \left| X(r) \right| ^{2} {\widetilde {Y}}(\theta ) ^{2}  \right. 
\nonumber \\ & \qquad
\left. + \dfrac{1}{g(r)^{2}} \left| X'(r) \right| ^{2} {\widetilde {Y}}(\theta ) ^{2}
+ \dfrac{4}{r^{2}} \left| X(r) \right| ^{2} {\widetilde {Y}}'(\theta ) ^{2} \right] 
+ \left[ \frac{1}{2} \left( 4\xi - 1 \right)  \mu ^{2} + \xi  R \left( 2 \xi - \frac{3}{10}\right) \right] \left| X(r) \right| ^{2} {\widetilde {Y}}(\theta ) ^{2} .
\end{align} 
\end{subequations}
In (\ref{eq:SETcomp1}), the radial functions $X(r)$ depend on the frequency $\omega $ and the azimuthal quantum number $p\in {\mathbb{Z}}$, while the angular functions ${\widetilde {Y}}(\theta )$ depend on the quantum number $m=-\ell , -\ell+ 1 , \ldots , \ell - 1, \ell $, the spin $s=-p/2$ (\ref{eq:spin}) and the quantum number $\ell = |s|, |s|+1, |s|+2, \ldots $.

We now use the addition theorems for the spin-weighted spherical harmonics from App.~\ref{sec:addition} (\ref{eq:addition1}) to perform the sum over $m$ in each of the SET components.
We define new quantities $t_{\mu \nu }$ by
\begin{equation}
    \sum _{m=-\ell }^{\ell } {}^{\aleph}T_{\mu \nu } = \frac{2\ell + 1}{4\pi } t_{\mu \nu },
\end{equation}
where ${}^{\aleph}T_{\mu \nu }$ are the components given in (\ref{eq:SETcomp1}).
A further simplification arises from the fact that, for our particular modes (\ref{eq:Xforms}), we have $\Im \left\{  X^{*}(r)X'(r) \right\} =0 $.
Then we have
\begin{subequations}
\label{eq:SETcomp2}
\begin{align}
    t_{tt} & =  
    \omega ^{2} \left| X(r) \right| ^{2} 
    +\frac{2\xi}{g(r)^{2}} \left[ f(r)f'(r) - h(r) \Omega (r) \frac{d}{dr} \left\{ h(r)\Omega (r) \right\}  \right] \Re \left\{  X^{*}(r)X'(r) \right\} 
    -\left[ f(r)^{2}-h(r)^{2}\Omega (r)^{2}\right]{\widetilde {{\mathfrak {Z}}}}(r)  
    ,
    \\
    t_{tr} & = 0
    ,
    \\
    t_{t\theta } & = 0,
    \\
    t_{t\varphi } & = \left\{ 
    \frac{1}{2}p\omega \left| X(r) \right| ^{2}  
   + \frac{\xi h(r)}{2g(r)^{2}} \left[h(r) \Omega '(r) + 2h'(r) \Omega (r) \right] \Re \left\{  X^{*}(r)X'(r) \right\} 
    -\frac{1}{2} h(r)^{2} \Omega (r){\widetilde {{\mathfrak {Z}}}}(r)   \right\}  \cos \theta 
    ,
    \\
    t_{t\psi } & = 
    -p\omega \left| X(r) \right| ^{2}
    +\frac{\xi h(r)}{g(r)^{2}} \left[h(r) \Omega '(r) + 2h'(r) \Omega (r) \right] \Re \left\{  X^{*}(r)X'(r) \right\}  
    - h(r)^{2} \Omega (r) {\widetilde {{\mathfrak {Z}}}}(r)    
    ,
    \\
    t_{rr} & = 
    \left( 1 - 2\xi \right) \left| X'(r) \right| ^{2}
    -2\xi  g(r)^{2}
    \left[ 
    \frac{1}{r^{2}h(r)^{2}}\left\{ p^{2}r^{2}  + \left( 4\lambda + r^{2}\nu ^{2}\right) h(r)^{2} \right\} 
    -  \frac{1}{f(r)^{2}}\left\{  \omega - p\Omega (r) \right\}  ^{2}
    \right] \left| X(r) \right| ^{2} 
    \nonumber \\ & \qquad
    +2\xi \left[ \frac{f'(r)}{f(r)}+ \frac{2h(r) +rh'(r)}{rh(r)} \right] \Re \left\{  X^{*}(r)X'(r) \right\}
    + g(r)^{2}{\widetilde {{\mathfrak {Z}}}}(r)  
    ,
    \\
    t_{r\theta } & = 0
    ,
    \\
    t_{r\varphi } & =0
    ,
    \\
    t_{r\psi } & =0
    ,
    \\
    t_{\theta \theta } & = 
    \frac{1}{8}\left( 1 - 4\xi \right)\left( 4\ell ^{2} + 4\ell - p^{2}\right)  \left| X(r) \right| ^{2} 
    +2\xi  \lambda \left| X(r) \right| ^{2} 
    -\frac{\xi r}{2g(r)^{2}}\Re \left\{  X^{*}(r)X'(r) \right\} 
    + \frac{1}{4}r^{2} {\widetilde {{\mathfrak {Z}}}}(r)   
    ,
    \\
    t_{\theta \varphi } & = 0,
    \\
    t_{\theta \psi } & = 0,
    \\
    t_{\varphi \varphi } & = 
     \frac{1}{2}{\mathfrak{C}} \left| X(r) \right| ^{2} {\widetilde {Y}}(\theta ) ^{2}  
    -\frac{\xi }{2g(r)^{2}}\left[ r\sin ^{2}\theta + h(r)h'(r) \cos^{2}\theta  \right] \Re \left\{  X^{*}(r)X'(r) \right\}
    +\frac{1}{4}\left[ r^{2}\sin ^{2} \theta + h(r)^{2}\cos ^{2} \theta  \right] {\widetilde {{\mathfrak {Z}}}}(r)   
    ,
    \\
    t_{\varphi \psi } & = -\left\{ 
    \frac{1}{2}p^{2} \left| X(r) \right| ^{2}  
   +\frac{\xi h(r)h'(r)}{g(r)^{2}} \Re \left\{  X^{*}(r)X'(r) \right\}  
    -\frac{1}{2}h(r)^{2}{\widetilde {{\mathfrak {Z}}}}(r) \right\}   \cos \theta  
    ,
    \\
    t_{\psi \psi } & = 
     p^{2} \left| X(r) \right| ^{2} 
    -\frac{2\xi h(r)h'(r)}{g(r)^{2}} \Re \left\{  X^{*}(r)X'(r) \right\}
    + h(r)^{2}{\widetilde {{\mathfrak {Z}}}}(r)  
    ,
\end{align}
where ${\mathfrak{C}}$ is given by (\ref{eq:frakC1}) and
\begin{multline}
{\widetilde {{\mathfrak {Z}}}}(r)  
= \left\{ 
\dfrac{p^{2}}{h(r)^{2}} 
+ \dfrac{1}{2r^{2}}  \left( 4\ell ^{2} + 4\ell -p^{2} \right)
-  \dfrac{1}{f(r)^{2}}\left[ \omega - p\Omega (r) \right] ^{2}
+ \frac{1}{2} \left( 4\xi - 1 \right)  \mu ^{2} + \xi  R \left( 2 \xi - \frac{3}{10}\right)
\right\} \left| X(r) \right| ^{2} 
 \\  
+ \dfrac{1}{g(r)^{2}} \left| X'(r) \right| ^{2} 
+ \dfrac{1}{2r^{2}} \left( 4\ell ^{2} + 4\ell - p^{2}\right) \left| X(r) \right| ^{2},
\end{multline}
\end{subequations}
and we have simplified using the result (\ref{eq:lambda}).  
It can be seen from (\ref{eq:SETcomp2}) that all the dependence on the angle $\theta $ is now determined in closed form. 

We now wish to compare the components (\ref{eq:SETcomp2}) with the form of the SET (\ref{eq:SETsimp1}) derived from symmetry principles. 
Using the metric (\ref{eq:metric}) to lower the indices on (\ref{eq:SETsimp1}), we find
\begin{subequations}
\label{eq:SETsimpdown}
    \begin{align}
    \langle {\hat {T}}_{tt} \rangle & = 
   \left[ f(r)^{2}-h(r)^{2}\Omega (r)^{2} \right] ^{2}  {\mathcal {F}}^{tt}(r)  
   +2 h(r)^{2}\Omega (r) \left[ f(r)^{2}-h(r)^{2}\Omega (r)^{2} \right] {\mathcal {F}}^{t\psi }(r) 
   + h(r)^{4}\Omega (r)^{2} {\mathcal {F}}^{\psi \psi }(r)
   \nonumber \\ & \qquad + \frac{1}{4}h(r)^{4}\Omega (r)^{2}{\mathcal {F}}^{\theta \theta }(r)
   ,
    \\
    \langle {\hat {T}}_{tr} \rangle & = 
    - g(r)^{2} \left[ f(r)^{2}-h(r)^{2}\Omega (r)^{2} \right] {\mathcal {F}}^{tr}(r) - g(r)^{2}h(r)^{2}\Omega (r) {\mathcal {F}}^{r\psi }(r)
    ,
    \\
    \langle {\hat {T}}_{t\theta } \rangle & = 0, 
    \\
    \langle {\hat {T}}_{t\varphi } \rangle & = 
    \frac{1}{2} h(r)^{2} \bigg\{ 
    \Omega (r) \left[ f(r)^{2}- h(r)^{2}\Omega (r) ^{2} \right] {\mathcal {F}}^{tt}(r)  
    -\left[ f(r)^{2}- 2h(r)^{2}\Omega (r) ^{2} \right] {\mathcal {F}}^{t\psi }(r) 
    -h(r)^{2}\Omega (r) {\mathcal {F}}^{\psi \psi }(r)
   \nonumber \\ & \qquad  
     - \frac{1}{4} h(r)^{2} \Omega (r) {\mathcal {F}}^{\theta \theta }(r) 
     \bigg\} \cos \theta 
    ,
    \\
    \langle {\hat {T}}_{t\psi } \rangle & = h(r)^{2} \bigg\{ 
    \Omega (r) \left[ f(r)^{2}- h(r)^{2}\Omega (r) ^{2} \right] {\mathcal {F}}^{tt}(r)  
    -\left[ f(r)^{2}- 2h(r)^{2}\Omega (r) ^{2} \right] {\mathcal {F}}^{t\psi }(r) 
    -h(r)^{2}\Omega (r) {\mathcal {F}}^{\psi \psi }
    \nonumber \\ &  \qquad 
     - \frac{1}{4} h(r)^{2} \Omega (r) {\mathcal {F}}^{\theta \theta }(r) 
    \bigg \} 
     ,
    \\
    \langle {\hat {T}}_{rr} \rangle & = g(r)^{4}{\mathcal {F}}^{rr}(r)
    ,
    \\
    \langle {\hat {T}}_{r\theta } \rangle & = 0,
    \\
    \langle {\hat {T}}_{r\varphi } \rangle & =\frac{1}{2} g(r)^{2}h(r)^{2} \left\{  {\mathcal {F}}^{r\psi }(r) - \Omega (r) {\mathcal {F}}^{tr}(r) \right\}  \cos \theta ,
    \\
    \langle {\hat {T}}_{r\psi } \rangle & = g(r)^{2}h(r)^{2} \left\{  {\mathcal {F}}^{r\psi }(r) - \Omega (r) {\mathcal {F}}^{tr}(r) \right\} ,
    \\
    \langle {\hat {T}}_{\theta \theta } \rangle & = \frac{r^{4}}{16}{\mathcal {F}}^{\theta \theta }(r), 
    \\
    \langle {\hat {T}}_{\theta \varphi } \rangle & = 0,
    \\
    \langle {\hat {T}}_{\theta \psi } \rangle & = 0, 
    \\
    \langle {\hat {T}}_{\varphi \varphi } \rangle & = 
    \frac{1}{4}h(r)^{4} \left\{ \Omega (r)^{2} {\mathcal {F}}^{tt}(r)  - 2\Omega (r) {\mathcal {F}}^{t\psi }(r) + {\mathcal {F}}^{\psi \psi }(r)\right\} \cos ^{2} \theta 
    + \frac{1}{16} \left[ h(r)^{4}\cos ^{2} \theta + r^{4} \sin ^{2}\theta  \right] {\mathcal {F}}^{\theta \theta }(r) ,
    \\
    \langle {\hat {T}}_{\varphi \psi } \rangle & = 
    \frac{1}{2}h(r)^{4} \left\{ \Omega (r)^{2} {\mathcal {F}}^{tt}(r)  - 2\Omega (r) {\mathcal {F}}^{t\psi }(r) + {\mathcal {F}}^{\psi \psi }(r)\right\} \cos  \theta 
    + \frac{1}{8} h(r)^{4} {\mathcal {F}}^{\theta \theta }(r) \cos \theta ,
    \\
    \langle {\hat {T}}_{\psi \psi } \rangle & =  h(r)^{4} \left\{ \Omega (r)^{2} {\mathcal {F}}^{tt}(r)  - 2\Omega (r) {\mathcal {F}}^{t\psi }(r) + {\mathcal {F}}^{\psi \psi }(r)\right\}  
    +\frac{1}{4} h(r)^{4} {\mathcal {F}}^{\theta \theta }(r),
\end{align}
\end{subequations}
from which it is clear that the dependence on the angle $\theta $ in all components is of the same form.
Let ${\mathfrak {F}}^{\bullet \bullet}$ be the classical mode contribution to ${\mathcal {F}}^{\bullet \bullet} $ arising from the scalar field mode (\ref{eq:mode}), with the sum over $m$ completed.
To find the ${\mathfrak {F}}^{\bullet \bullet}$, it is simplest to use the inverse metric (\ref{eq:inversemetric}) to raise both indices on the components in (\ref{eq:SETcomp2}) and then compare with (\ref{eq:SETsimp1}). 
This gives
\begin{subequations}
\label{eq:frakF}
    \begin{align}
        {\mathfrak {F}}^{tt}(r) & =
        \frac{1}{f(r)^{4}}\left[ \omega - p\Omega (r) \right]^{2}\left| X(r) \right| ^{2}
        + \frac{2\xi f'(r)}{f(r)^{3}g(r)^{2}}\Re \left\{  X^{*}(r)X'(r) \right\} 
        -\frac{1}{f(r)^{2}} {\widetilde {\mathfrak {Z}}}(r) 
        ,
        \\
        {\mathfrak {F}}^{tr}(r) & = 0
        ,
        \\ 
        {\mathfrak {F}}^{t\psi }(r) & = 
        \frac{1}{f(r)^{4}h(r)^{2}} \left\{ \omega - p\Omega (r) \right\}  \left\{ pf(r)^{2}+h(r)^{2}\Omega (r) \left[ \omega - p\Omega (r) \right]  \right\}  \left| X(r) \right| ^{2}
        \nonumber \\ & \qquad 
         -\frac{\xi }{f(r)^{3}g(r)^{2}} \left\{ f(r) \Omega '(r) - 2f'(r) \Omega (r)  \right\} \Re \left\{  X^{*}(r)X'(r) \right\} 
        -\frac{\Omega (r)}{f(r)^{2}}{\widetilde {\mathfrak {Z}}}(r) 
        ,
        \\
        {\mathfrak {F}}^{rr} (r) & = 
        \frac{1}{g(r)^{4}}\left( 1 -2\xi \right) \left| X'(r) \right| ^{2}
         - 2\xi \left[ 
       \frac{1}{g(r)^{2}} \left\{ \frac{4\lambda }{r^{2}} + \nu ^{2} + \frac{p^{2}}{h(r)^{2}} -\frac{\left[ \omega - p\Omega (r)\right] ^{2}}{f(r)^{2}}  \right\}  \left| X(r) \right| ^{2}
       \right. \nonumber \\ & \qquad \left. 
        -\frac{1}{g(r)^{4}} \left\{  \frac{2}{r} + \frac{f'(r)}{f(r)} + \frac{h'(r)}{h(r)} \right\}   \Re \left\{  X^{*}(r)X'(r) \right\} 
        \right]
        +\frac{1}{g(r)^{2}} {\widetilde {\mathfrak {Z}}}(r) 
        ,
        \\
        {\mathfrak {F}}^{r\psi }(r) & = 0
        ,
        \\
        {\mathfrak {F}}^{\theta \theta } (r) & = 
        \frac{2}{r^{4}}\left( 1 -4\xi \right) \left[ 4\ell ^{2}+ 4\ell - p^{2} \right] \left| X(r) \right| ^{2}
         -8\xi  \left[
        \frac{1}{r^{3}g(r)^{2}}\Re \left\{  X^{*}(r)X'(r) \right\} 
        - \frac {4\lambda }{r^{4}}\left| X(r) \right| ^{2}
        \right]
         + \frac{4}{r^{2}}{\widetilde {\mathfrak {Z}}}(r) 
        ,
        \\
        {\mathfrak {F}}^{\psi \psi }(r)  & = 
         \frac{1}{f(r)^{4}h(r)^{4}} \left\{ 
        p f(r)^{2} + h(r)^{2}\Omega (r)  \left[ \omega - p \Omega (r)  \right]
        \right\}  ^{2} \left| X(r) \right| ^{2}
        \nonumber \\ & \qquad 
        - \frac{2\xi }{g(r)^{2}}\left\{  \frac{h'(r)}{h(r)^{3}} + \frac{\Omega (r)\Omega '(r)}{f(r)^{2}} - \frac{f'(r) \Omega (r)^{2}}{f(r)^{3}} - \frac{1}{r^{3}}  \right\}  \Re \left\{  X^{*}(r)X'(r) \right\} 
        + \left[  \frac{1}{h(r)^{2}}- \frac{\Omega (r)^{2}}{f(r)^{2}} \right] {\widetilde {\mathfrak {Z}}}(r) 
        ,
    \end{align}
\end{subequations}
where the above expressions will need to be multiplied by an overall factor of $\left(2\ell + 1 \right)/4\pi $ in the final mode sums. 
\end{widetext}

\bibliography{adS5}

\end{document}